\documentclass[preprintnumbers,amsmath,amssymb,floatfix]{revtex4}
\usepackage{graphicx,color}
\usepackage{float}
\usepackage{wrapfig}
\usepackage{bm}
\usepackage{mathrsfs}
\usepackage{color}
\usepackage{slashed}
\usepackage{dcolumn}
\usepackage{extarrows}
\usepackage[colorlinks,linkcolor=blue,citecolor=green]{hyperref}
\newcommand{\dd}{\textrm{\,d}}

\begin{document}

\title{Analysis on hadron spectra in heavy-ion collisions with a new non-extensive approach}

\author{Ke-Ming Shen}
\email{shen_keming.ecut@hotmail.com}
\affiliation{School of Science, East China University of Technology, Nanchang 330013, China}
\affiliation{Wigner Research Centre for Physics of Hungarian Academy of Sciences, Budapest, Hungary}

\date{\today}

\begin{abstract}

The transverse momentum spectra of identified charged hadrons stemming from high energy collisions at different beam energies are described by a new non-extensive distribution, the Kaniadakis $\kappa$-distribution, with respect to the constraints in non-extensive quantum statistics. 
All fittings are also compared with the Tsallis distributions as well as the usual Boltzmann-Gibbs one. $\chi^2/ndf$ is also used to test the fitting goodness of all functions.
Our results show that these different non-extensive approaches can be well applied in high energy collisions rather than the classical one. 
The Kaniadakis statistics is typically better applied into such systems with both positive and negative particles considered. 
This provides an alternative non-extensive view to study high energy physics. 
Analysis on the fitting parameters are present as well. 
The similar relationships of all functions remind us of the further understanding of the non-extensivity.

\end{abstract}

\maketitle

\section{Introduction}

Nowadays more and more attentions have been paid on the analysis of the transverse momentum ($p_T$) spectra in proton-proton and heavy-ion collisions in the non-extensive approach~\cite{A-07,A-10,PHENIX-11,K-10,A-11,Aa-11,C-11}. 
As a basic quantity measured in experiments, $p_T$ spectrum can reveal useful information on the dynamics of the colliding systems. 
However, it has been realized that data on many single-particle distributions show a power-law than exponential behavior, which one does not expect from the usual statistical models based on Boltzmann\,--\,Gibbs (BG) statistics~\cite{UBB-11,LWL-13,WW-12,WW-13,RW-14,K-18}.
In addition, people find it inadequate to apply tools from BG statistical physics in high energy collisions when it is far from the thermodynamical limit of equilibrium, such as the number of particles is much smaller than the Avogadro number ($N\ll N_A$) and fluctuation effects strongly influence the final-state particle energy distributions~\cite{BBBUT-17}.

More specifically, due to the high multiplicities produced in high energy collisions, even in the elementary proton-proton ($pp$) collisions, one could use the statistical models to study the mechanism correspondingly~\cite{H-65, GGP-11, KTSC-17}.
The usual BG statistics, on the other hand, could not describe the identified particle spectra at the experimental groups at high $p_T$ region.
There have been numerous descriptions of single inclusive hadron spectra via models encoding modified statistics, such as Tsallis distributions and variants thereof~\cite{A-07,A-10,PHENIX-11,K-10,A-11,Aa-11,C-11,UBB-11,LWL-13,WW-12,WW-13,RW-14,BBBUT-17}. 
It introduces the spectra described by
\begin{eqnarray}
E\frac{\dd^3\sigma}{\dd^3p}\propto\exp_q (-\frac{m_T-m}{T})
\label{q-spec}
\end{eqnarray}
with the transverse mass $m_T=\sqrt{m^2+p_T^2}$ including the rest mass, $m$, of particle and the generalized Tsallis $q-$exponential distribution \cite{T-88},
\begin{eqnarray}
\exp_q(x):=[1+(1-q)x]^{\frac{1}{1-q}} ~,
\label{q-exp}
\end{eqnarray}
which is easily proved to recover the normal exponential form when $q\to 1$. 
In this work we study the hadron spectra within the $m_T-m$ scaling.
This form has shown nice fits to the spectra of identified hadrons and charged hadrons over a wide range~\cite{Shen-2019}.
Note that the term $E=m_T-m$ in Eq.(\ref{q-spec}) instead of $p_T$ itself is considered to account for various charged hadrons~\cite{WW-12} and the effects over a large $p_T$ range~\cite{Ab-07}.
This $q$-exponential was firstly proposed by V. Pareto in 1896~\cite{P-96} to study the distribution of wealth, and then promoted by C. Tsallis~\cite{T-88} in connection with the non-extensive entropy thermodynamically.
From 2002 the so-called superstatistics was proposed by C. Beck and became well applied into many cases~\cite{Beck-02}, which gave out another generalized probability distribution function as $\exp^q_q(x)$ based on the Tsallis $q$-exponential distribution. 

Nevertheless, further constraints arise in the non-extensive quantum statistics when both positive and negative particles are considered, such as the generalized KMS relation from particle-hole CPT symmetry~\cite{BSZ-15}. 
It means that a missing negative energy particle state is equivalent with the corresponding positive energy hole state, namely, the connection to the canonical thermodynamical weight factor should satisfy
\begin{eqnarray}
f(E)\cdot f(-E)=1~.
\end{eqnarray}
The $q-$exponential function of Eq.(\ref{q-exp}), however, does not generally follow this relation,
\begin{eqnarray}
\exp_q(x)\cdot \exp_q(-x)&&=[1+(1-q)x]^{\frac{1}{1-q}}\cdot 
[1-(1-q)x]^{\frac{1}{1-q}} \nonumber \\
&&\neq 1 ~.
\label{KMS}
\end{eqnarray}
Therefore, special attention should be paid when investigating on the applications of the $p_T$ spectra of Bosons and Fermions in high energy physics, especially fittings on the spectra of $X+\bar{X}$ (particles and their anti-particles) stemming from both elementary and heavy-ion collisions.

In order to deal with such a situation, G. Kaniadakis~\cite{K-01} considered this symmetry and gave out another non-extensive approach with the deformed $\kappa$-exponential function,
\begin{eqnarray}
\exp_{\kappa}(x):=[\sqrt{1+(\kappa x)^2}+\kappa x]^{\frac{1}{\kappa}} ~.
\label{k-exp}
\end{eqnarray}
This $\kappa -$deformed non-extensive statistics has also been used in various kinds of
fields \cite{K-01,CHM-07,OBT-15,K-09}. 

This paper is organized as follows: in the 2nd Section, we will briefly introduce the framework of Kaniadakis' $\kappa$ statistics.
Using this generalized $\kappa$-exponential distribution, we demonstrate the $\kappa$ fittings to the $p_T$ spectra in both $pp$ collisions and heavy-ion collisions at different energies in Section III.
Section III also shows the results of Tsallis' distribution $\exp_q(x)$ and Beck's $\exp_q^q(x)$ together with the BG case as comparisons. 
The fitting parameters are analyzed as well.
Finally we close our paper with the summary and outlook in Section IV.


\section{$\kappa-$Statistics}

For the description of relativistic plasmas \cite{K-01, K-09, Ka-09}, G. Kaniadakis firstly proposed the $\kappa-$deformed exponential function, cf. Eq.(\ref{k-exp}), and its inverse function is given as
\begin{eqnarray}
\ln_{\kappa}(x):=\frac{x^\kappa -x^{-\kappa}}{2\kappa} ~.
\label{k-ln}
\end{eqnarray}
Accordingly, the Kaniadakis entropy is given as
\begin{eqnarray}
S_\kappa =-k_B\sum_i p_i \ln_\kappa p_i
\label{k-entropy}
\end{eqnarray}
which has the standard properties of BG entropy: it is thermodynamically stable, Lesche stable, and obeys the Khinchin axioms of continuity, maximality, expandability and generalized additivity. \cite{K-01}

Here we will list some basic properties as follows:
\begin{eqnarray}
\exp_{0}(x)=\exp (x)
\label{e70-2}
\end{eqnarray}
\begin{eqnarray}
\frac{\dd}{\dd x}\exp_{\kappa}(x)>0
\label{e70-4}
\end{eqnarray}
\begin{eqnarray}
\exp_{\kappa}(x)\exp_{\kappa}(-x)=1
\label{e70-9}
\end{eqnarray}
Consider its power law asymptotic behaviour
\begin{eqnarray}
\exp_{\kappa}(x) \stackrel{x\rightarrow \pm \infty}{\sim} | 2\kappa x |^{\pm 1/ {| \kappa |}}
\label{e70-12}
\end{eqnarray}
which owns the similar property as Tsallis $q$-exponential function. 
This reminds us to apply it into physical systems where the Tsallis $q-$exponential has been used especially the power-law tail of $p_T$ spectra in high energy collisions~\cite{Shen-2019}.
We will then investigate this $\kappa-$deformed non-extensive distribution on the $p_T$ spectra of identified hadrons and charged particles in various kinds of collisions at different energies as well as the Tsallis and BG distributions.


\section{Results and Discussion}

In order to investigate the hadron spectra in high energy physics, one has to disentangle hard QCD and soft collective effects and test whether the results agree with the thermal assumption. 
In this work we focus on fitting the transverse momentum spectra within the most well-applied $m_T-m$ scaling:
\begin{eqnarray}
\frac{\dd^2 N}{N_{ev}2\pi p_T\dd y\dd p_T}=f(m_T-m)~.
\end{eqnarray}
Our aim is to figure out whether this $\kappa$ statistics well applies into high energy collisions, while investigating the differences with the other two (Tsallis and Beck) non-extensive approaches and the usual BG one.
Specifically, we analyzed hadron spectra from proton-proton, proton-lead and lead-lead collisions at different energies within the following four different fittings expressions:
\begin{eqnarray}
f_{Ts}&&=A_1\cdot \left(1+\frac{m_T-m}{n_1T_1}\right)^{-n_1} ~,\nonumber \\
f_{Be}&&=A_2\cdot\left(1+\frac{m_T-m}{n_2T_2}\right)^{-n_2-1} ~,\nonumber \\
f_{Ka}&&=A_3\cdot \left[\sqrt{1+(\frac{m_T-m}{n_3T_3})^2}+\frac{m_T-m}{n_3T_3}\right]^{-n_3} ~,\nonumber \\
f_{BG}&&=A_4\cdot \exp\left(-\frac{m_T-m}{T_4}\right)~,
\label{fit-fun}
\end{eqnarray}
where the 1st is the previous Tsallis $q-$exponential function with $n_1=\frac{1}{q-1}$, the 2nd is the superstatistical one (or the Beck distribution) with $n_2=\frac{1}{q-1}$, the 3rd is the Kappa distribution with $n_3=\frac{1}{\kappa}$ and the 4th the classical BG form. 
We considered all of these parameters free, the normalization parameter, $A_i$, the fitting temperature, $T_i$, and the non-extensive parameter, $n_i$, ($i=1,2,3,4$).
For the fit procedure, the minimal chi-square method~\cite{Gabor-2018} was used to fit all the present experimental data~\cite{pp900-1, pp900-2, pp7-1,pp7-2,pA-1, pA-2, pA-3,AA-1, AA-2, AA-3, AA-4} within the mathematica program.
Both the statistical and systematic uncertainties were considered for data sets.
Note that the Beck non-extensive distribution could be connected with the Tsallis one by a change in the power index~\cite{Shen-2019}:
\begin{eqnarray}
\frac{q}{1-q}=\frac{1}{1-q}+1~.
\label{TBq}
\end{eqnarray}
This is why the power index of the Beck function is denoted as $-n_2-1$.
Nevertheless, it is worthy to firstly analyze both of them for comparisons and better understanding theoretically and experimentally.

\subsection{$pp$ collisions}

\begin{table}[htb]
\caption{Fitting range [GeV/c] of $p_T$ in hadron spectra in $pp$ collisions~\cite{pp900-1, pp900-2, pp7-1,pp7-2} for different charged particles:}
\scalebox{1.2}[1.3]{
\begin{tabular}{c c c c}
\hline
\hline
particles &  mass [$GeV/c^2$] & 900 GeV & 7 TeV \\
\hline
$\pi$ & 0.140 & 0.1-2.6 & 0.1-20 \\

$K$ & 0.494 &  0.2-2.4 & 0.2-20 \\

$K_S^0$ & 0.498 & 0.2-3.0 &  ~~ \\

$p$ & 0.938 &  0.35-2.4  & 0.3-20 \\

$\Lambda$ & 1.116 &  0.6-3.5 & ~~ \\

$\Xi$ &  1.321 &    &  0.6-8.5 \\

$\Omega$ &  1.672 &    &  0.8-5.0 \\
\hline
\hline
\end{tabular}
}
\label{tabpppT}
\end{table}

In high energy physics, $pp$ (proton-proton) collision is recognized as the elementary process and has also been performed and measured under different energies within Tsallis $q-$exponential distributions \cite{TXRBWX-09,WWCT-15}.
In the present paper we will firstly pay attention on the fittings within different kinds of non-extensive approaches on the $p_T$ spectra for $pp$ collisions at different energies.

In this section, we focus on the fittings of transverse momentum distributions not only for $\frac{\pi^++\pi^-}{2}$, $\frac{K^++K^-}{2}$, $\frac{p+\bar{p}}{2}$ in $pp$ collisions at 900 GeV and 7 TeV, but also for $K_S^0$, $\frac{\Lambda+\bar{\Lambda}}{2}$ at 900 GeV and $\frac{\Xi^++\Xi^-}{2}$, $\frac{\Omega^++\Omega^-}{2}$ at 7 TeV. 
Data are taken from the ALICE Collaboration~\cite{pp900-1, pp900-2, pp7-1,pp7-2}.
Detailed fitting $p_T$ ranges are listed in Table \ref{tabpppT}.
Note that a large body of data points from both LHC and RHIC have been well analyzed within the Tsallis non-extensive approach~\cite{entropy-2017}.
In this work we have also fitted plenty of data from different experimental groups.
This paper, on the other hand, lists the results and parameters from the ALICE Collaborations when considering both positive and negative particles at various kinds of beam energies.

All the fitting parameters by these four different functions are tabulated in Table~\ref{tabpppar} for fitting all kinds of data obtained in $pp$ collisions at both 900 GeV and 7 TeV. 
The same values of normalization constants, $A_1$ and $A_2$, indeed tell us that there is no big difference between the first two non-extensive functions.
Their non-extensive parameters, on the other hand, do follow the above connection, $n_1=n_2+1$, cf. Eq.(\ref{TBq}).
Note that for the BG fitting results, we set its non-extensive parameters as $n_4=10^{10}$ since theoretically it should be infinity.
All the results are shown without the corresponding error bars for simplicity. 

\begin{table}[htb]
\caption{Fitting parameters of all four fitting functions on $p_T$ spectra in $pp$ collisions (for the BG case the non-extensive parameter $1/n_4$ vanishes so that we set $n_4=10^{10}$)}
\scalebox{1.15}[1.2]{
\begin{tabular}{c c c c c c c c c c c c c c}
\hline
\hline
$\sqrt{s}$ & hadron & $A_1$ & $A_2$ & $A_3$ & $A_4$ & $T_1$ & $T_2$ & $T_3$ & $T_4$ & $n_1$ & $n_2$ & $n_3$ & $n_4$ \\
\hline
~ & $\pi$ & 5.294 & 5.294 & 4.669 & 2.812 & 0.1265  &  0.1449  & 0.1532 & 0.2183 & 7.894 & 6.894 & 5.011 & $10^{10}$ \\

~ & $K$ & 0.208 & 0.208 & 0.188 & 0.133 & 0.1598 & 0.1915 & 0.1939 & 0.2686 & 6.039 & 5.039 & 3.791 & $10^{10}$ \\

900 GeV & $p$ & 0.051 & 0.051 & 0.047 & 0.038 & 0.1833 & 0.2121 & 0.2131 & 0.2635 & 7.371 & 6.371 & 4.126 & $10^{10}$ \\

~ & $\Lambda$ & 0.022 & 0.022 & 0.019 & 0.014 & 0.2224 & 0.2473 & 0.2602 & 0.3229 & 9.926 & 8.926 & 5.565 & $10^{10}$ \\

~ & $K^0_S$ & 0.199  & 0.199 & 0.169 & 0.093 & 0.1657 & 0.1965 & 0.2088 & 0.3085 & 6.373 & 5.373 & 4.225 & $10^{10}$ \\
\hline
~ & $\pi$ & 12.274 & 12.274 & 11.689 & 9.408 & 0.2921 & 0.3295 & 0.3282 & 0.4114 & 8.809 & 7.809 & 4.592 & $10^{10}$ \\

~ & $K$ & 0.883  & 0.883 & 0.839 & 0.653 & 0.4208 & 0.4783 & 0.4820 & 0.6245 & 8.319 & 7.319 & 4.702 & $10^{10}$ \\

7 TeV & $p$ & 0.313  & 0.313 & 0.304 & 0.269 & 0.4612 & 0.4932 & 0.4965 & 0.5594 & 15.378 & 14.378 & 6.604 & $10^{10}$ \\

~ & $\Xi$ & 0.00188  & 0.00188 & 0.00165 & 0.00107 & 0.3426 & 0.3798 & 0.4074 & 0.5499 & 10.211 & 9.211 & 6.225 & $10^{10}$ \\

~ & $\Omega$ & 0.000107  & 0.000107 & 0.0000995 & 0.0000828 & 0.4295 & 0.4634 & 0.4769 & 0.5427 & 13.666 & 12.666 & 6.421 & $10^{10}$ \\
\hline
\hline
\end{tabular}
}
\label{tabpppar}
\end{table}

\begin{table}[htb]
\caption{Values of $\chi^2/ndf$ of all four fitting functions on $p_T$ spectra in $pp$ collisions}
\scalebox{1.2}[1.2]{
\begin{tabular}{c c c c c c c}
\hline
\hline
$\sqrt{s}$ & particles & NDF & $f_{Ts}$ & $f_{Be}$ & $f_{Ka}$ & $f_{BG}$ \\
\hline
~ & $\pi$ & 33 & 0.281407 & 0.281407 & 1.35515 & 110.815 \\

~ & $K$ & 27 & 0.174877 & 0.174877 & 0.182194 & 8.04702 \\

900 GeV & $p$ & 24 & 0.372409 & 0.372409 & 0.534296 & 3.49085 \\

~ & $\Lambda$ & 9 & 0.347503 & 0.347503 & 0.493316 & 3.20131 \\

~ & $K^0_S$ & 16  & 0.736263 & 0.736263 & 1.69305 & 23.1269 \\
\hline
~ & $\pi$ & 41 & 0.968075 & 0.968075 & 12.0626 & 1316.27 \\

~ & $K$ & 48  & 0.420192 & 0.420192 & 3.13449 & 520.19 \\

7 TeV & $p$ & 46  & 0.448063 & 0.448063 & 1.81175 & 254.338 \\

~ & $\Xi$ & 18  & 0.235202 & 0.235202 & 0.334564 & 25.7687 \\

~ & $\Omega$ & 8  & 0.436939 & 0.436939 & 0.563252 & 1.10363 \\
\hline
\hline
\end{tabular}
}
\label{tabppf}
\end{table}

\begin{figure*}[htb]
\scalebox{1}[1]{
\includegraphics[width=0.45\linewidth]{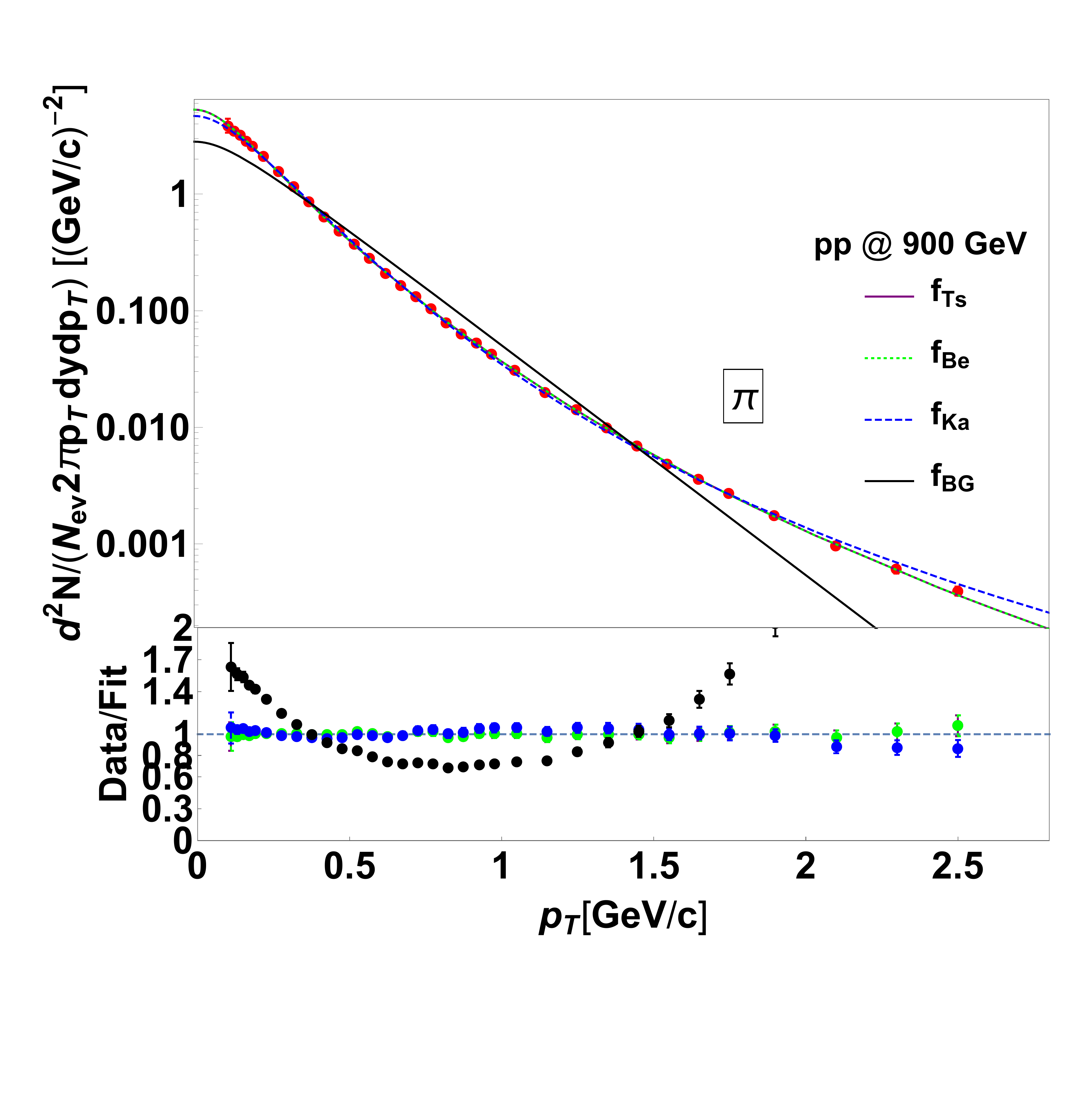}
\includegraphics[width=0.45\linewidth]{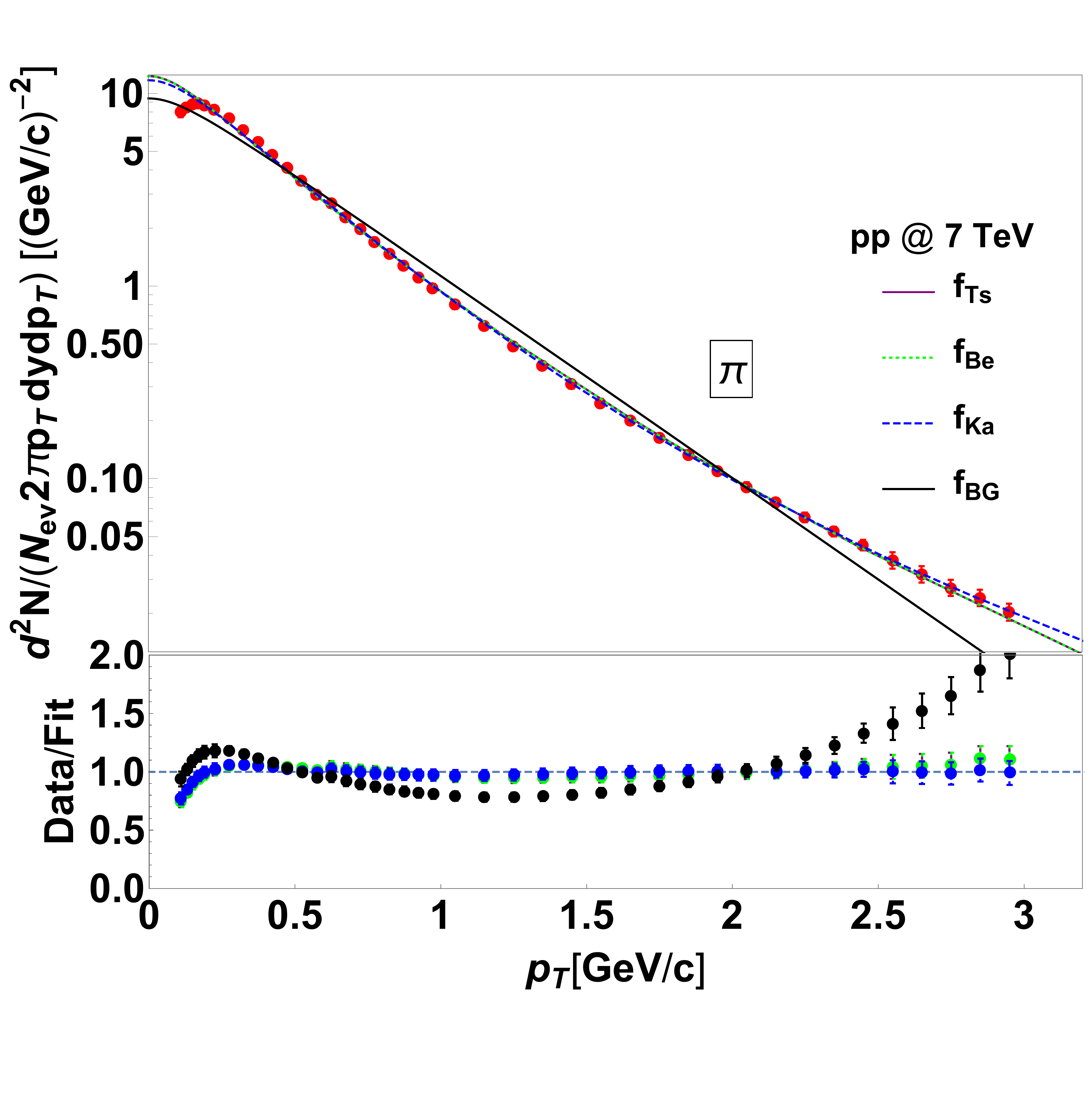}
}
\scalebox{1}[1]{
\includegraphics[width=0.45\linewidth]{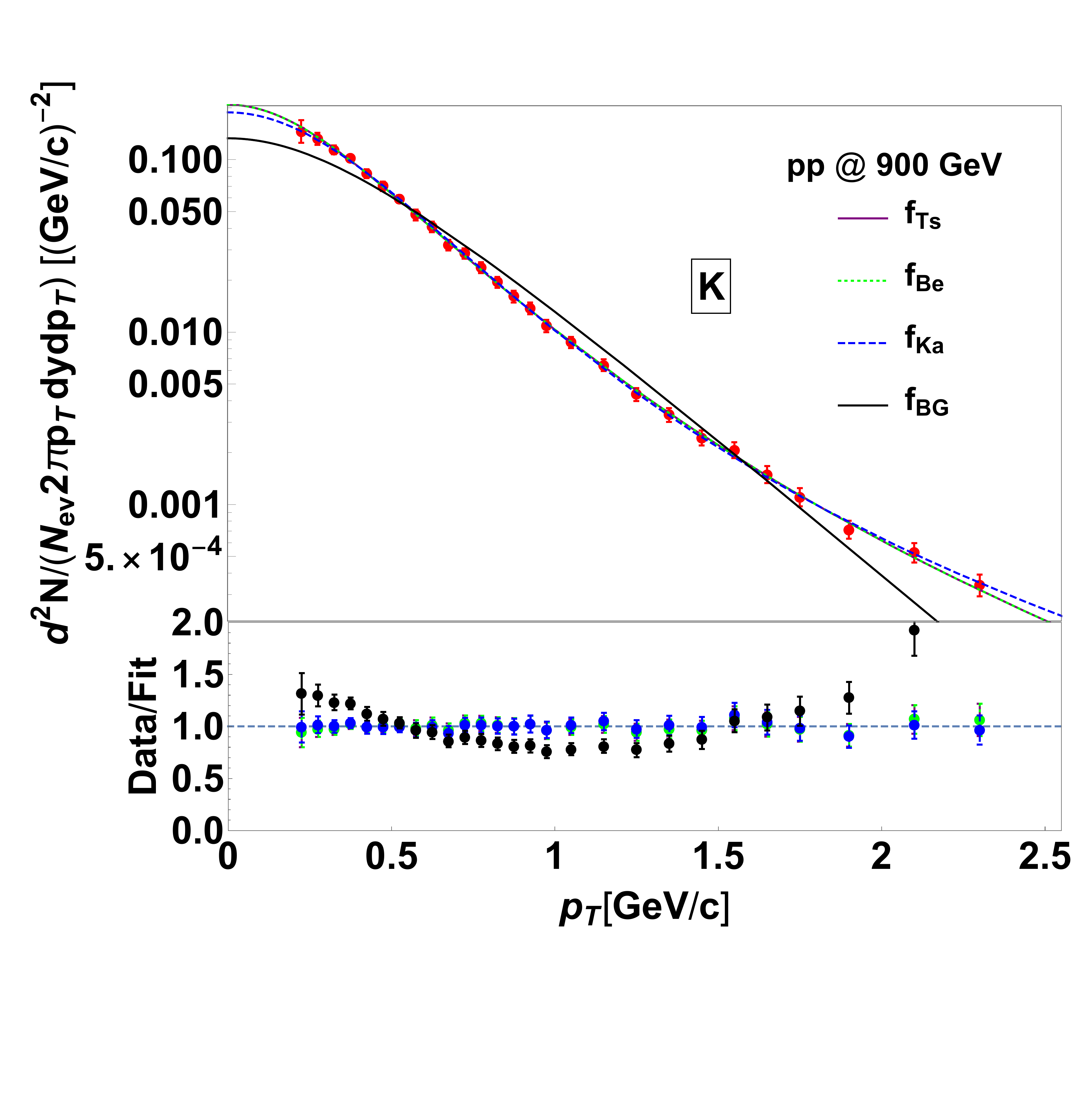}
\includegraphics[width=0.45\linewidth]{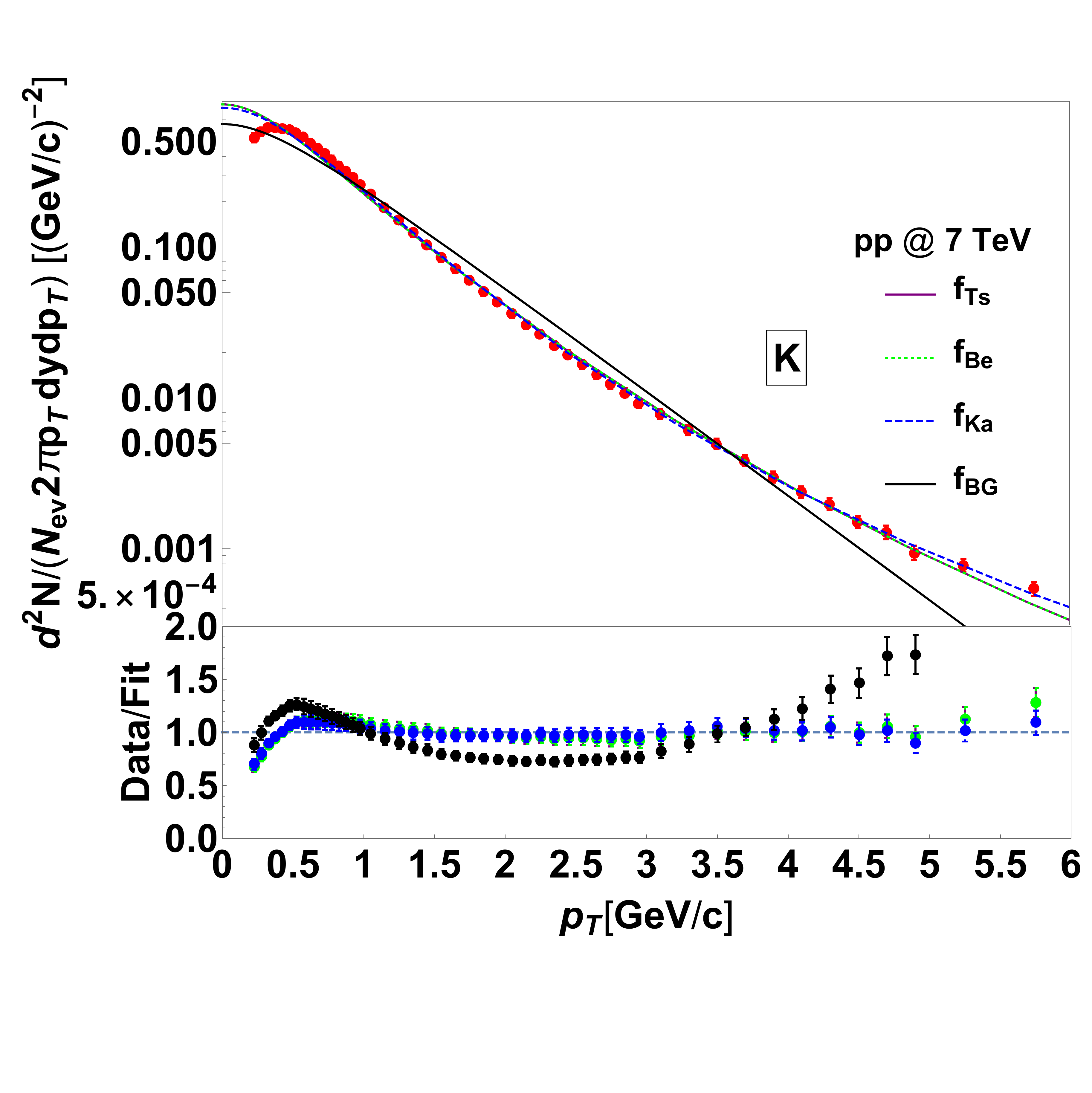}
}
\scalebox{1}[1]{
\includegraphics[width=0.45\linewidth]{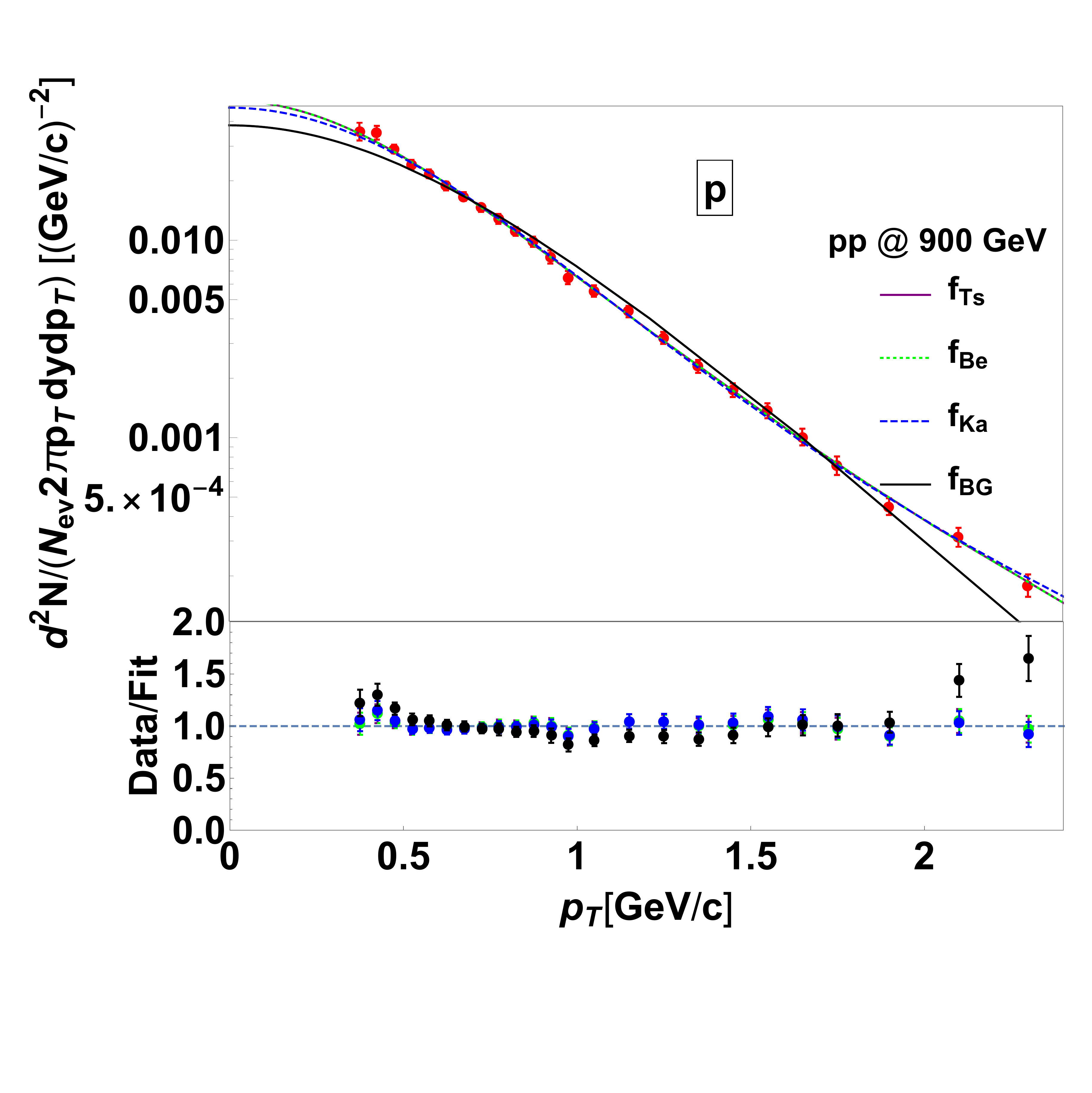}
\includegraphics[width=0.45\linewidth]{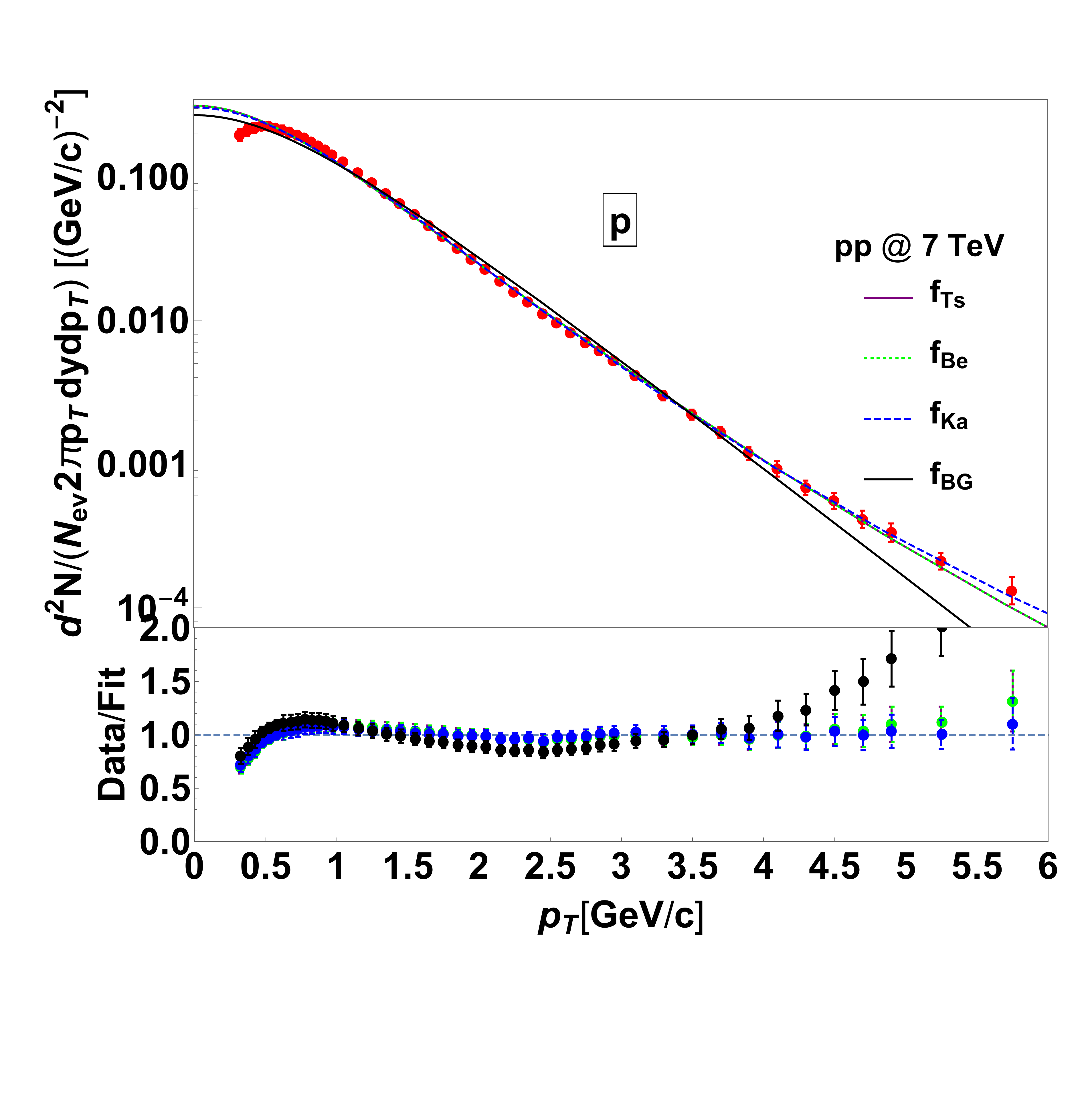}
}
  \vspace*{-19mm}
\caption{Fittings on $p_T$ spectra of pions, kaons and protons in $pp$ collisions at 900 GeV and 7 TeV respectively by the four different functions in Eq.(\ref{fit-fun}). The $p_T$ ranges are listed in Table \ref{tabpppT}. The lower panel lists the ratio of data and fitting results.}
\label{figpT1}
\end{figure*}


\begin{figure*}[htb]
\scalebox{1}[1]{
\includegraphics[width=0.45\linewidth]{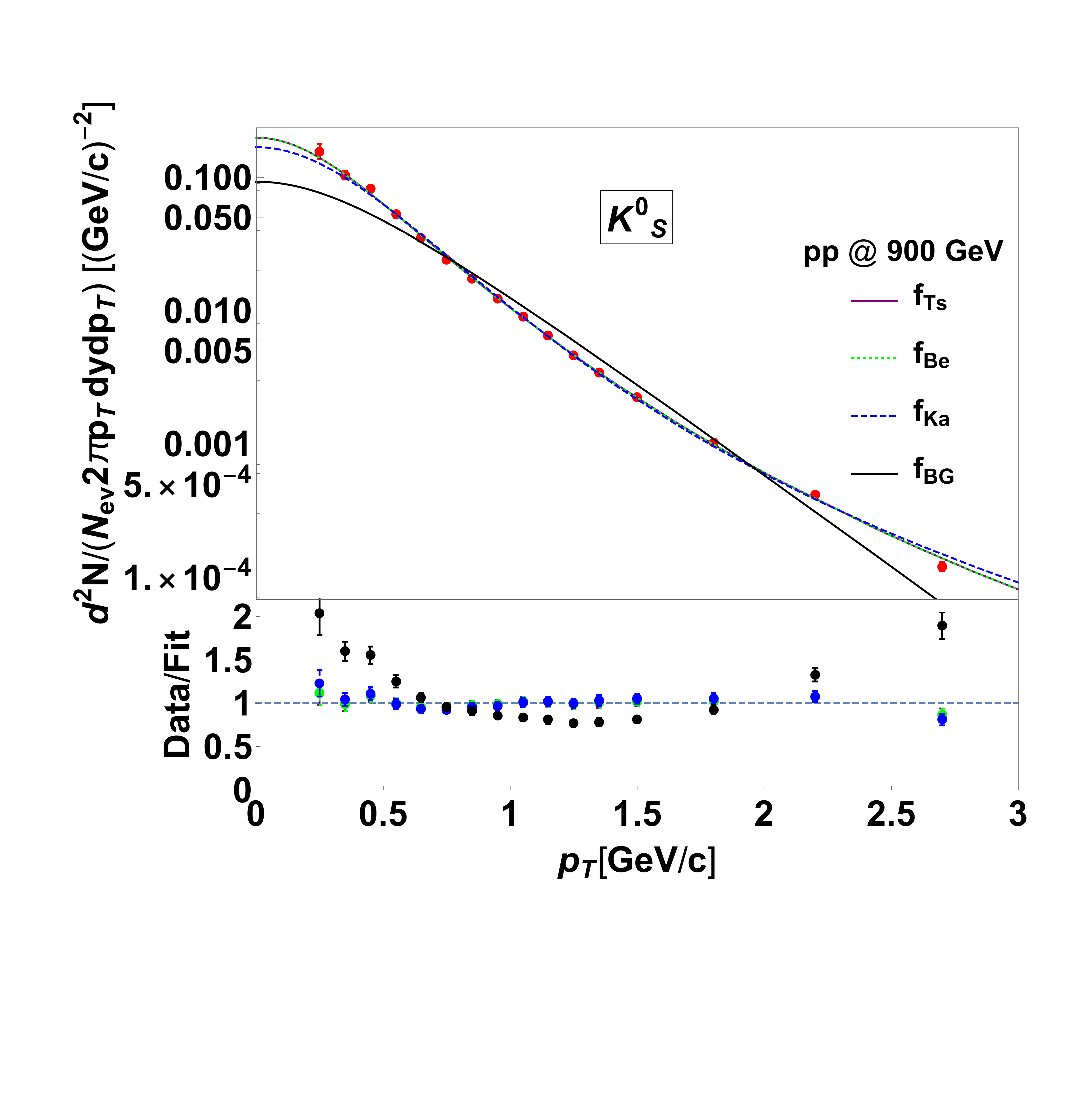}
\includegraphics[width=0.45\linewidth]{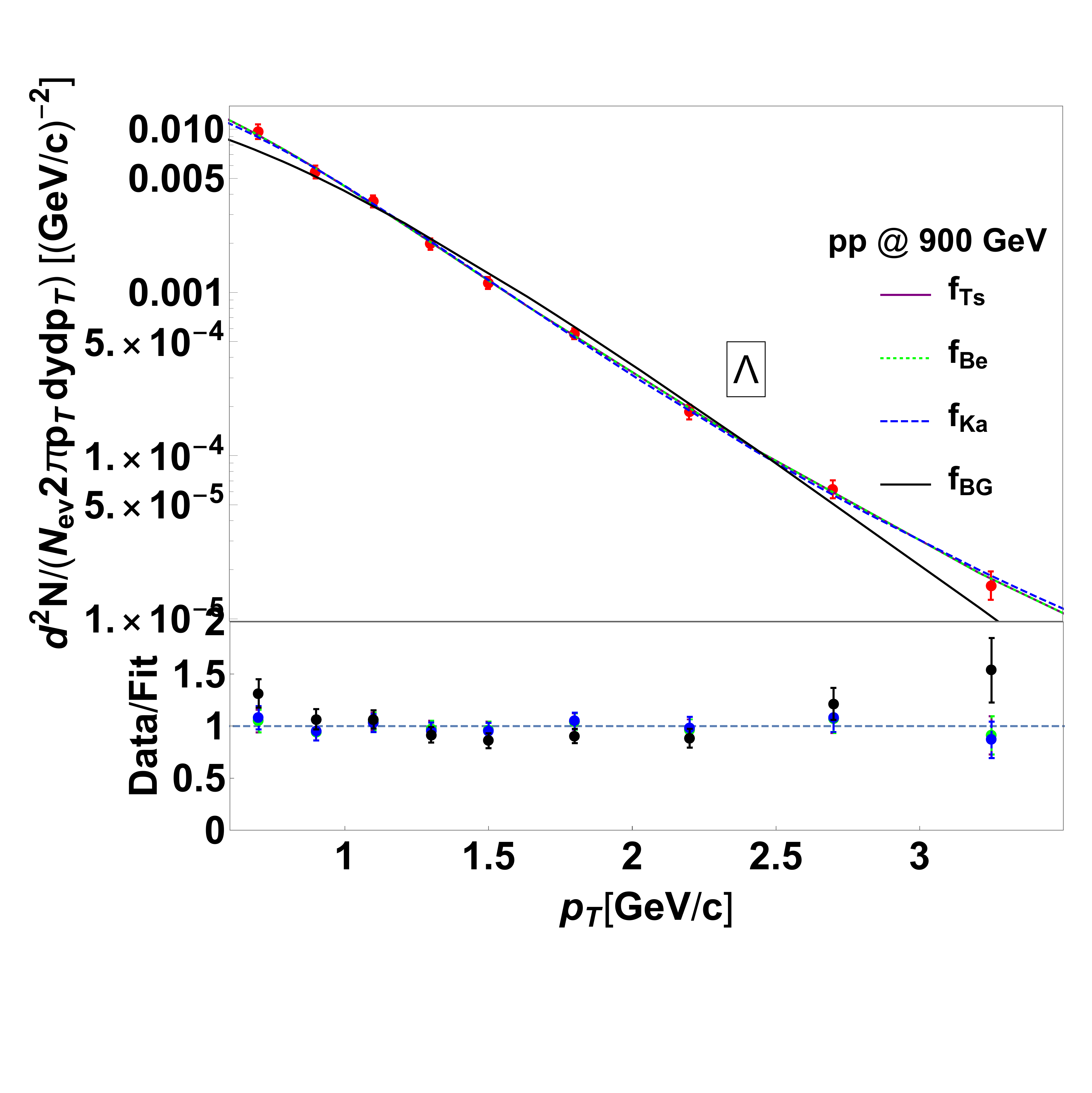}
}
\scalebox{1}[1]{
\includegraphics[width=0.45\linewidth]{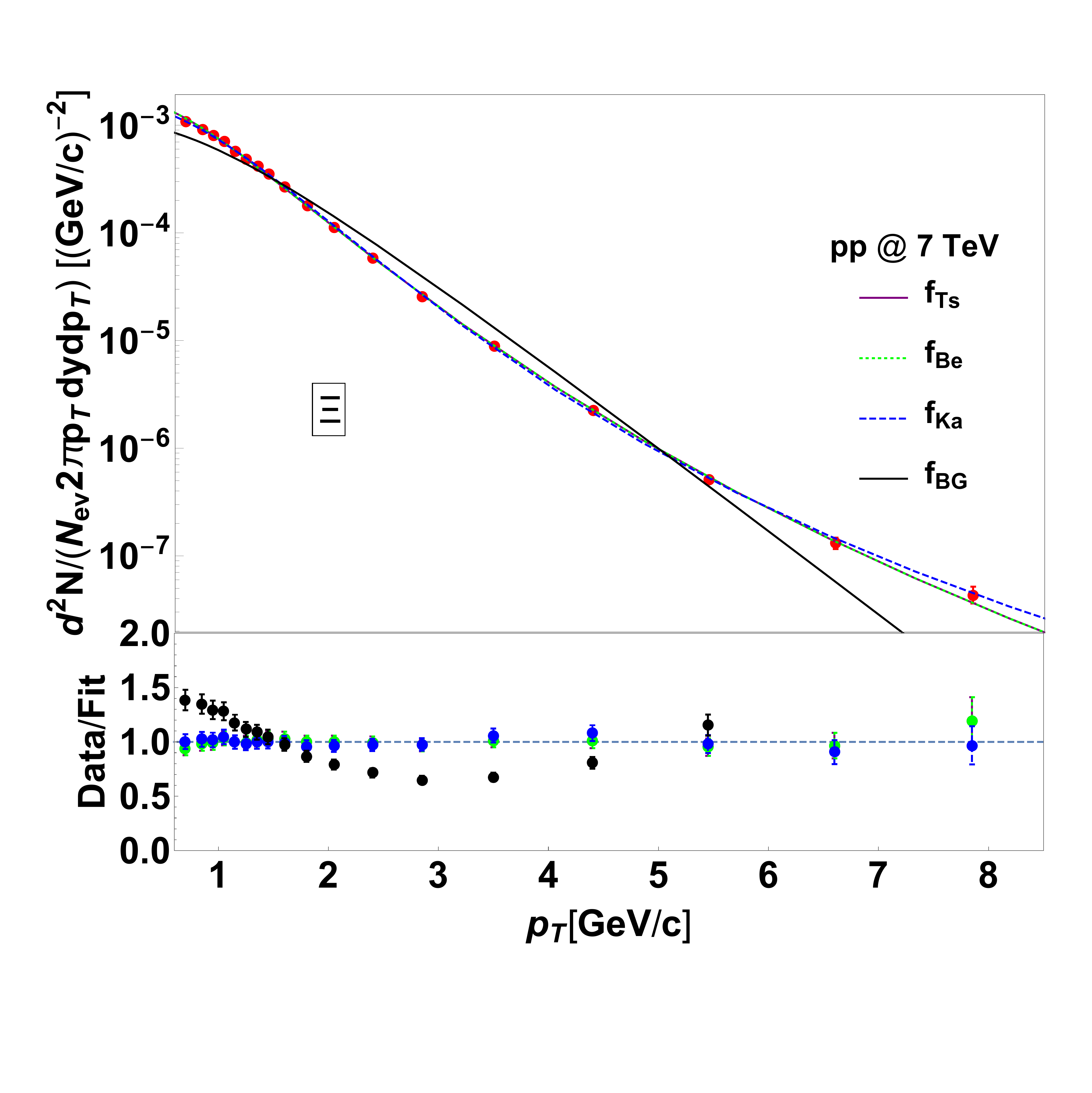}
\includegraphics[width=0.45\linewidth]{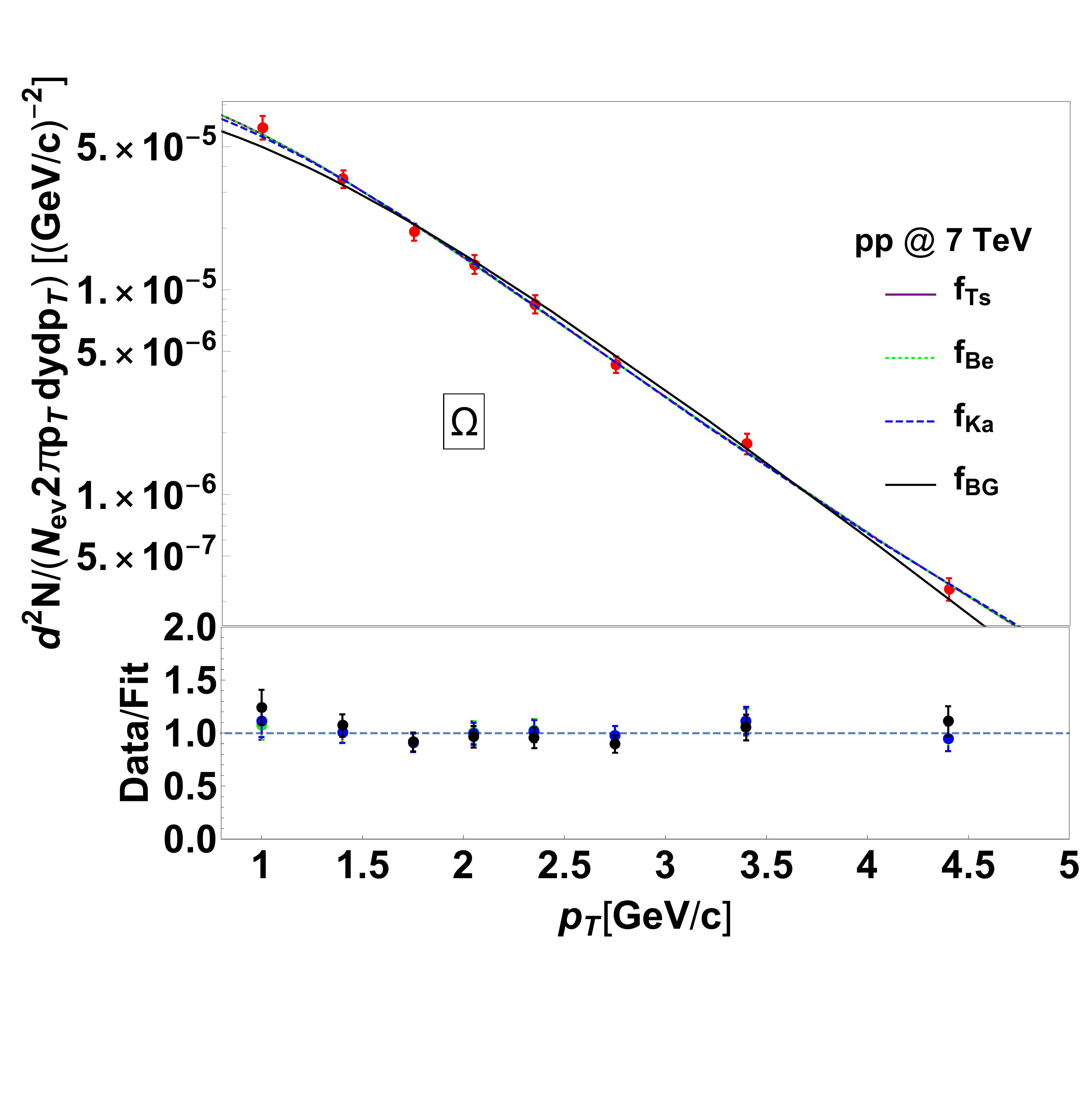}
}
\caption{Fittings on $p_T$ spectra of $K_S^0$ and $(\Lambda +\bar{\Lambda})/2$ at 900 GeV and $(\Xi^+ +\Xi^-)/2$, $(\Omega+\bar{\Omega})/2$ at 7 TeV in $pp$ collisions by the four different functions in Eq.(\ref{fit-fun}), and the $p_T$ range are listed in Table \ref{tabpppT}. The lower panel lists the ratio of data and fitting results.}
\label{figpT2}
\end{figure*}

\begin{figure*}[htb]
\scalebox{1}[1]{
\includegraphics[width=0.45\linewidth]{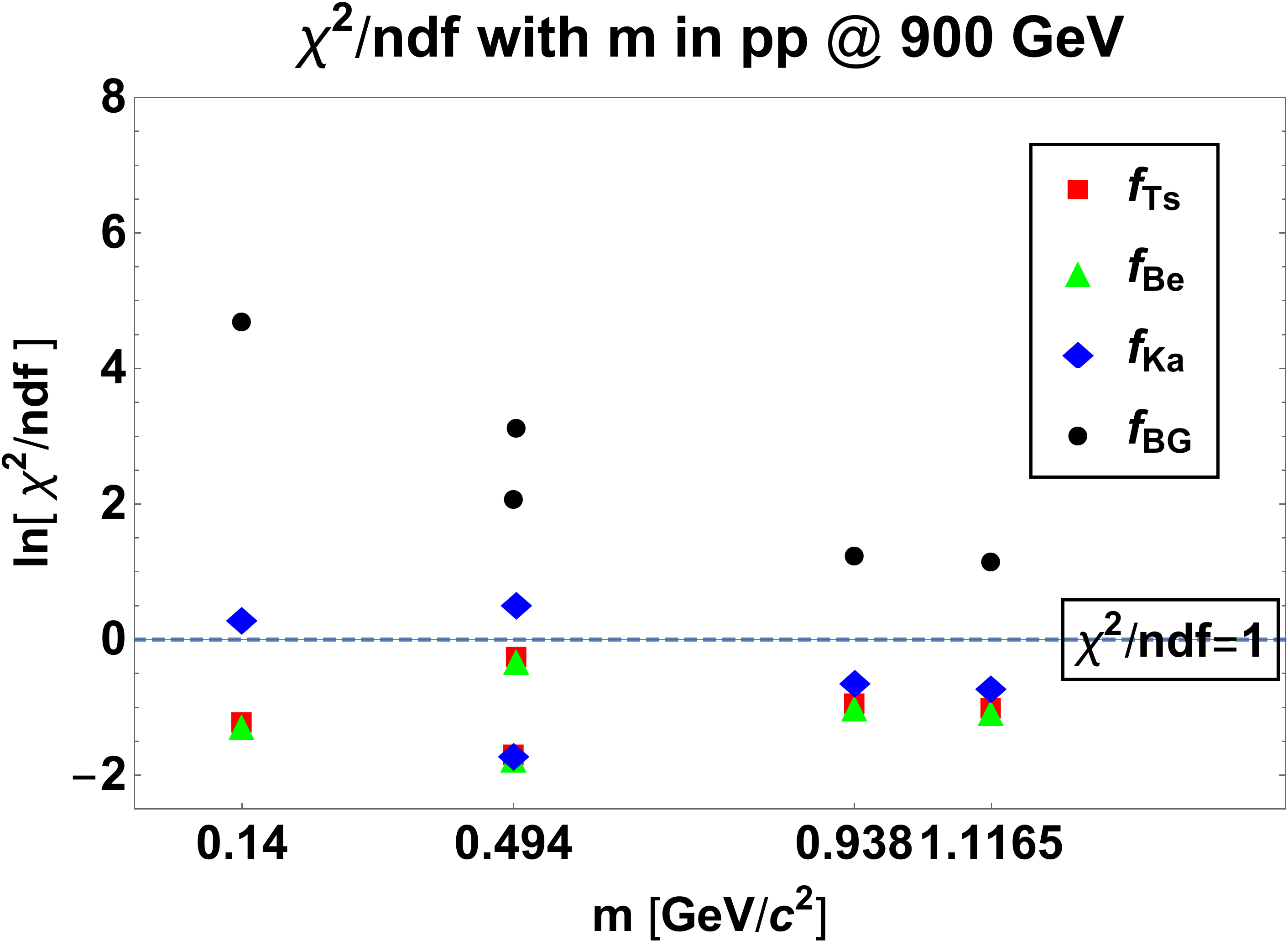}
\includegraphics[width=0.45\linewidth]{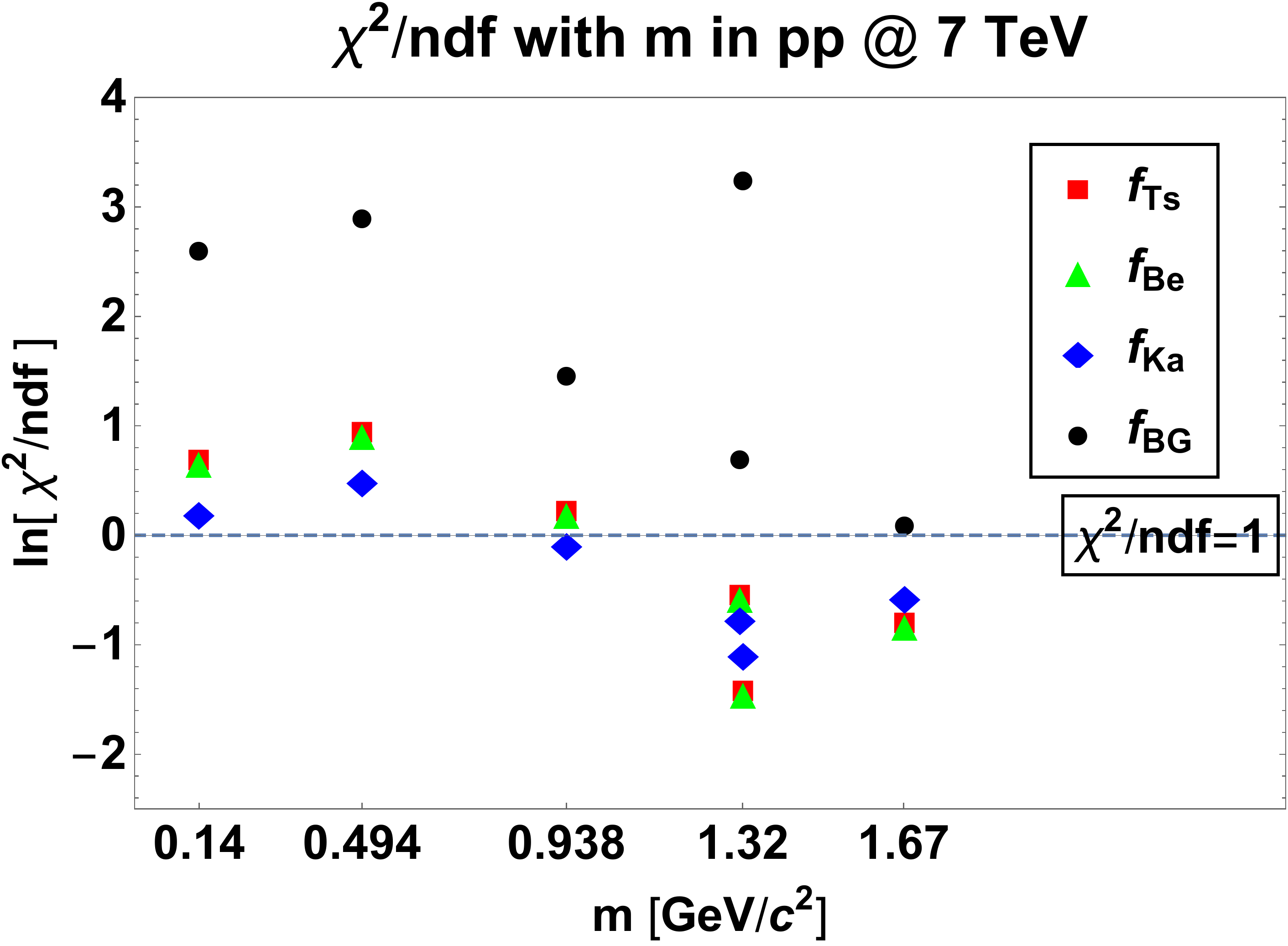}
}

\caption{$\chi^2/ndf$ of all four fittings on $p_T$ spectra of identified hadrons in $pp$ collisions.  (Left: $\frac{\pi^++\pi^-}{2}$, $\frac{K^++K^-}{2}$, $K_S^0$, $\frac{p+\bar{p}}{2}$ and $\frac{\Lambda+\bar{\Lambda}}{2}$ at 900 GeV; right: $\frac{\pi^++\pi^-}{2}$, $\frac{K^++K^-}{2}$, $\frac{p+\bar{p}}{2}$, $\frac{\Xi^++\Xi^-}{2}$ and $\frac{\Omega^++\Omega^-}{2}$ at 7 TeV). The fitting error $\chi^2/ndf =1$ is also given as a baseline. More details are seen in Table \ref{tabppf}.}
\label{figxi1}
\end{figure*}


We see that, as shown in Fig.\ref{figpT1} and Fig.\ref{figpT2}, the BG distribution ($f_{BG}$) indeed fails describing the $p_T$ spectra over a wide $p_T$ range given as Table \ref{tabpppT}, especially in the high $p_T$ part.
For the other three non-extensive approaches, $f_{Ts}$, $f_{Be}$ and $f_{Ka}$, there seems no big difference on the fittings of various spectra.
The Kaniadakis non-extensive statistics, similar to the Tsallis and Beck ones, is proved to be an alternative tool to investigate the hadron spectra in $pp$ collisions, while the usual BG form not.

In order to further investigate their discrepancies, error analysis on the relevant fittings at $900$ GeV and $7$ TeV are shown in Fig.\ref{figxi1}. 
All corresponding parameters are summarized in Table \ref{tabppf}.
We could see that the values of $\chi^2/ndf$ of the first two, Tsallis and Beck distributions, behave similar when applying into the hadron spectra fittings in $pp$ collisions.
This agrees with the discussions above, that Beck formula is one of the generalization of the Tsallis $q$-distribution.
On the other hand, the $\chi^2/ndf$ values of the Kaniadakis distribution, $f_3$, show that this new non-extensive approach can also be a good tool in the analysis on hadron spectra in the elementary collisions.
The first two $q$-non-extensive approaches, however, lead to the smallest values of $\chi^2/ndf$.
This is due to the fact that, in $pp$ collisions, there are relatively smaller multiplicities and lower particle-hole CPT symmetry effects.
One realizes, furthermore, that for all the non-extensive fittings, their values of $\chi^2/ndf$ are around the ideal value, $\chi^2/ndf=1$, except for the formula $f_{BG}$ which gives the worst results of all.

Checking the fitting parameters, $T$ and $n$, we observe that these formulas obtain quite different parameters although the Tsallis and Beck distributions share the same fit goodness, $\chi^2/ndf$.
This evokes further investigations on the connections between these two similar approaches.
We also examine the connections of the fitting temperature, $T$, and non-extensive parameter, $n$, by all three non-extensive functions for various hadron spectra fittings in the same $pp$ collision.

\begin{table}[htb]
\caption{Values of $T_B$ and $T_n$ of all three non-extensive functions on the linear relations between $T$ and $1/n$ in $pp$ collisions}
\scalebox{1.2}[1.2]{
\begin{tabular}{c c c c c}
\hline
\hline
$\sqrt{s}$ & parameters & $f_{Ts}$ & $f_{Be}$ & $f_{Ka}$ \\
\hline
~ & $T_B$ & 0.30924 & 0.30402 & 0.40972  \\

900 GeV & $T_n$ & -0.90457 & -0.56757 & -0.81827  \\

\hline
~ & $T_B$ & 0.56206 & 0.57242 & 0.63677  \\

7 TeV & $T_n$ &  -2.1003 & -1.6249 & -1.4063 \\

\hline
\hline
\end{tabular}
}
\label{tabppTq}
\end{table}

\begin{figure*}[htb]
\scalebox{1}[1]{
\includegraphics[width=0.45\linewidth]{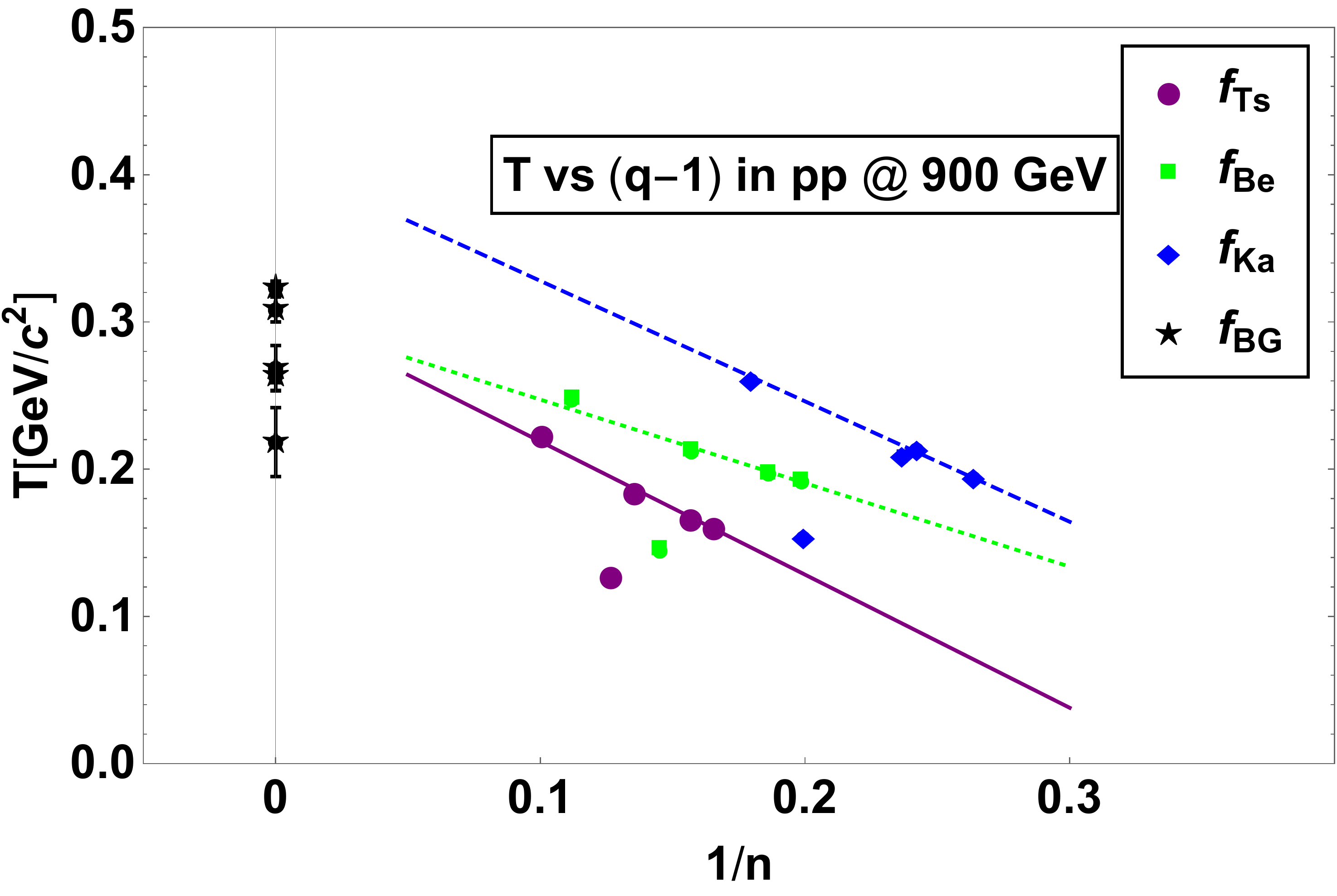}
\includegraphics[width=0.45\linewidth]{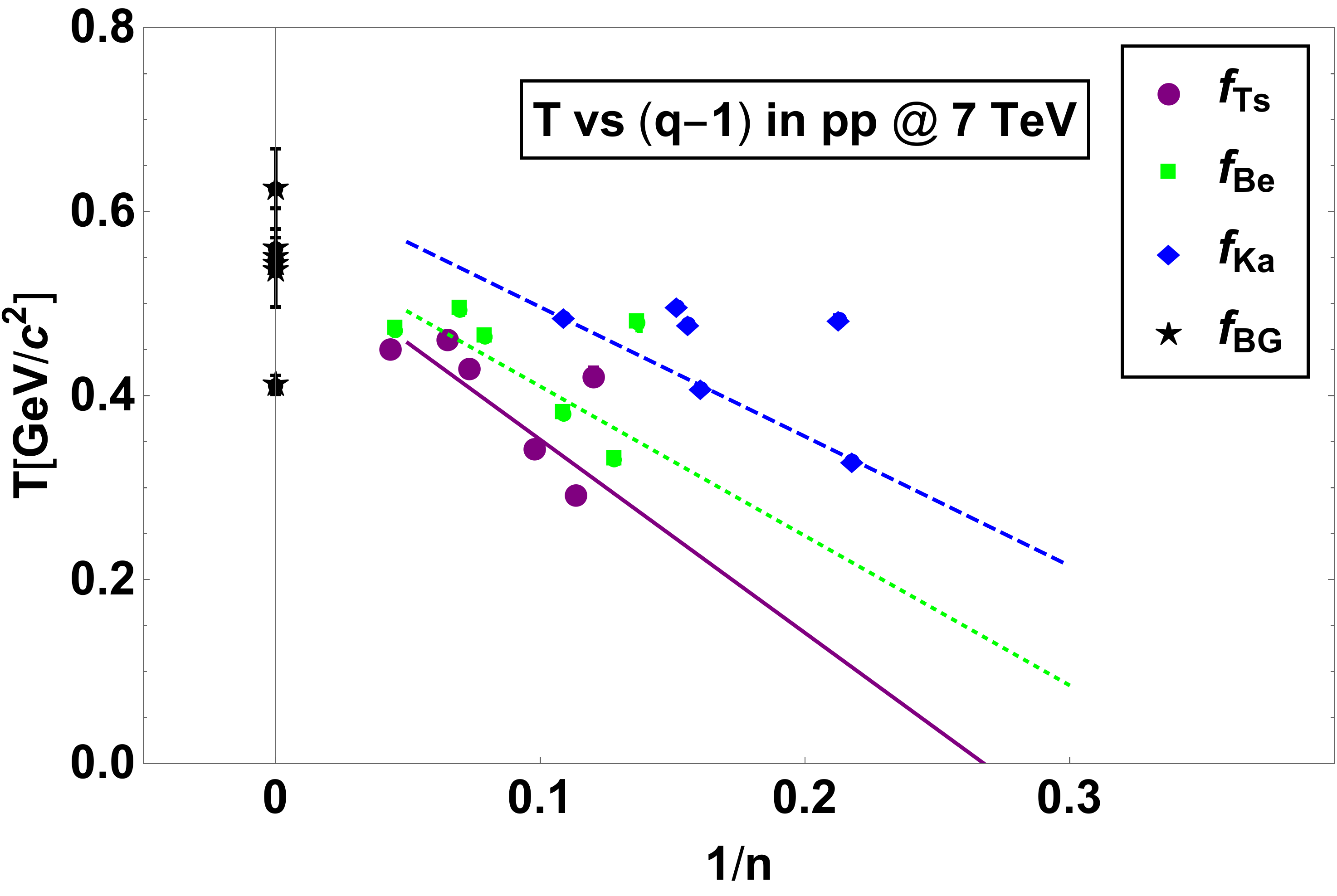}
}
\caption{Distributions of the inverse slope parameter, $T$, with the non-extensive parameter, $1/n$, for all kinds of charged particles in $pp$ collisions at $\sqrt{s}=900$ GeV and 7 TeV respectively. The black stars of $1/n=0$ are for the BG results by $f_{BG}$. We could see the obvious linear combination, cf. Eq.(\ref{Tqlin}), for each fitting formula, where the exception stands for the results of pions.}
\label{figTq1}
\end{figure*}

As shown in Fig.\ref{figTq1}, it indicates that for all fitting results of different hadron species in $pp$ collisions at the same beam energy from these three non-extensive formulas, the inverse slope parameter, $T$, shows a slightly linear dependence on the non-extensive parameter $1/n$ ($1/n=q-1$ for $f_{Ts}$ and $f_{Be}$, $1/n=1/\kappa$ for $f_{Ka}$):
\begin{eqnarray}
T=T_B+T_n\cdot \frac{1}{n}~.
\label{Tqlin}
\end{eqnarray}
Here $T_B$ denotes the limiting values  for $1/n\to 0$, namely, the usual BG case and $T_n$ is the non-extensive slope parameter of the linear dependence.
In Table \ref{tabppTq} we present the exact fitting results for $T$ and $1/n$ in $pp$ collisions at 900 GeV and 7 TeV.
Values of the limiting temperature, $T_B$, are shown to be larger for the Kaniadakis distribution than the others.
The temperature obtained by the BG function, $f_{BG}$, increases independently of the non-extensive parameter $n$ since $1/n=0$ for the BG case.

\begin{figure}[htb]
\vspace{-12mm}
\scalebox{1}[1]{
\includegraphics[width=0.35\linewidth]{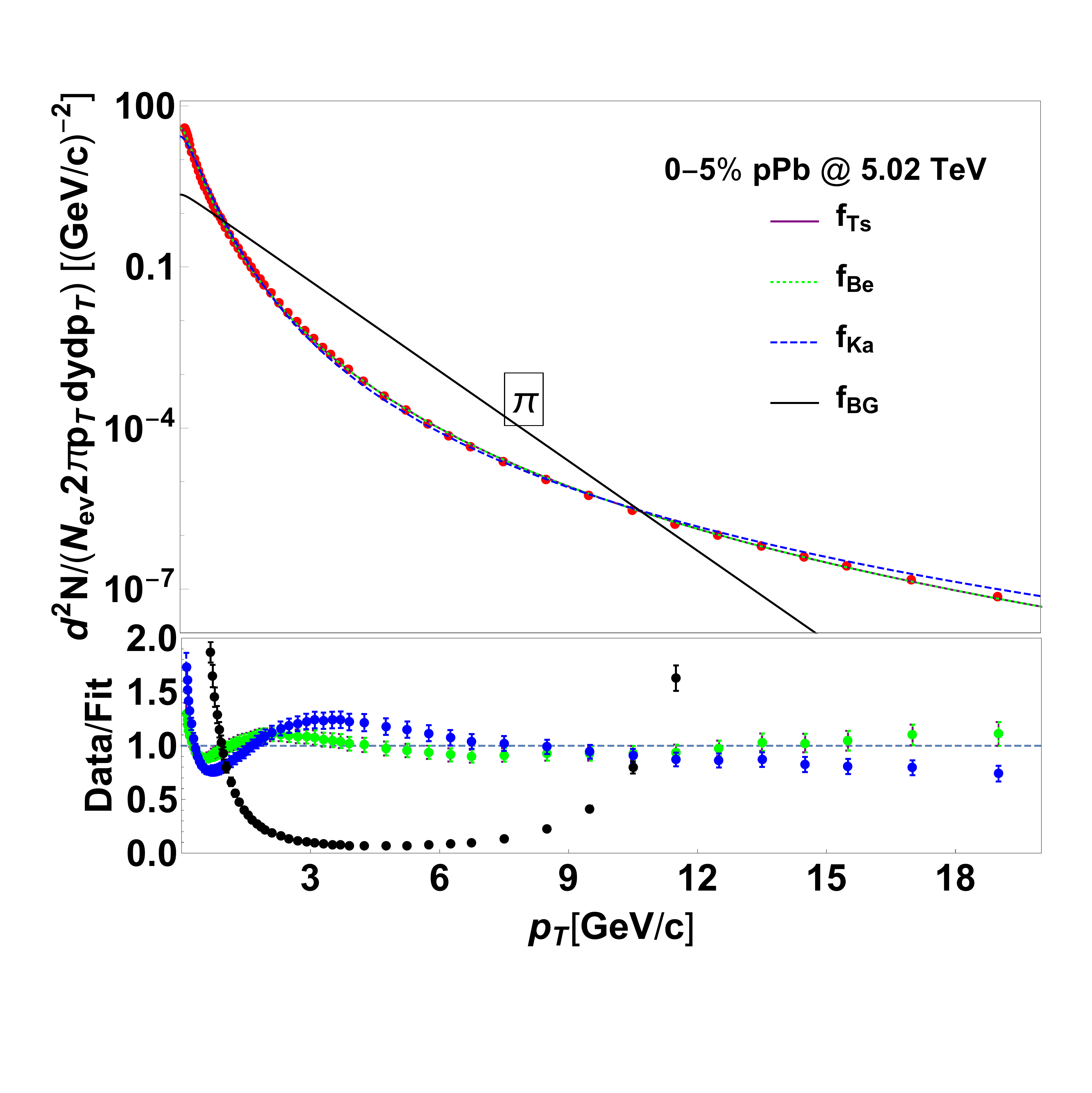}
\includegraphics[width=0.35\linewidth]{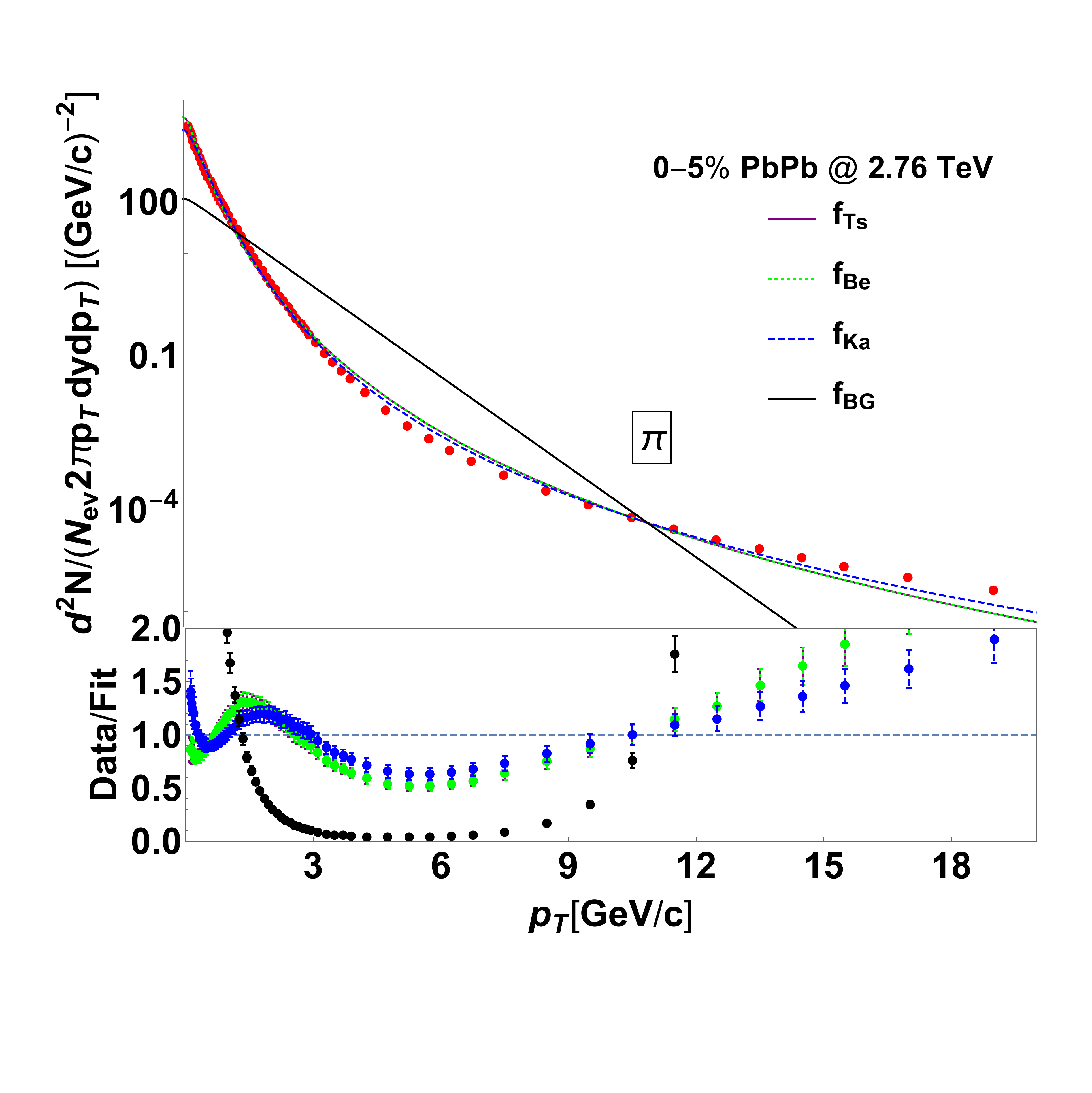}
}
\scalebox{1}[1]{
\includegraphics[width=0.35\linewidth]{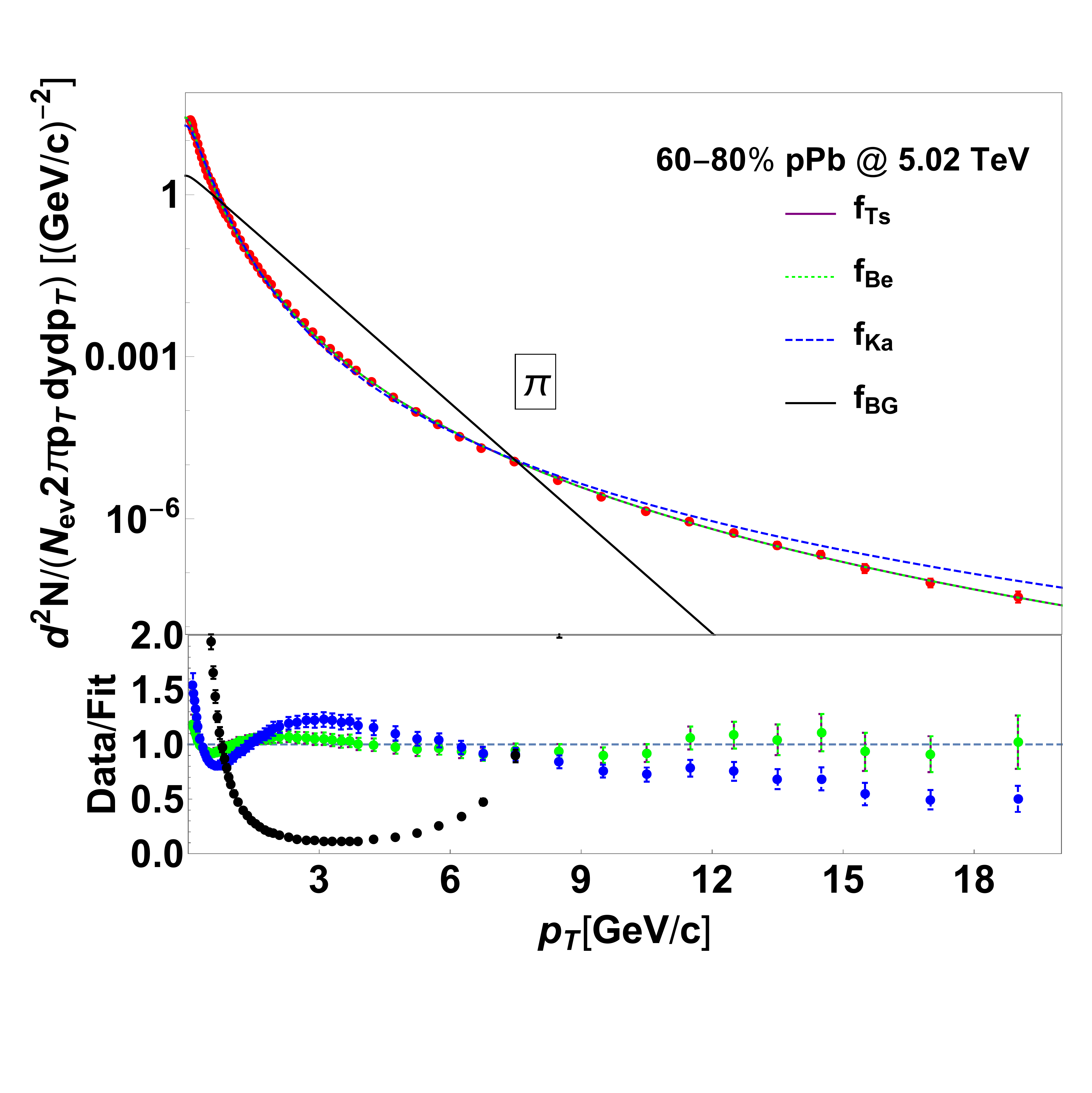}
\includegraphics[width=0.35\linewidth]{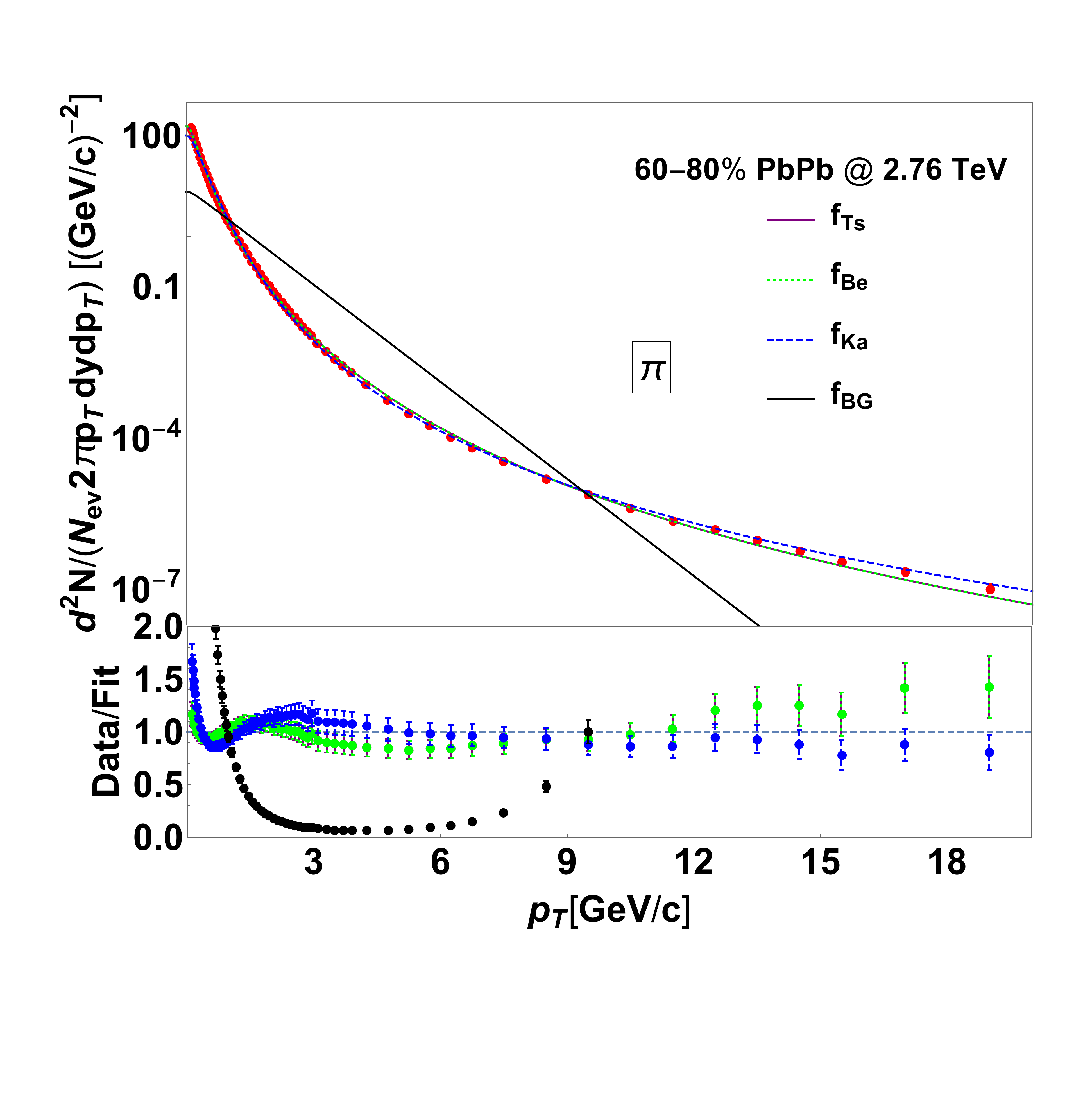}
}
\vspace{-12mm}
\caption{Fittings on $p_T$ spectra of pions in $pPb$ (left) and $PbPb$ (right) collisions at 5.02 and 2.76 TeV respectively by the four different functions in Eq.(\ref{fit-fun}). Ratio of data and fittings is also listed. All spectra are fitted in the range of $p_T$ given in Table \ref{tabAApT}.}
\label{figpT3}
\end{figure}

\begin{figure}[htb]
\vspace{-12mm}
\scalebox{1}[1]{
\includegraphics[width=0.35\linewidth]{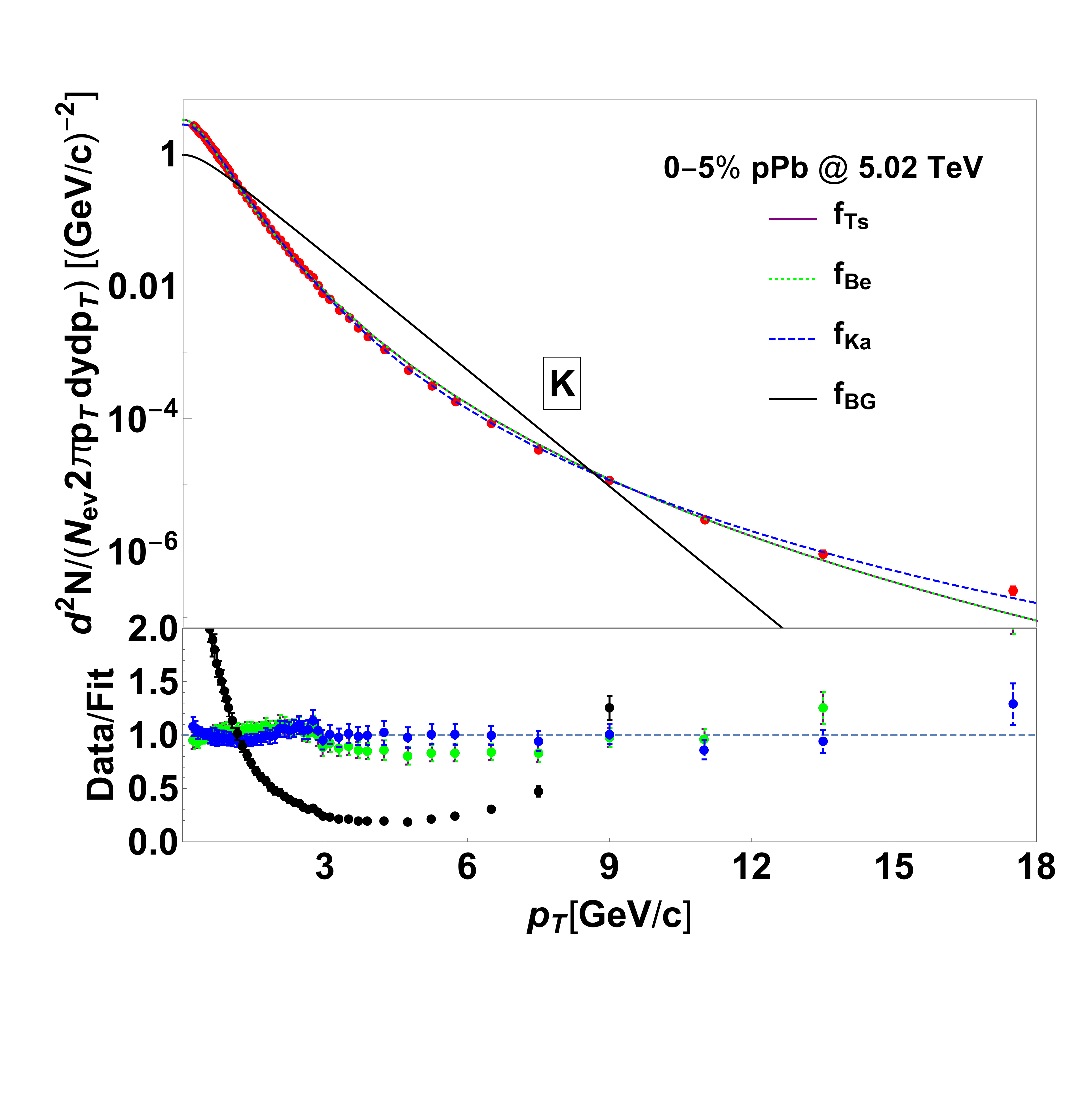}
\includegraphics[width=0.35\linewidth]{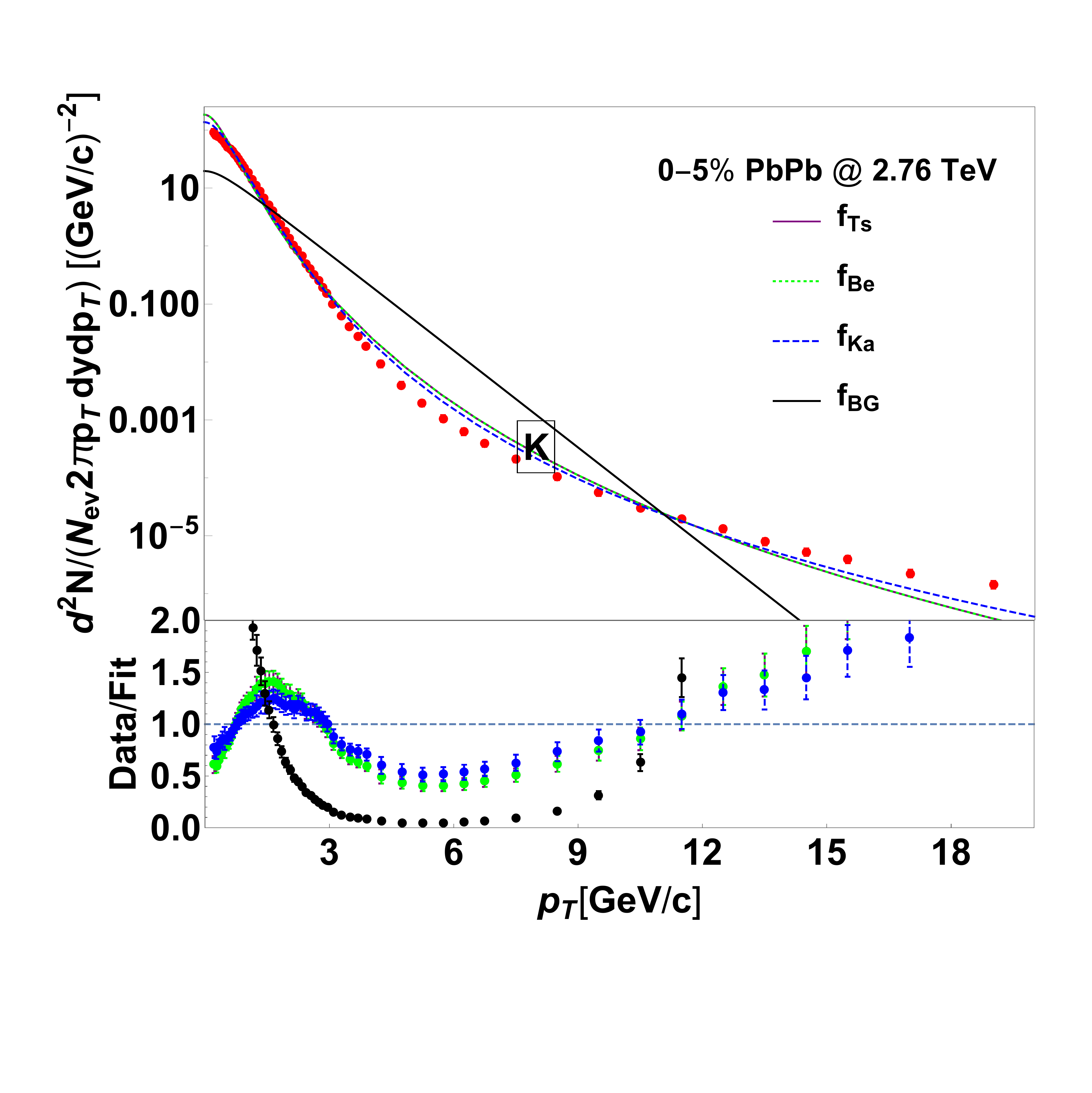}
}
\scalebox{1}[1]{
\includegraphics[width=0.35\linewidth]{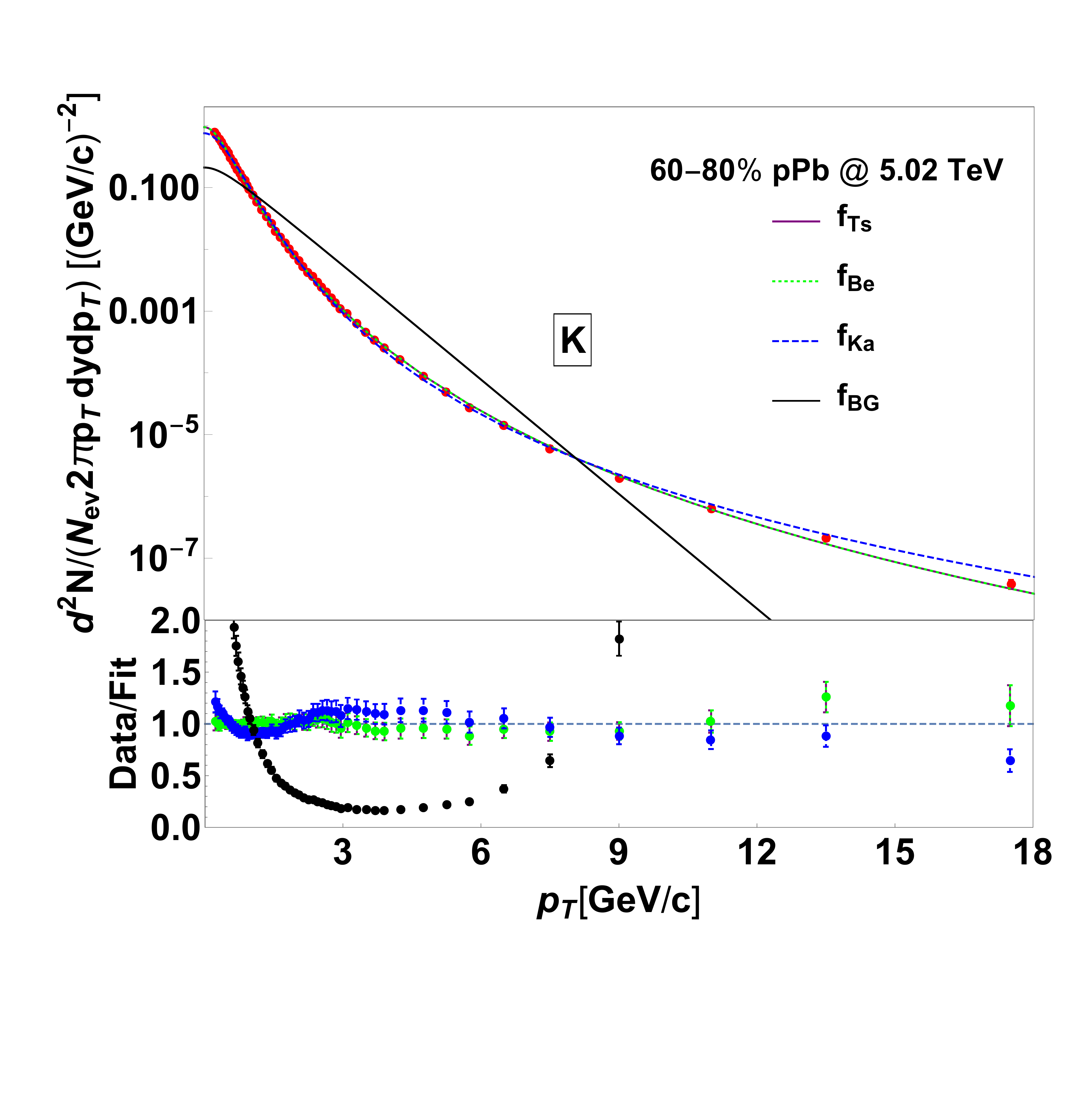}
\includegraphics[width=0.35\linewidth]{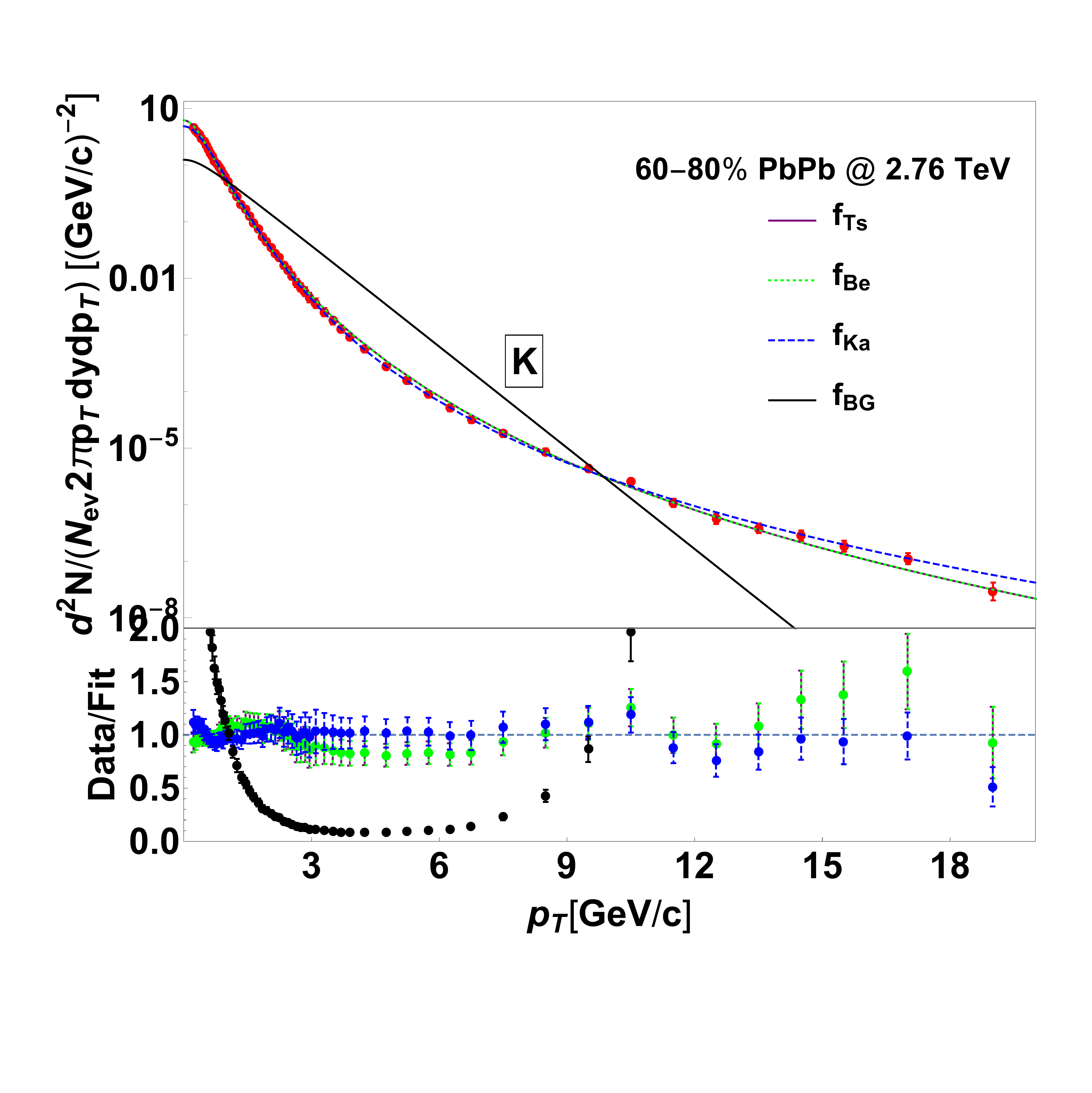}
}
\vspace{-12mm}
\caption{Fittings on $p_T$ spectra of kaons in $pPb$ (left) and $PbPb$ (right) collisions at 5.02 and 2.76 TeV respectively by the four different functions in Eq.(\ref{fit-fun}). Ratio of data and fittings is also listed. All spectra are fitted in the range of $p_T$ given in Table \ref{tabAApT}.}
\label{figpT4}
\end{figure}

\begin{figure*}[htb]
\scalebox{1}[1]{
\includegraphics[width=0.35\linewidth]{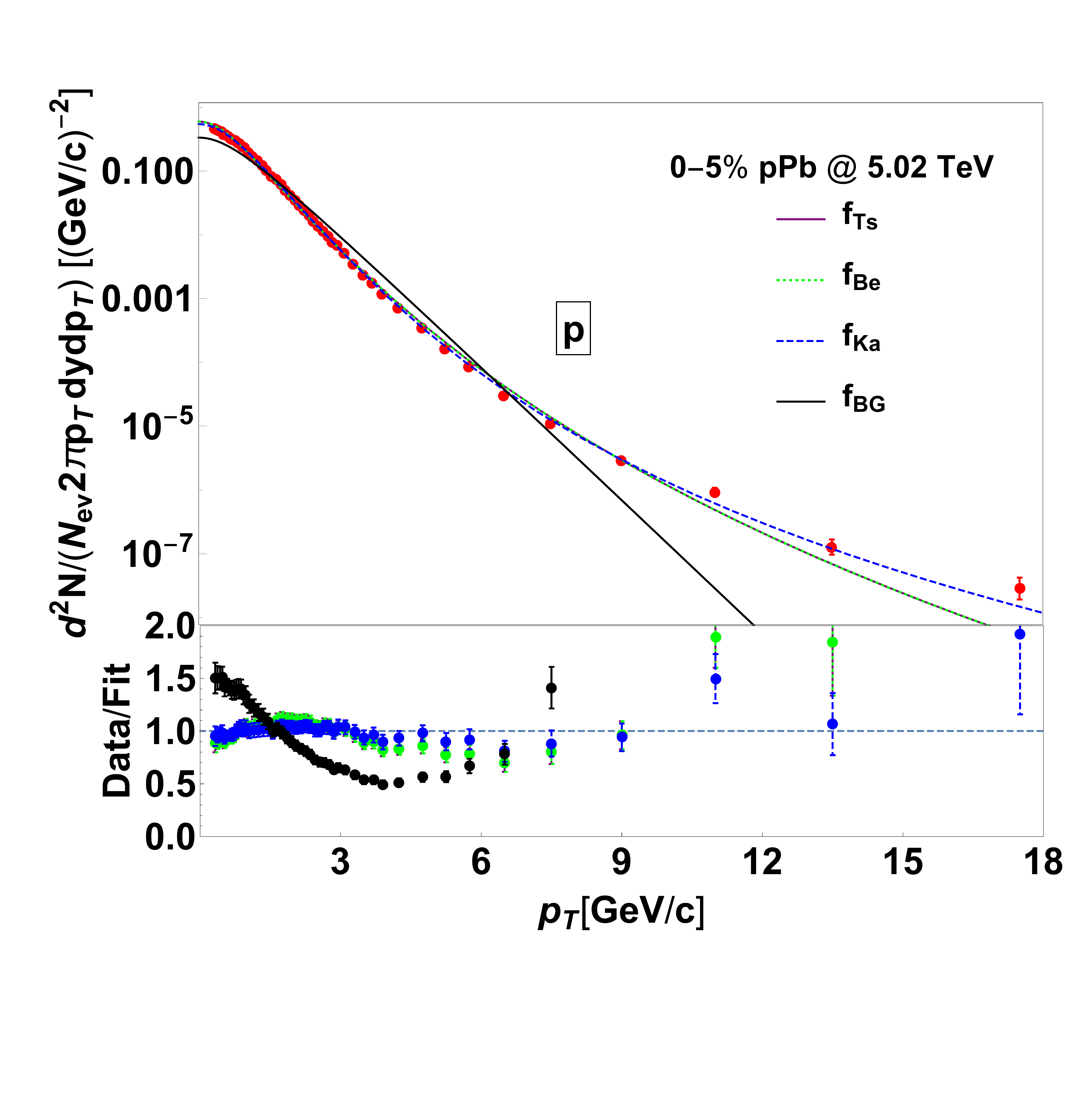}
\includegraphics[width=0.35\linewidth]{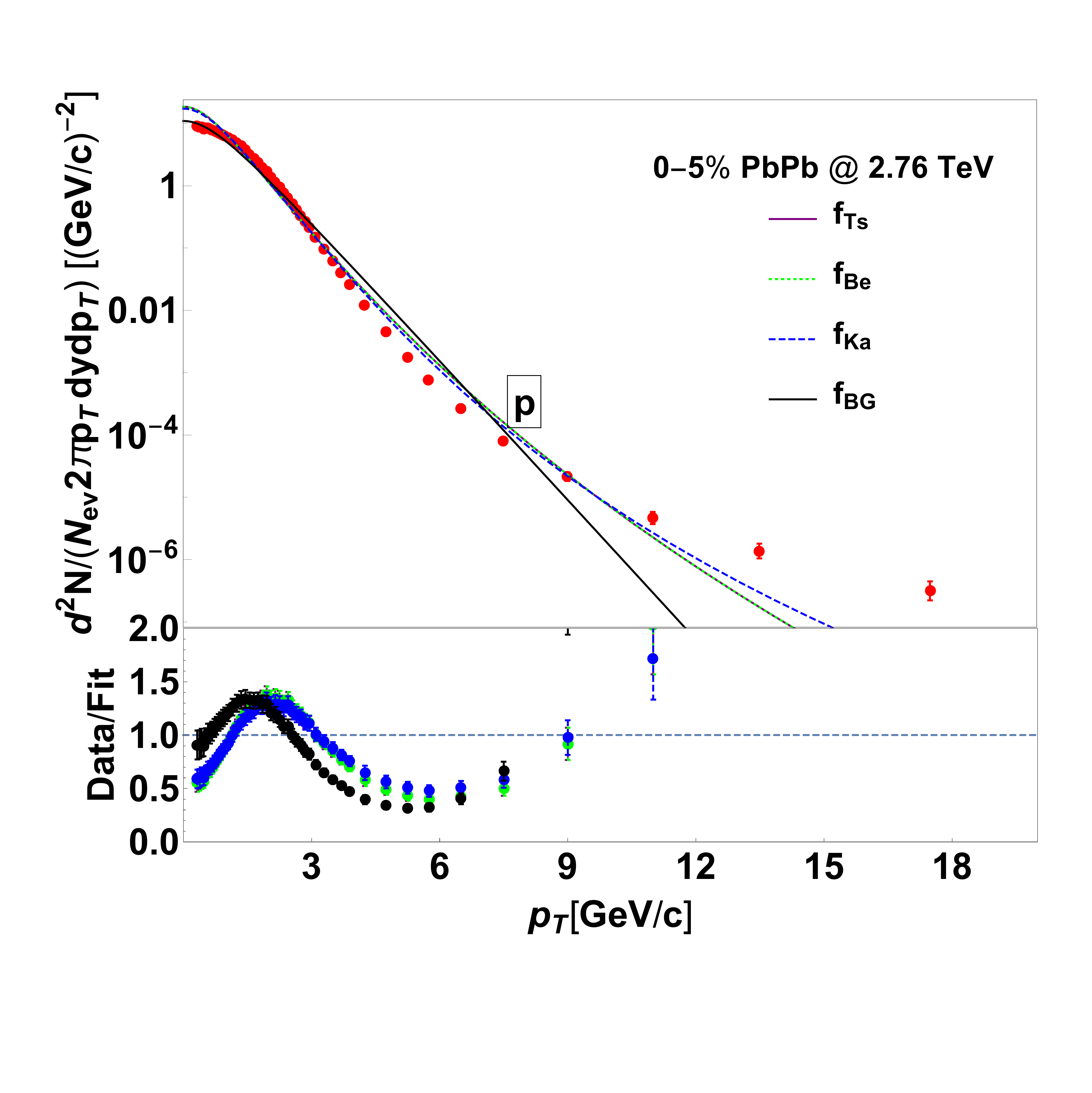}
}
\scalebox{1}[1]{
\includegraphics[width=0.35\linewidth]{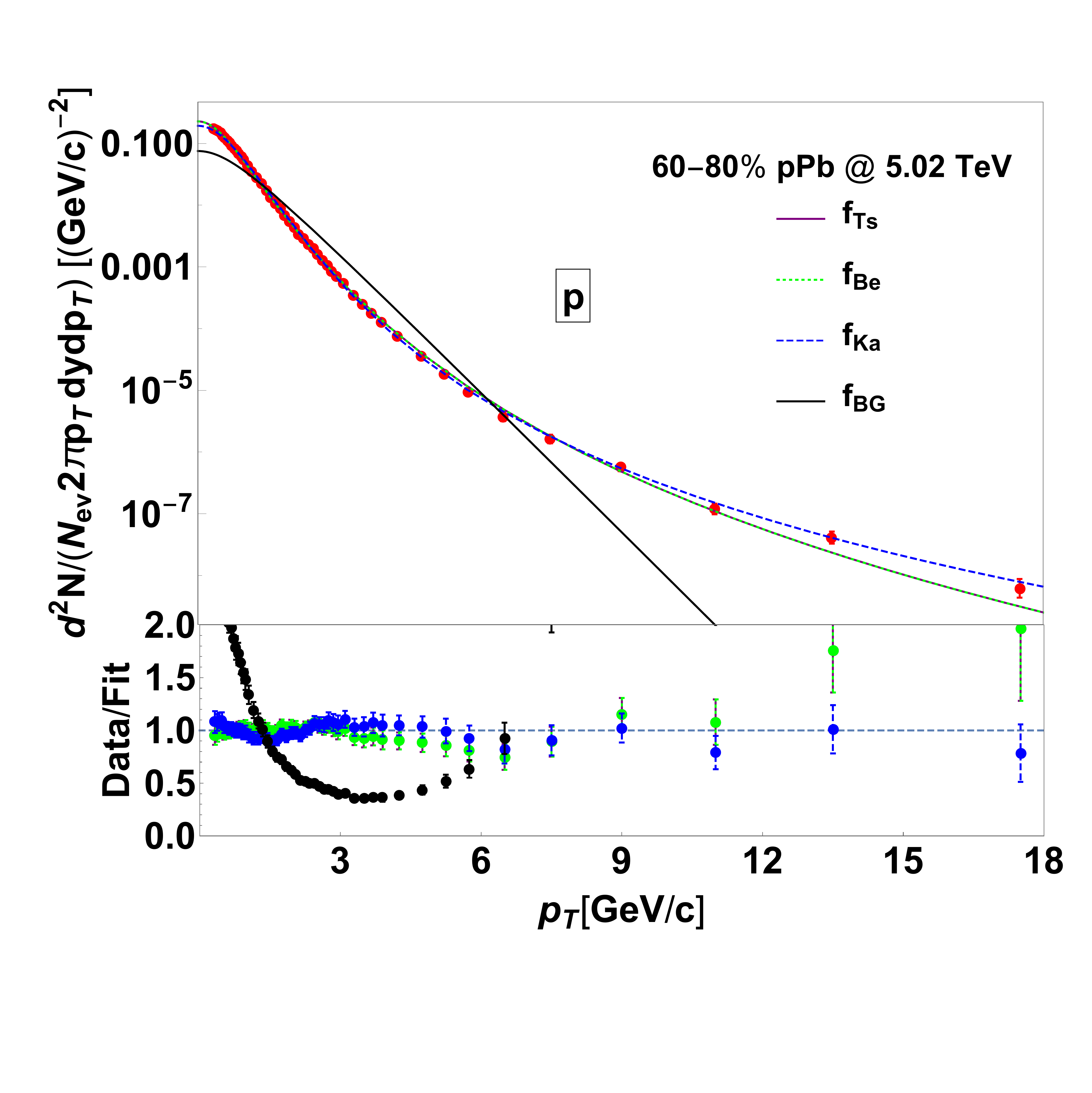}
\includegraphics[width=0.35\linewidth]{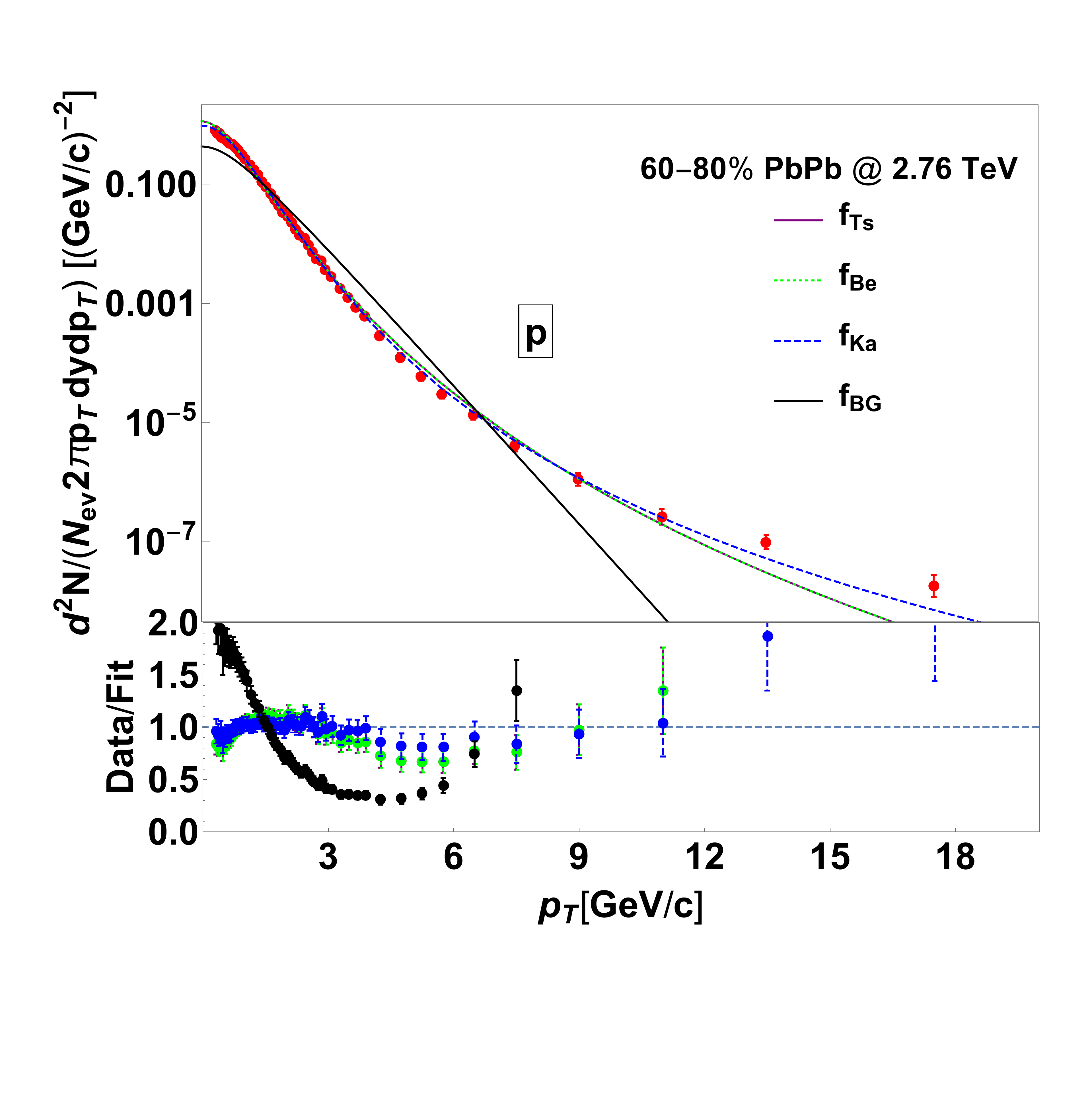}
}
\vspace{-12mm}
\caption{Fittings on $p_T$ spectra of protons in $pPb$ (left) and $PbPb$ (right) collisions at 5.02 and 2.76 TeV respectively by the four different functions in Eq.(\ref{fit-fun}). Ratio of data and fittings is also listed. All spectra are fitted in the range of $p_T$ given in Table \ref{tabAApT}.}
\label{figpT5}
\end{figure*}


\begin{figure*}[htb]
\scalebox{1}[1]{
\includegraphics[width=0.4\linewidth]{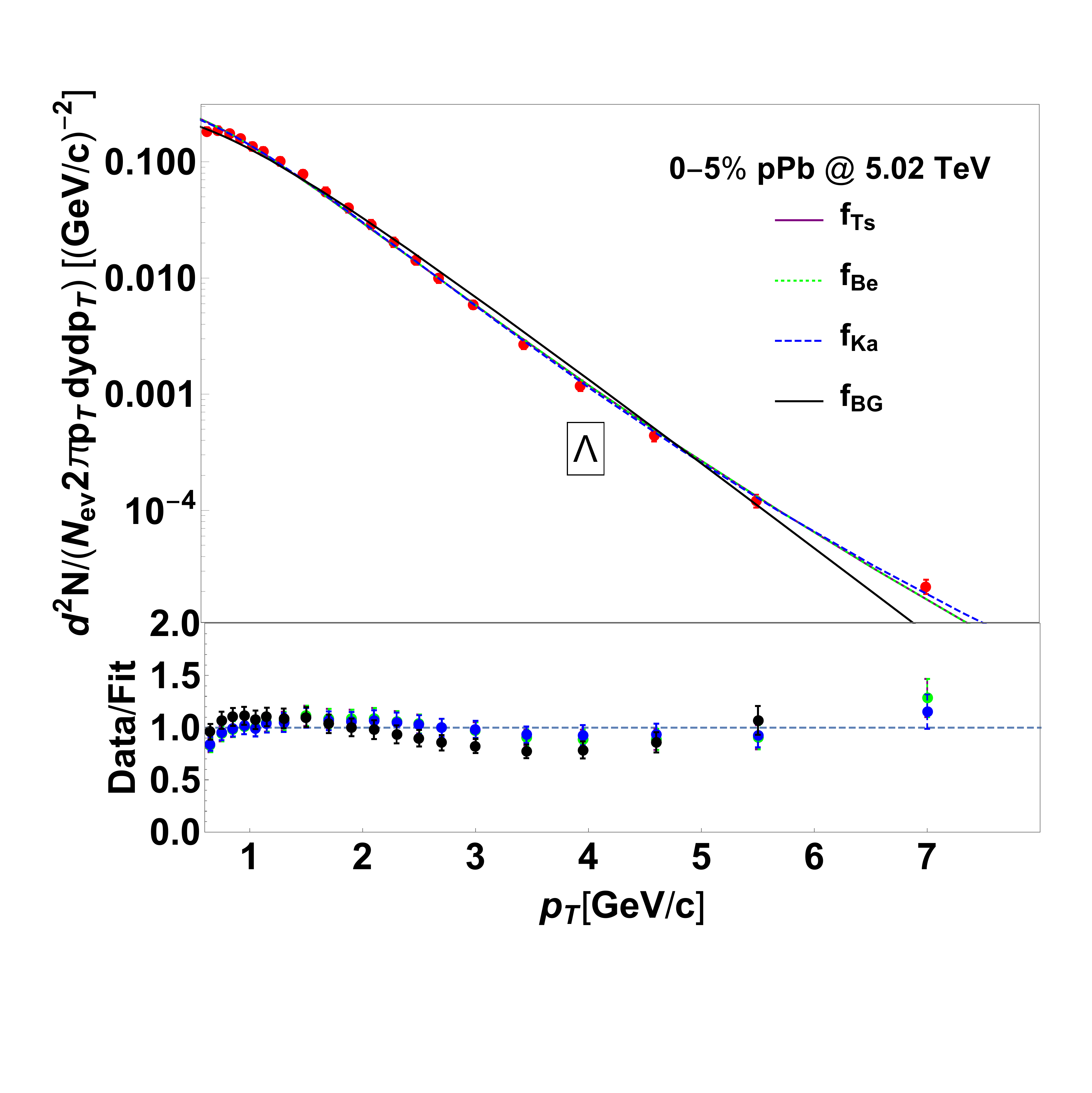}
\includegraphics[width=0.4\linewidth]{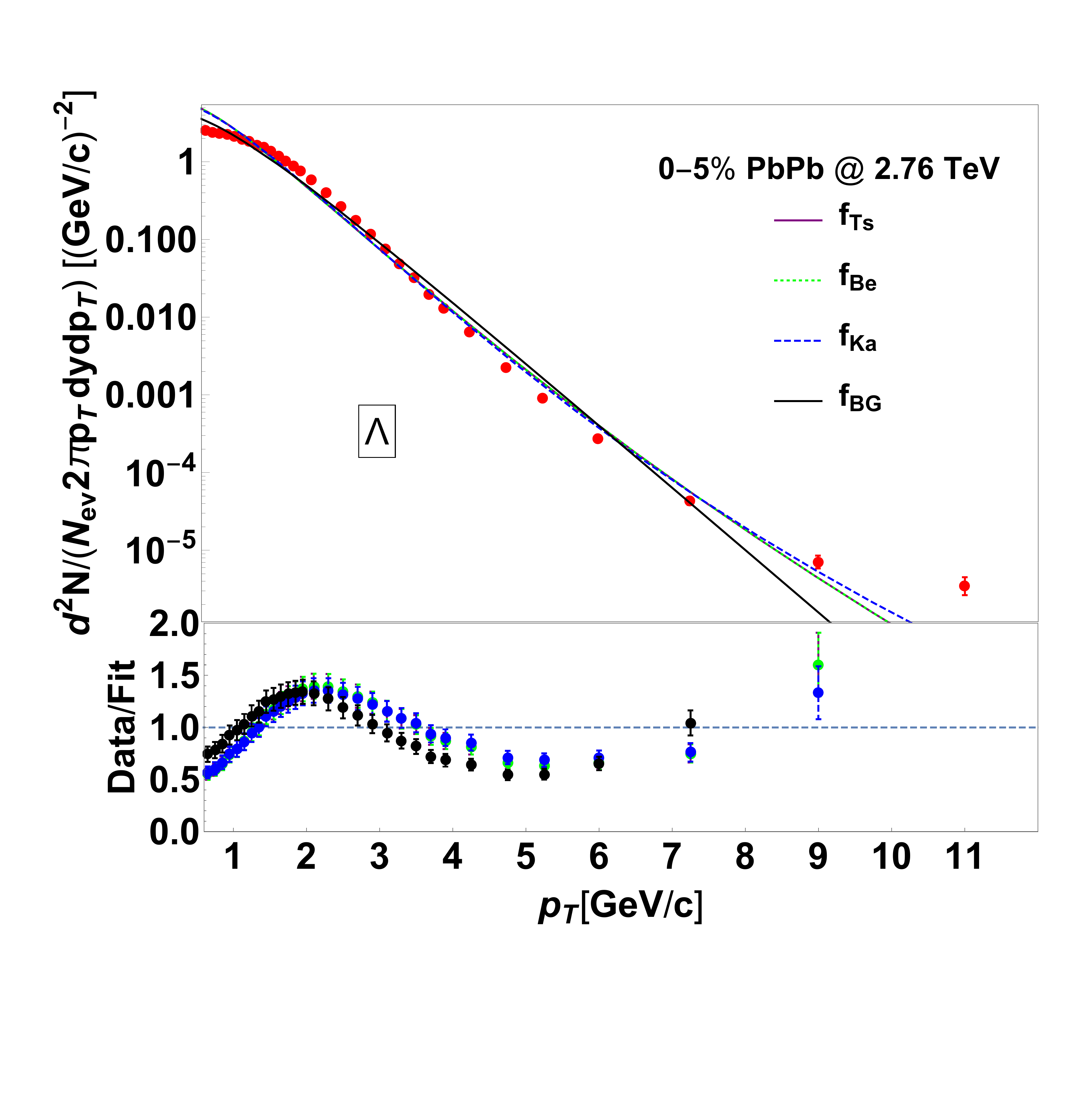}
}
\scalebox{1}[1]{
\includegraphics[width=0.4\linewidth]{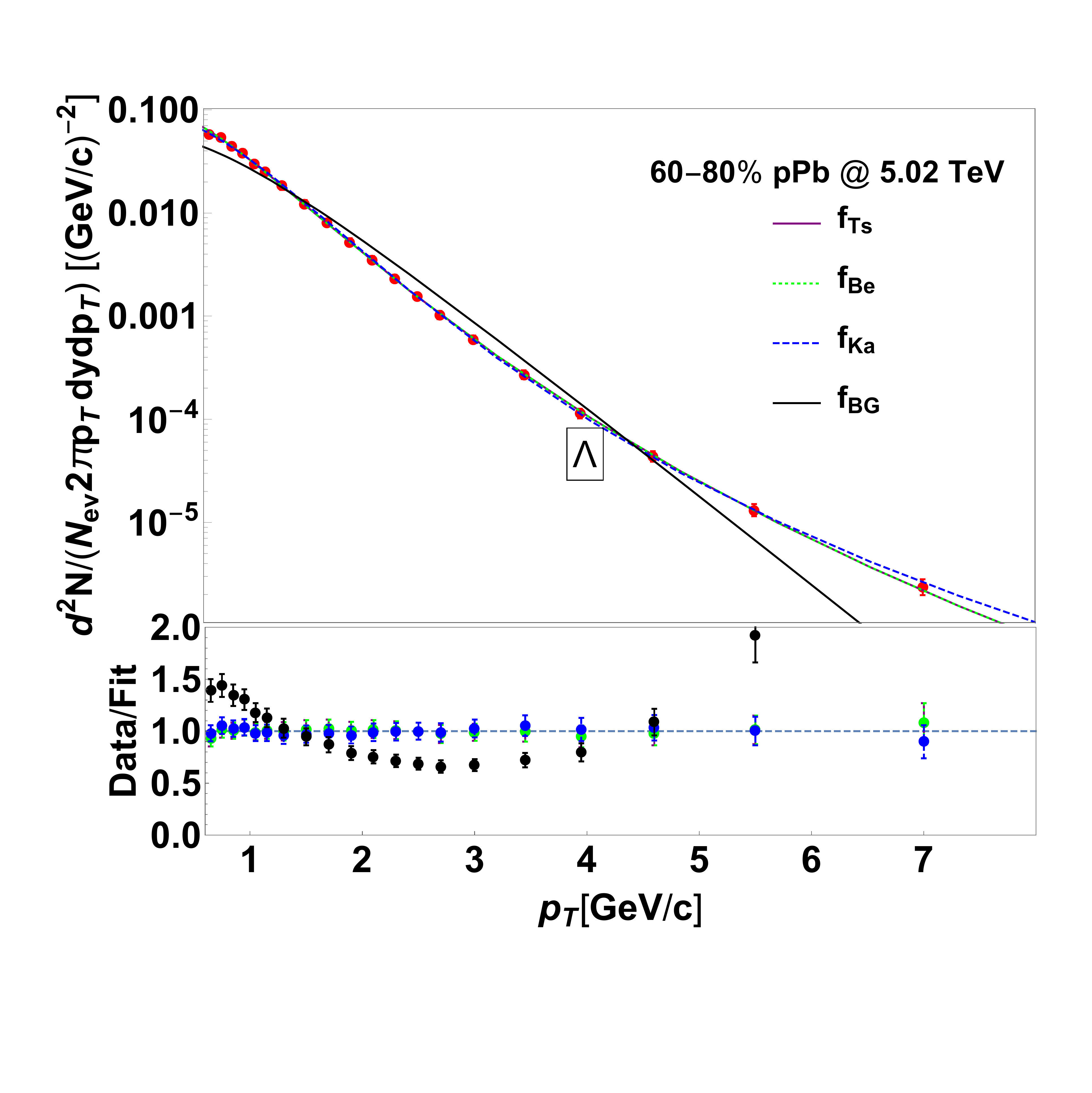}
\includegraphics[width=0.4\linewidth]{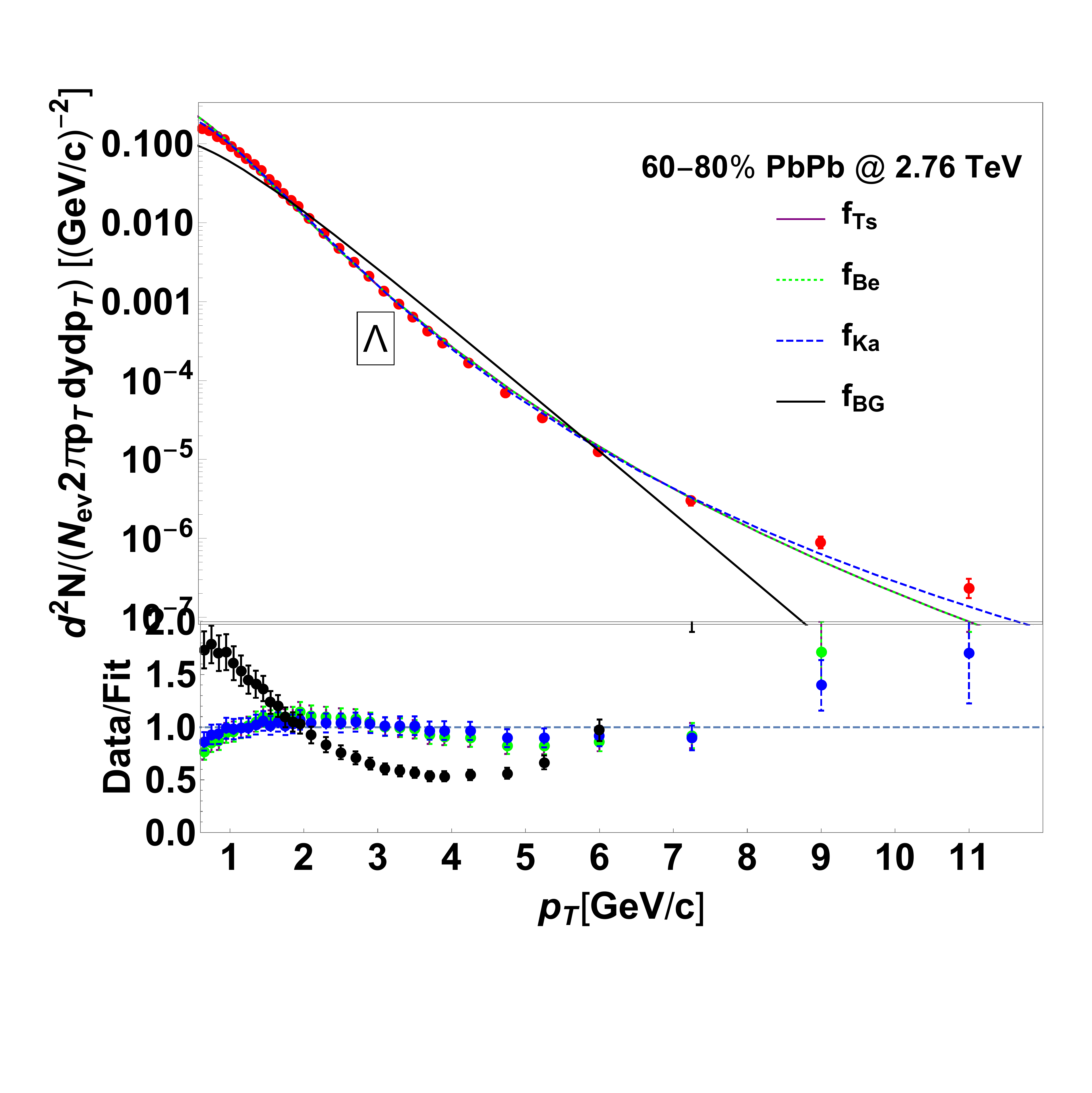}
}
\vspace{-12mm}
\caption{Fittings on $p_T$ spectra of $\Lambda$ in $pPb$ (left) and $PbPb$ (right) collisions at 5.02 and 2.76 TeV respectively by the four different functions in Eq.(\ref{fit-fun}). Ratio of data and fittings is also listed. All spectra are fitted in the range of $p_T$ given in Table \ref{tabAApT}.}
\label{figpT6}
\end{figure*}

\begin{figure*}[htb]
\scalebox{1}[1]{
\includegraphics[width=0.4\linewidth]{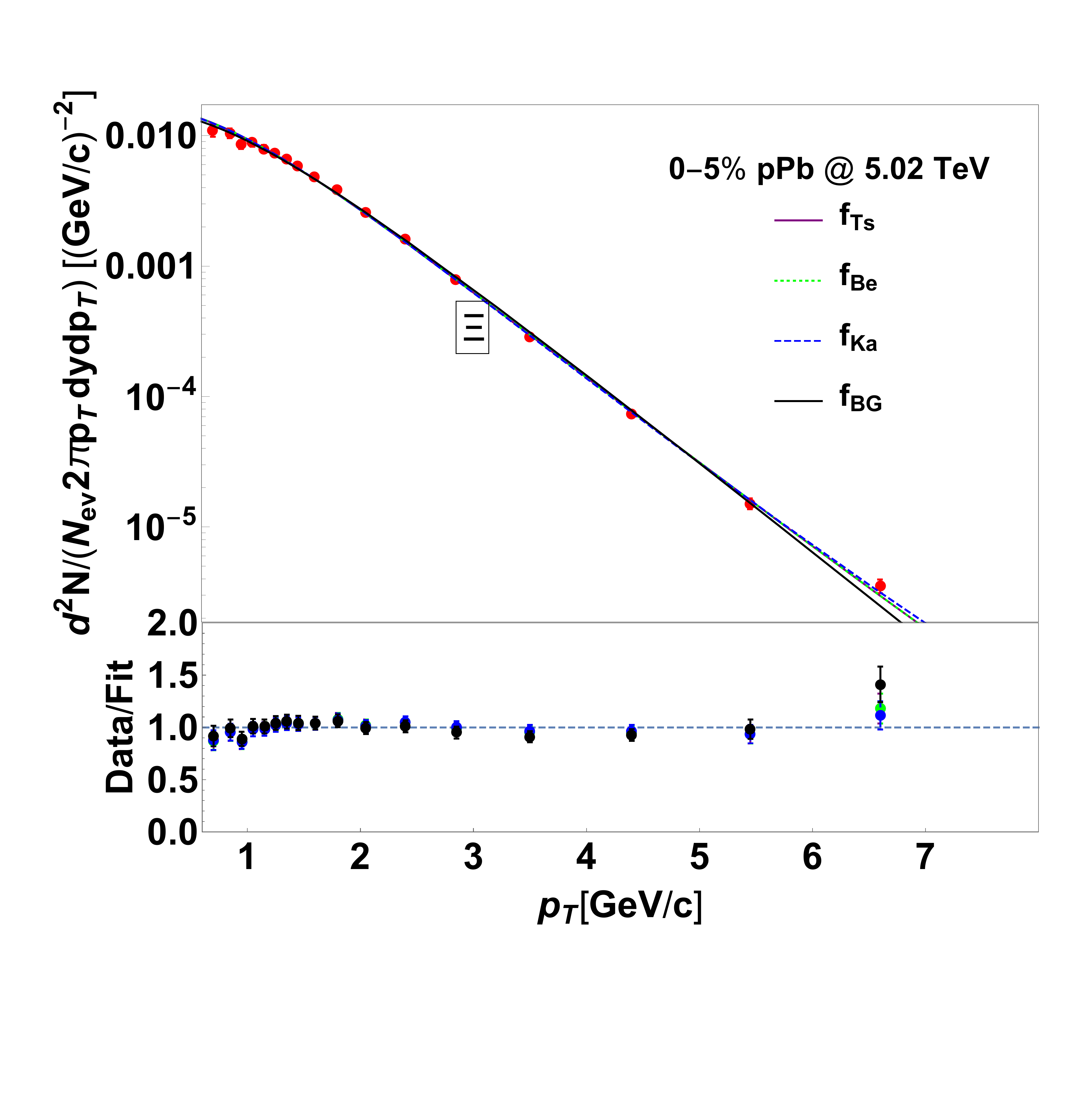}
\includegraphics[width=0.4\linewidth]{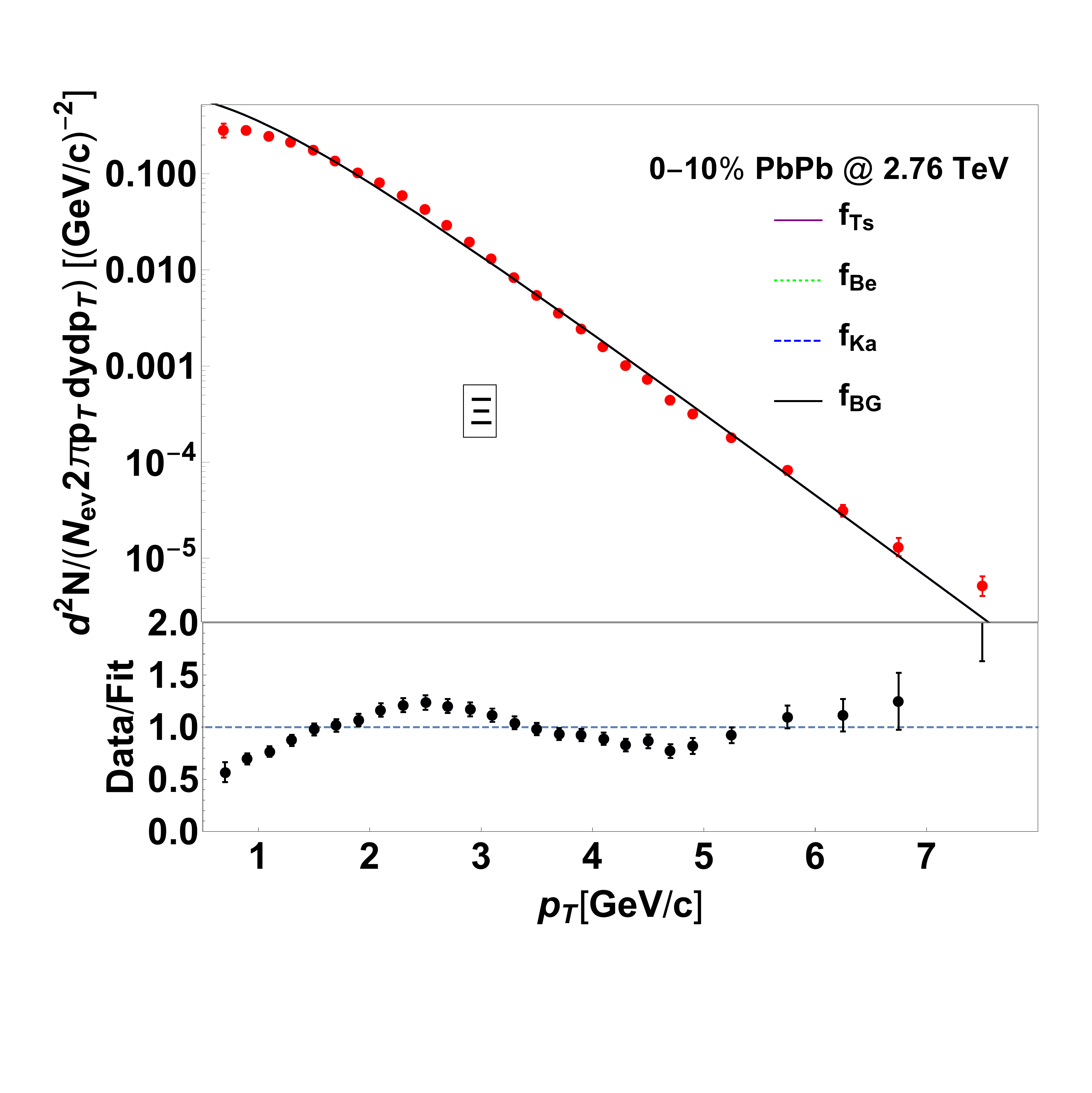}
}
\scalebox{1}[1]{
\includegraphics[width=0.4\linewidth]{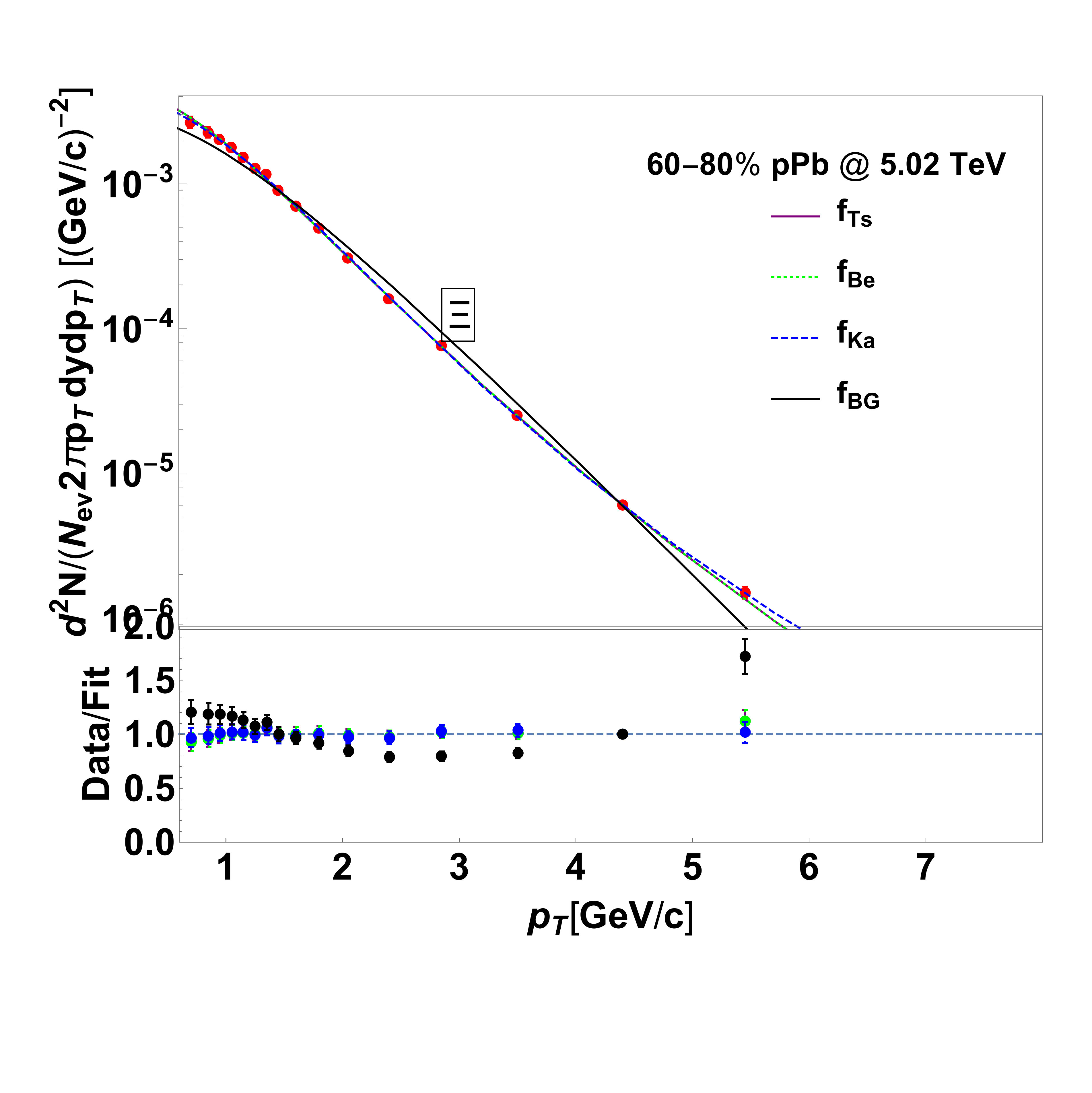}
\includegraphics[width=0.4\linewidth]{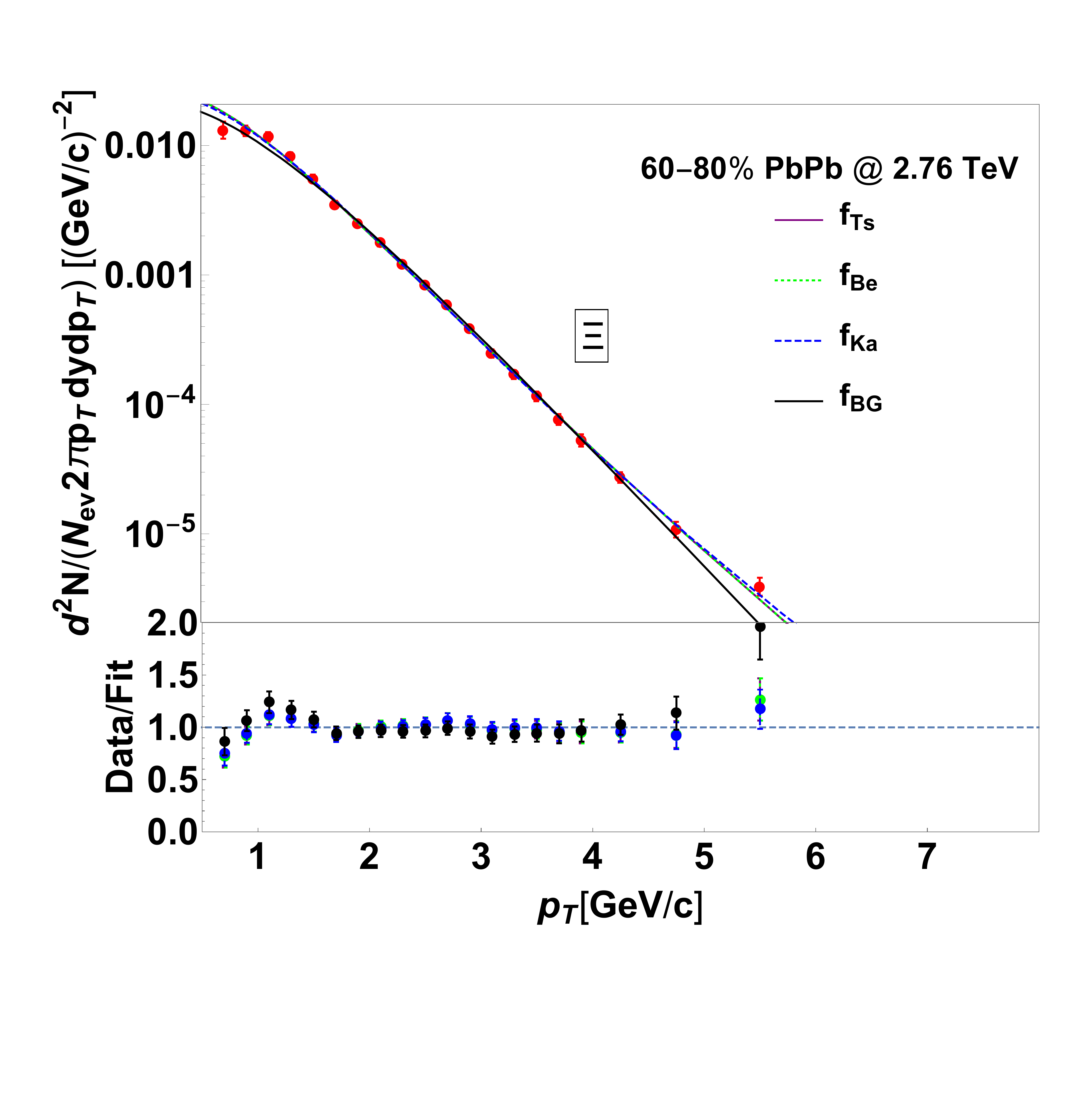}
}
\vspace{-12mm}
\caption{Fittings on $p_T$ spectra of $\Xi$ in $pPb$ (left) and $PbPb$ (right) collisions at 5.02 and 2.76 TeV respectively by the four different functions in Eq.(\ref{fit-fun}). Ratio of data and fittings is also listed. All spectra are fitted in the range of $p_T$ given in Table \ref{tabAApT}.}
\label{figpT7}
\end{figure*}

\begin{figure*}[htb]
\scalebox{1}[1]{
\includegraphics[width=0.4\linewidth]{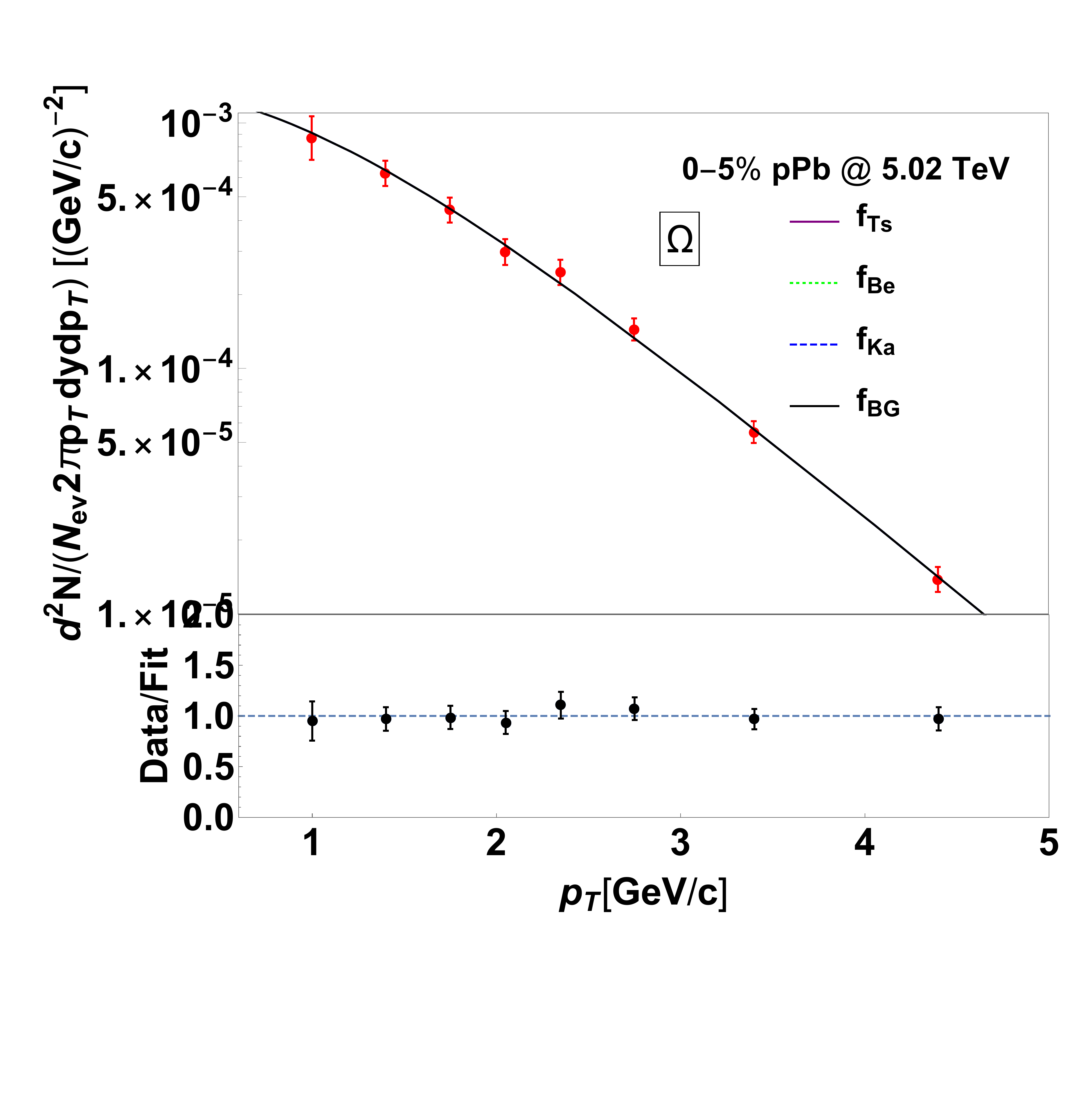}
\includegraphics[width=0.4\linewidth]{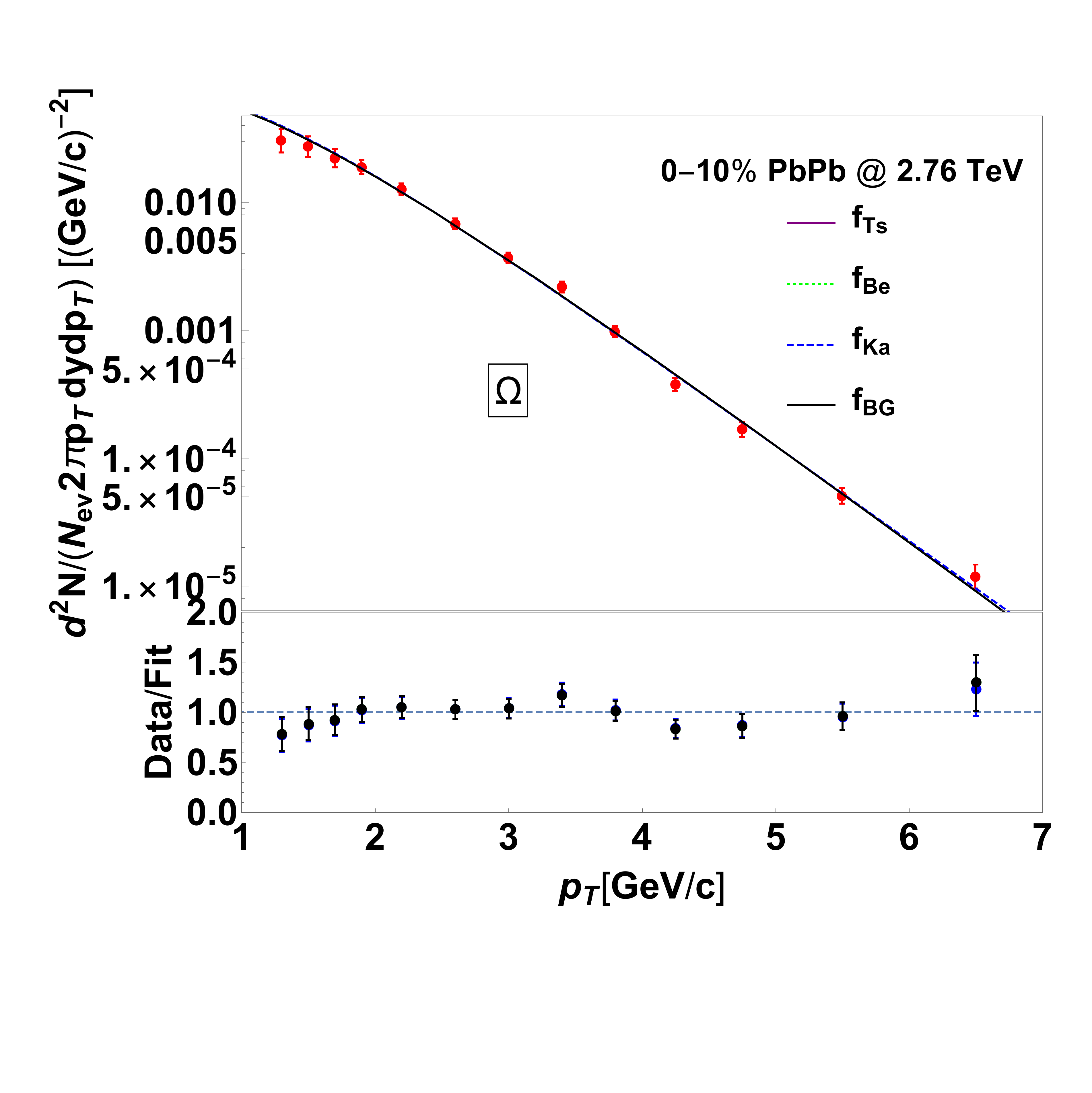}
}
\scalebox{1}[1]{
\includegraphics[width=0.4\linewidth]{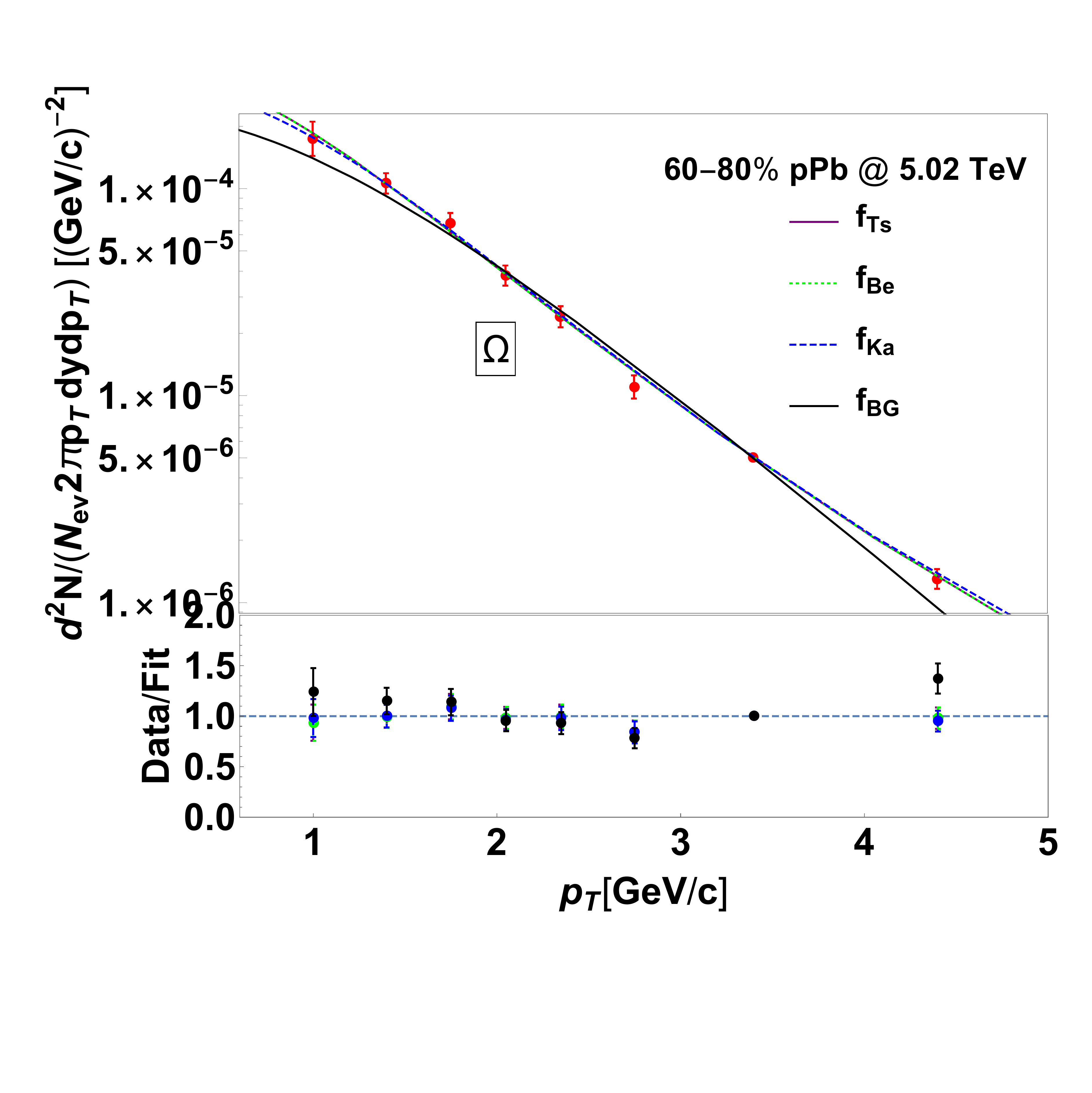}
\includegraphics[width=0.4\linewidth]{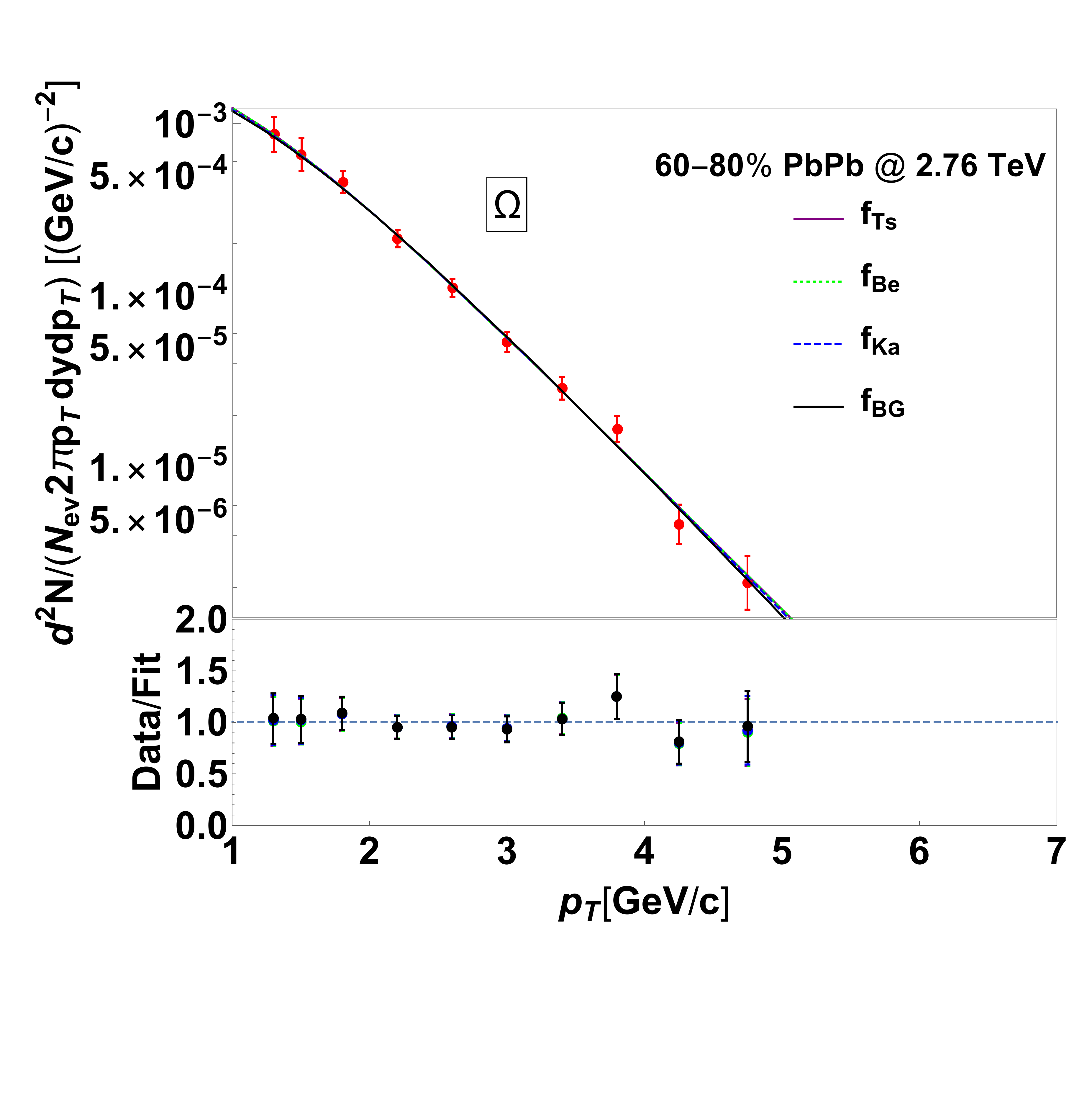}
}
\vspace{-12mm}
\caption{Fittings on $p_T$ spectra of $\Omega$ in $pPb$ (left) and $PbPb$ (right) collisions at 5.02 and 2.76 TeV respectively by the four different functions in Eq.(\ref{fit-fun}). Ratio of data and fittings is also listed. All spectra are fitted in the range of $p_T$ given in Table \ref{tabAApT}.}
\label{figpT8}
\end{figure*}

\subsection{Heavy-ion collisions}

\begin{table}[h]
\caption{Fitting range [GeV/c] of $p_T$ in the spectra in $pPb$ at 5.02 TeV and $PbPb$ at 2.76 TeV collisions for different charged particles:}
\scalebox{1.2}[1.2]{
\begin{tabular}{c c c c}
\hline
\hline
particles &  mass [$GeV/c^2$] & $pPb$ & $PbPb$ \\
\hline
$\pi$ & 0.140 & 0.1-20 & 0.1-20 \\
$K$ & 0.494 &  0.2-20 & 0.2-20 \\
$p$ & 0.938 &  0.3-20  & 0.3-20 \\
$\Lambda$ & 1.116 &  0.6-8.0 & 0.6-8.0 \\
$\Xi$ &  1.321 & 0.6-7.2   &  0.6-8.0 \\
$\Omega$ &  1.672 &  0.8-5.0 &  1.2-6.0 \\
\hline
\hline
\end{tabular}
}
\label{tabAApT}
\end{table}
\begin{table*}[htb]
\caption{Fitting parameters of all four fitting functions on $p_T$ spectra of $0\sim 5\%$ and $60\sim 80\%$ in $pPb$ collisions at 5.02 TeV (for the BG case the non-extensive parameter $1/n_4$ vanishes so that we set $n_4=10^{10}$)}
\scalebox{1}[1.1]{
\begin{tabular}{c c c c c c c c c c c c c c}
\hline
\hline
hadron & centrality & $A_1$ & $A_2$ & $A_3$ & $A_4$ & $T_1$ & $T_2$ & $T_3$ & $T_4$ & $n_1$ & $n_2$ & $n_3$ & $n_4$ \\

\hline
~ & $0\sim 5\%$  & 38.728 & 38.728 & 26.963 & 2.229 & 0.1572 & 0.1836 & 0.2323 & 0.7778 & 6.9516 & 5.9516 & 5.8384 & $10^{10}$ \\

$\pi$ & $60\sim 80\%$  & 26.617 & 26.617 & 18.942 & 2.236 & 0.1404 & 0.1649 & 0.2039 & 0.6092 & 6.7219 & 5.7219 & 5.5284 & $10^{10}$ \\
\hline
~ & $0\sim 5\%$  & 3.2474 & 3.2474 & 2.7545 & 0.9547 & 0.2867 & 0.3269 & 0.3684 & 0.7388 & 8.1341 & 7.1341 & 6.0222 & $10^{10}$ \\

$K$ & $60\sim 80\%$  & 0.9474 & 0.9474 & 0.7566 & 0.2108 & 0.2185 & 0.2559 & 0.2960 & 0.6995 & 6.8412 & 5.8412 & 5.3341 & $10^{10}$ \\
\hline
~ & $0\sim 5\%$  & 0.5916  & 0.5916 & 0.5398 & 0.3318 & 0.3994 & 0.4329 & 0.4624 & 0.6213 & 12.923 & 11.923 & 7.812 & $10^{10}$ \\

$p$ & $60\sim 80\%$  & 0.2271 & 0.2271 & 0.1924 & 0.0751 & 0.2586 & 0.2936 & 0.3284 & 0.5721 & 8.390 & 7.390 & 6.028 & $10^{10}$ \\
\hline
~ & $0\sim 5\%$  & 0.3179 & 0.3179 & 0.3059 & 0.2584 & 0.4794 & 0.5027 & 0.5137 & 0.5808 & 21.539 & 20.539 & 9.087 & $10^{10}$ \\

$\Lambda$ & $60\sim 80\%$  & 0.1090  & 0.1090 & 0.0952 & 0.0592 & 0.3052 & 0.3394 & 0.3658 & 0.4958 & 9.947 & 8.947 & 6.183 & $10^{10}$ \\
\hline
~ & $0\sim 5\%$  & 0.0169  & 0.0169 & 0.0167 & 0.0158 & 0.5798 & 0.5902 & 0.5920 & 0.6182 & 56.737 & 55.737 & 13.726 & $10^{10}$ \\

$\Xi$ & $60\sim 80\%$  &  0.0045  & 0.0045 & 0.0041 & 0.0031 & 0.3754 & 0.4081 & 0.4253 & 0.5258 & 12.505 & 11.505 & 6.447 & $10^{10}$ \\
\hline
~ & $0\sim 5\%$  & 0.0014  & 0.0014 & 0.0014 & 0.0014 & 0.6627 & 0.6627 & 0.6627 & 0.6627 & 278307000 & 88721200 & 39944600 & $10^{10}$  \\

$\Omega$ & $60\sim 80\%$  &  0.00038  & 0.00038 & 0.00033 & 0.00023 & 0.3653 & 0.4187 & 0.4427 & 0.5518 & 7.841 & 6.841 & 4.644 & $10^{10}$ \\
\hline
\hline
\end{tabular}
}
\label{tabpApar}
\end{table*}

\begin{table}[htb]
\caption{Fitting parameters of all four fitting functions on $p_T$ spectra of $0\sim 5\%$ and $60\sim 80\%$ in $PbPb$ collisions at 2.76 TeV (for the BG case the non-extensive parameter $1/n_4$ vanishes so that we set $n_4=10^{10}$)}
\scalebox{1}[1.1]{
\begin{tabular}{c c c c c c c c c c c c c c}
\hline
\hline
hadron & centrality & $A_1$ & $A_2$ & $A_3$ & $A_4$ & $T_1$ & $T_2$ & $T_3$ & $T_4$ & $n_1$ & $n_2$ & $n_3$ & $n_4$ \\
\hline
~ & $0\sim 5\%$  & 4528.32 & 4528.32 & 2555.84 & 116.64 & 0.1479 & 0.1695 & 0.2258 & 0.7367 & 7.838 & 6.838 & 6.614 & $10^{10}$ \\

$\pi$ & $60\sim 80\%$  & 154.251 & 154.251 & 99.714 & 7.678 & 0.1425 & 0.1652 & 0.2096 & 0.6755 & 7.278 & 6.278 & 6.032 & $10^{10}$ \\
\hline
~ & $0\sim 5\%$  & 184.393 & 184.393 & 136.92 & 19.617 & 0.2280 & 0.2584 & 0.3112 & 0.7757 & 8.498 & 7.498 & 6.706 & $10^{10}$ \\

$K$ & $60\sim 80\%$  & 6.142 & 6.142  & 4.813 & 1.235 & 0.2143 & 0.2469 & 0.2891 & 0.7269 & 7.588 & 6.588 & 5.941 & $10^{10}$ \\
\hline
~ & $0\sim 5\%$  & 18.517  & 18.517 & 17.193 & 10.907 & 0.4243 & 0.4455 & 0.4688 & 0.5798 & 20.976 & 19.976 & 10.993 & $10^{10}$ \\

$p$ & $60\sim 80\%$  & 1.1397 & 1.1397  & 0.9659 & 0.4303 & 0.2910 & 0.3205 & 0.3556 & 0.5575 & 10.871 & 9.871 & 7.356 & $10^{10}$ \\
\hline
~ & $0\sim 5\%$  & 6.8431 & 6.8431 & 6.4938 & 4.7167 & 0.4315 & 0.4470 & 0.4631 & 0.5339 & 28.901 & 27.901 & 12.454 & $10^{10}$ \\

$\Lambda$ & $60\sim 80\%$  & 0.3579  & 0.3579 & 0.2884 & 0.1238 & 0.2995 & 0.3298 & 0.3712 & 0.5444 & 10.883 & 9.883 & 7.192 & $10^{10}$ \\
\hline
~ & $0\sim 10\%$  & 0.7014  & 0.7014 & 0.7014 & 0.7014 & 0.5001 & 0.5001 & 0.5001 & 0.5001 & 17931200 & 101584000 & 5246090 & $10^{10}$ \\

$\Xi$ & $60\sim 80\%$  &  0.0278  & 0.0278 & 0.02632  & 0.02206 & 0.3985 & 0.4142 & 0.4256 & 0.4656 & 26.361 & 25.361 & 10.169 & $10^{10}$ \\
\hline
~ & $0\sim 10\%$  & 0.0877  & 0.0877 & 0.0900 & 0.0877 & 0.5497 & 0.5497 & 0.5433 & 0.5497 & 35680800 & 46951700 & 31.1927 & $10^{10}$  \\

$\Omega$ & $60\sim 80\%$  &  0.0022  & 0.0022 & 0.0021 & 0.0021 & 0.4697 & 0.4761 & 0.4858 & 0.4924 & 74.436 & 73. 436 & 22.496 & $10^{10}$ \\
\hline
\hline
\end{tabular}
}
\label{tabAApar}
\end{table}

\begin{table}[htb]
\caption{Values of $\chi^2/ndf$ of all four fitting functions on $p_T$ spectra of $0\sim 5\%$ and $60\sim 80\%$ in $pPb$ collisions at 5.02 TeV:}
\scalebox{1}[1.1]{
\begin{tabular}{c c c c c c c}
\hline
\hline
 particles & centrality & NDF & $f_{Ts}$ & $f_{Be}$ & $f_{Ka}$ & $f_{BG}$ \\
\hline
~ & $0\sim 5\%$  & 58 & 2.46153 & 2.46153 & 12.6916 & 1102.45 \\

$\pi$ & $60\sim 80\%$  & 58 & 1.69764 & 1.69764 & 13.7005 & 1155.55 \\
\hline
~ & $0\sim 5\%$  & 51 & 1.93262 & 1.93262 & 0.460774 & 242.112 \\

$K$ & $60\sim 80\%$  & 51 & 0.3036 & 0.3036 & 2.15548 & 341.367 \\
\hline
~ & $0\sim 5\%$  & 49  & 2.90907 & 2.90907 & 0.647414 & 64.0674 \\

$p$ & $60\sim 80\%$  & 49 & 0.695845 & 0.695845 & 0.759807 & 161.188 \\
\hline
~ & $0\sim 5\%$  & 20 & 1.15437 & 1.15437 & 0.691625 & 4.29884 \\

$\Lambda$ & $60\sim 80\%$  & 20  & 0.114131 & 0.114131 & 0.141428 & 18.6023 \\
\hline
~ & $0\sim 5\%$  & 17  & 0.957339 & 0.957339 & 0.779487 & 1.28519 \\

$\Xi$ & $60\sim 80\%$  &  16  & 0.309287 & 0.309287 & 0.197367 & 8.3288 \\
\hline
~ & $0\sim 5\%$  & 8  & 0.357513 & 0.357513 & 0.357513 & 0.297928  \\

$\Omega$ & $60\sim 80\%$  &  8  & 0.493654 & 0.493654 & 0.527026 & 2.78109 \\
\hline
\hline
\end{tabular}
}
\label{tabpAf}
\end{table}

\begin{table}[htb]
\caption{Values of $\chi^2/ndf$ of all four fitting functions on $p_T$ spectra of $0\sim 5\%$ and $60\sim 80\%$ in $PbPb$ collisions at 2.76 TeV:}
\scalebox{1}[1]{
\begin{tabular}{c c c c c c c}
\hline
\hline
 particles & centrality & NDF & $f_{Ts}$ & $f_{Be}$ & $f_{Ka}$ & $f_{BG}$ \\
\hline
~ & $0\sim 5\%$  & 63 & 14.8666 & 14.8666 & 6.53005 & 812.613 \\

$\pi$ & $60\sim 80\%$  & 63 & 1.22957 & 1.22957 & 4.20393 & 579.611 \\
\hline
~ & $0\sim 5\%$  & 58 & 17.3383 & 17.3383 & 7.75522 & 369.32 \\

$K$ & $60\sim 80\%$  & 58 & 0.905461 & 0.905461 & 0.485381 & 176.925 \\
\hline
~ & $0\sim 5\%$  & 49  & 22.0948 & 22.0948 & 15.0886 & 59.2162 \\

$p$ & $60\sim 80\%$  & 49 & 2.63246 & 2.63246 & 0.619071 & 75.2498 \\
\hline
~ & $0\sim 5\%$  & 31 & 13.2339 & 13.2339 & 10.1799 & 21.4633 \\

$\Lambda$ & $60\sim 80\%$  & 31  & 2.14001 & 2.14001 & 0.631195 & 41.5449 \\
\hline
~ & $0\sim 10\%$  & 27  & 6.55693 & 6.55693 & 6.55693 & 6.29465 \\

$\Xi$ & $60\sim 80\%$  &  20  & 0.849086 & 0.849086 & 0.725044 & 2.11456 \\
\hline
~ & $0\sim 10\%$  & 13  & 1.03585 & 1.03585 & 1.02259 & 0.941685  \\

$\Omega$ & $60\sim 80\%$  &  10  & 0.466724 & 0.466724 & 0.475601 & 0.419032 \\
\hline
\hline
\end{tabular}
}
\label{tabAAf}
\end{table}

\begin{figure}[htb]
\scalebox{1}[1]{
\includegraphics[width=0.4\linewidth]{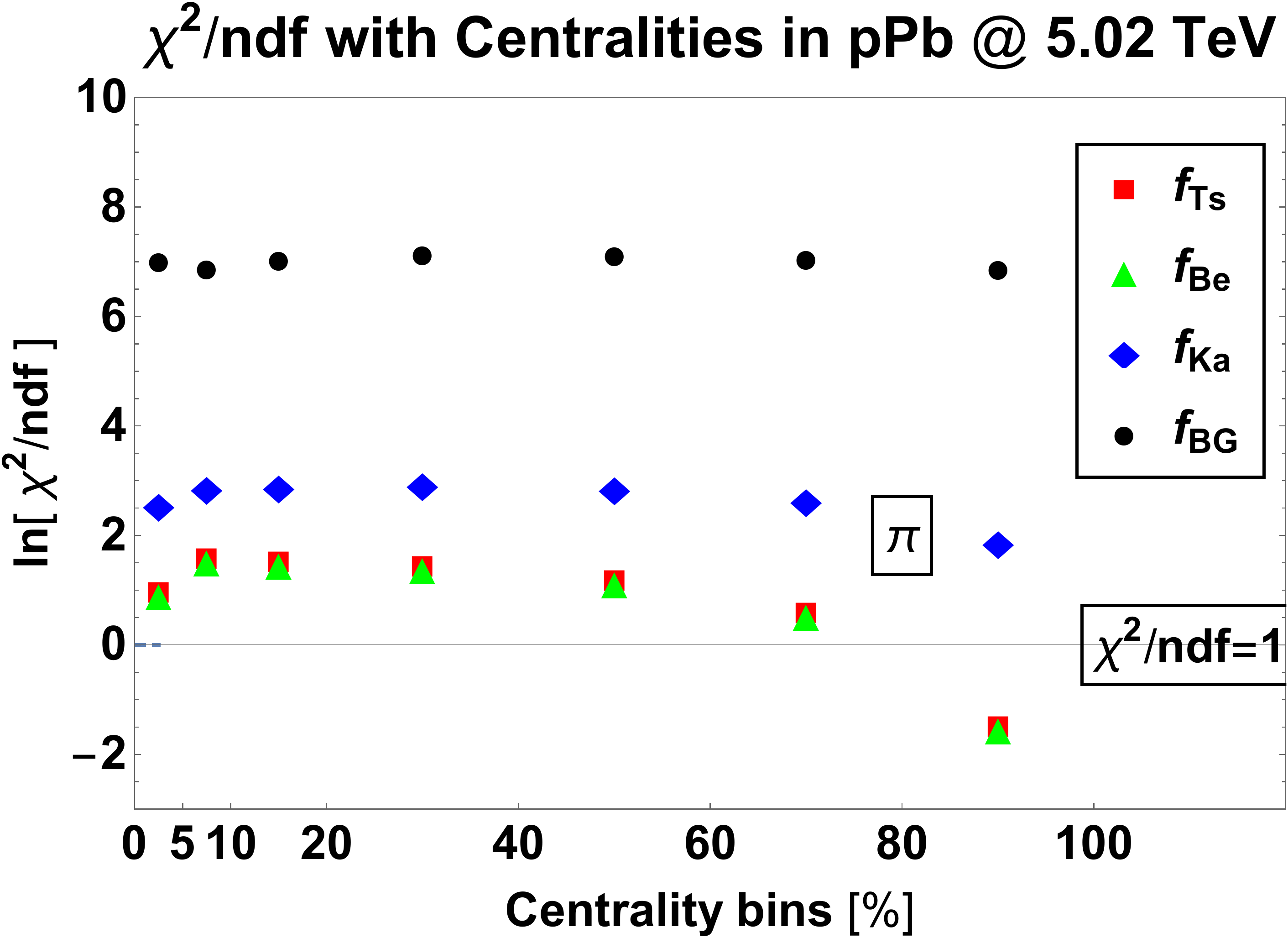}
\includegraphics[width=0.4\linewidth]{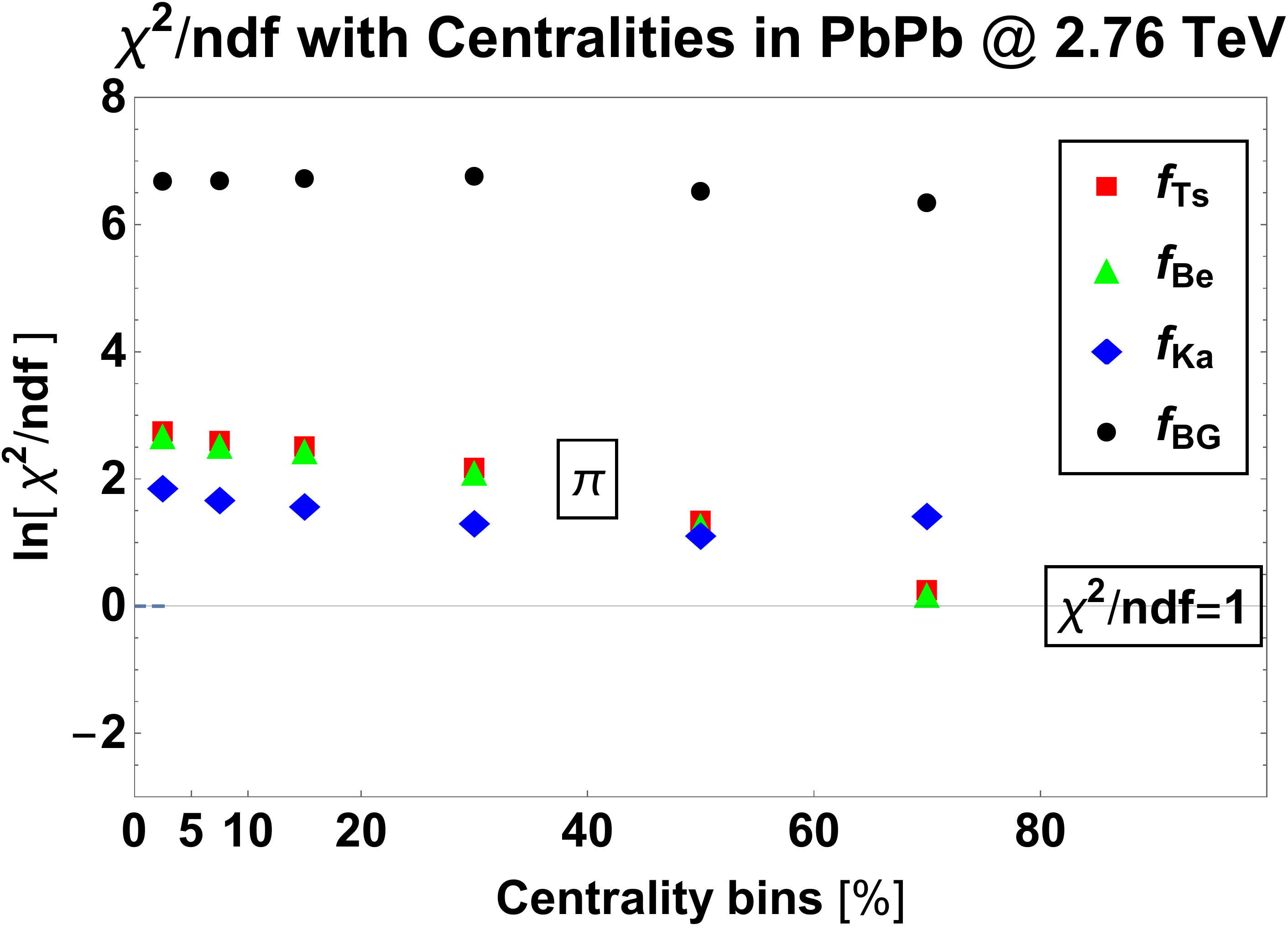}
}
\scalebox{1}[1]{
\includegraphics[width=0.4\linewidth]{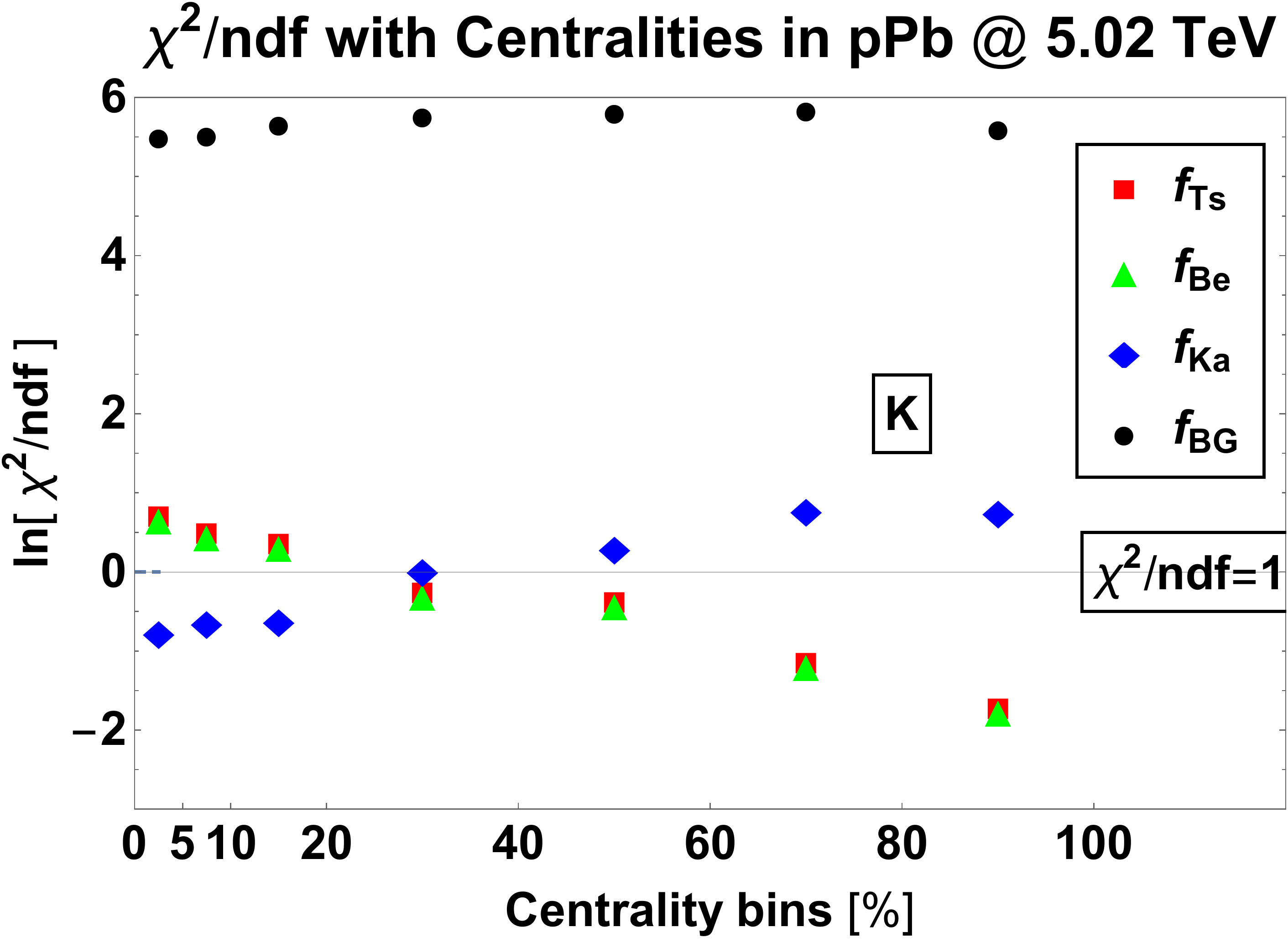}
\includegraphics[width=0.4\linewidth]{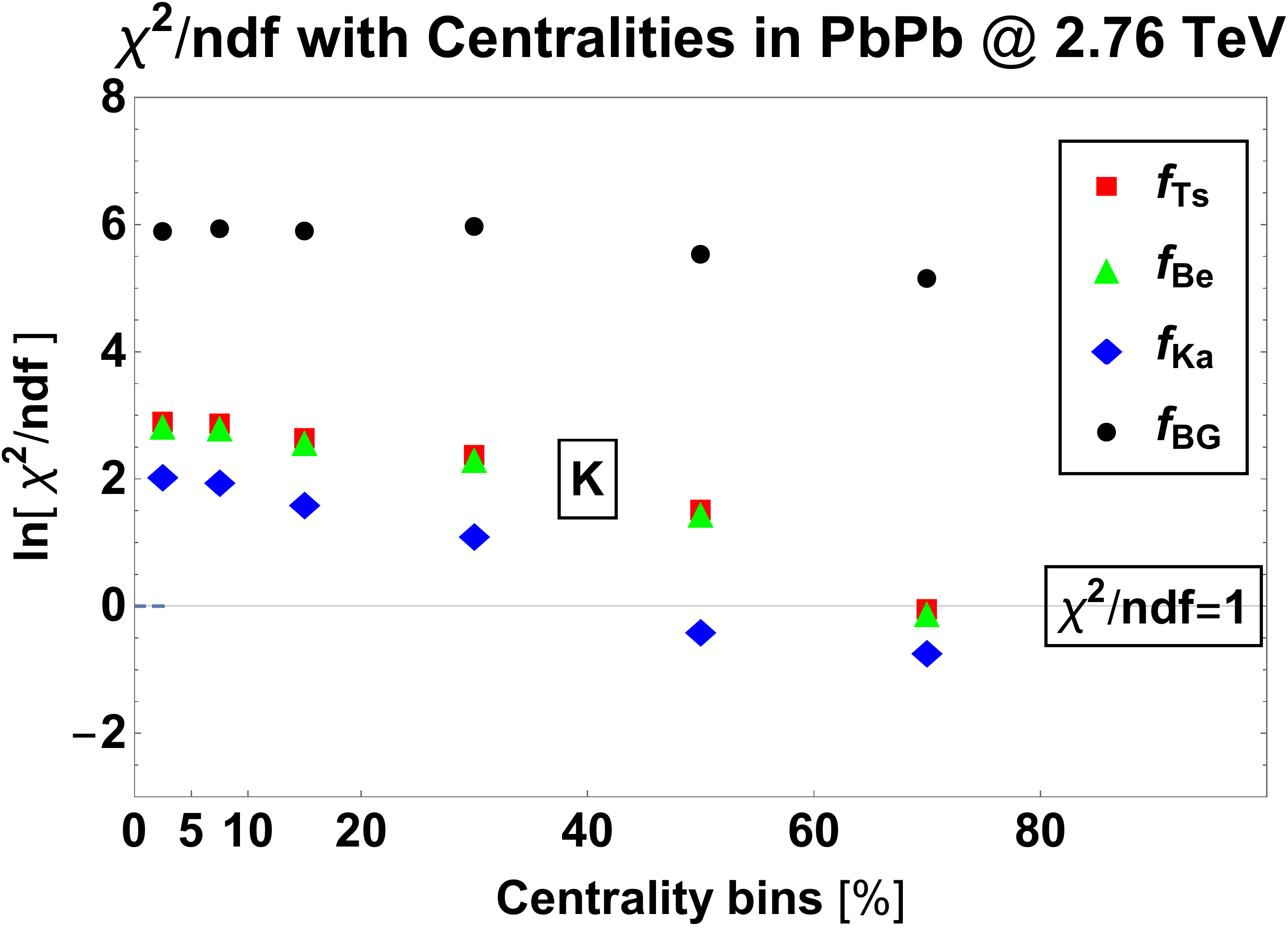}
}
\scalebox{1}[1]{
\includegraphics[width=0.4\linewidth]{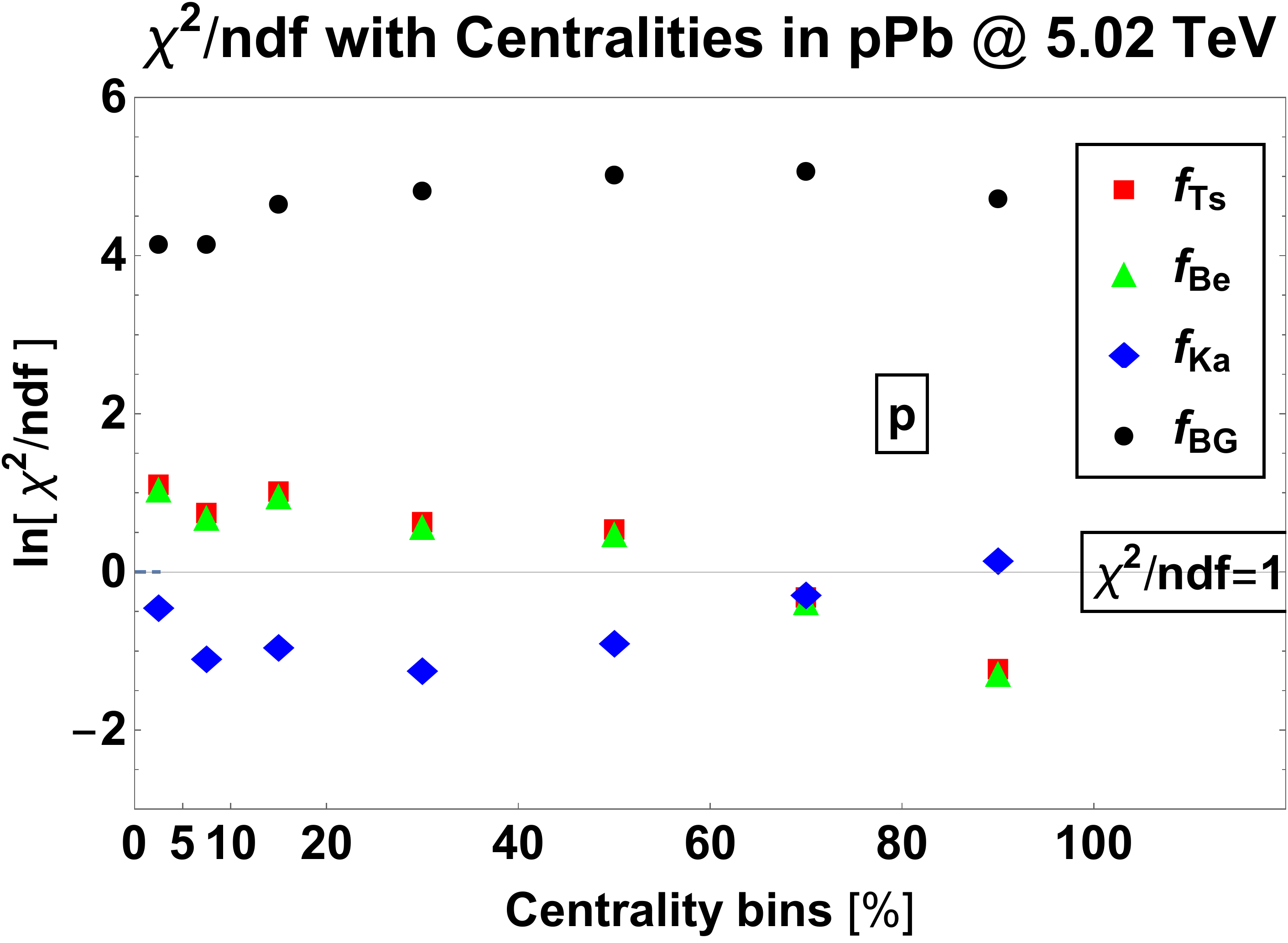}
\includegraphics[width=0.4\linewidth]{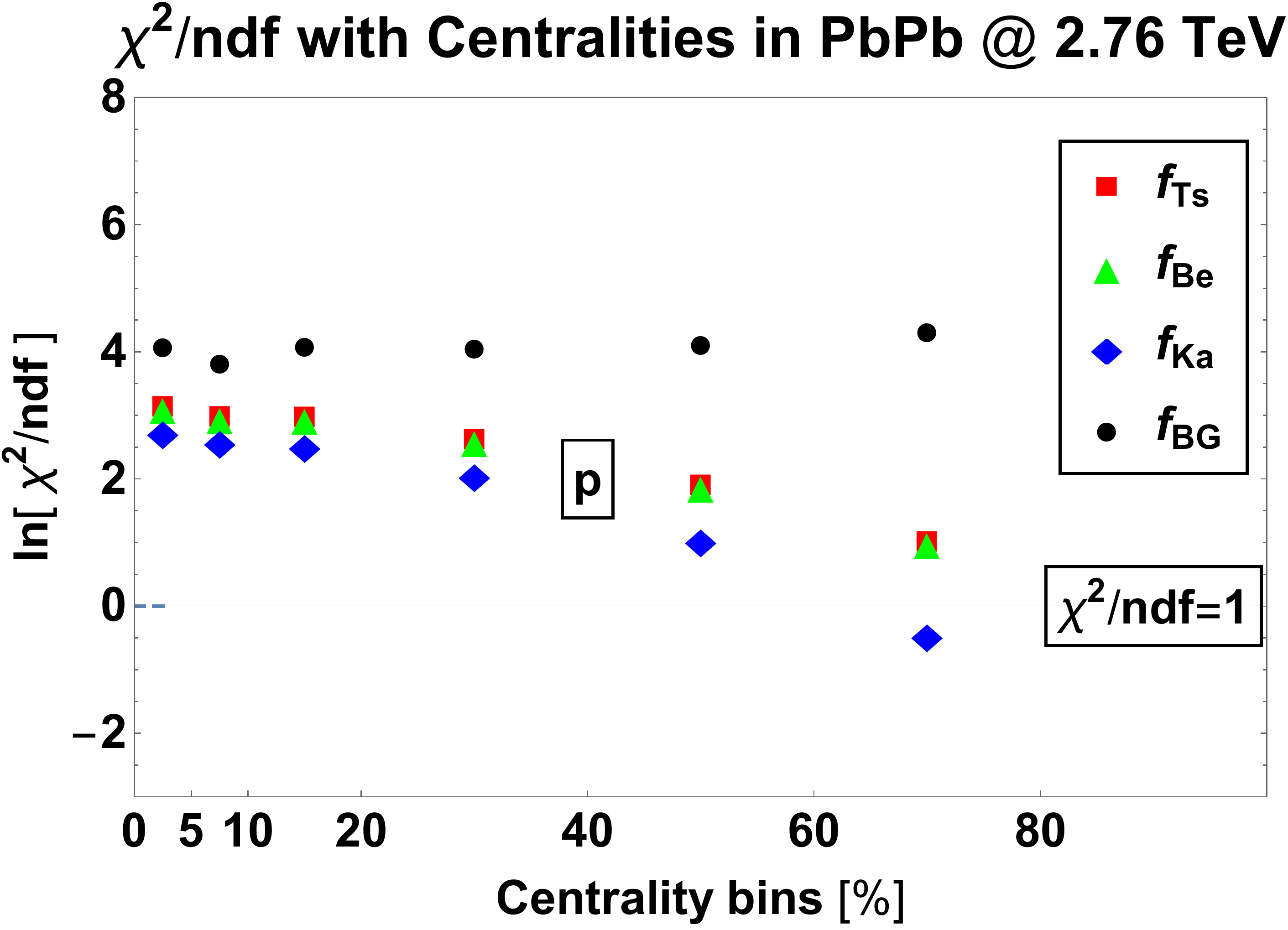}
}
\caption{$\chi^2/ndf$ of all four fittings on $p_T$ spectra of identified hadrons ($\pi$, $K$ and $p$) in $pPb$ at 5.02 TeV (left) and $PbPb$ at 2.76 TeV (right) collisions respectively with all centralities. The fitting error $\chi^2/ndf =1$ is also given as a baseline. Note that the $\chi^2/ndf$ are shown as its logarithmic values for comparisons. Exact values are also listed in Tables \ref{tabpAf} and \ref{tabAAf}. }
\label{figxi2}
\end{figure}

\begin{figure}[htb]
\scalebox{1}[1]{
\includegraphics[width=0.4\linewidth]{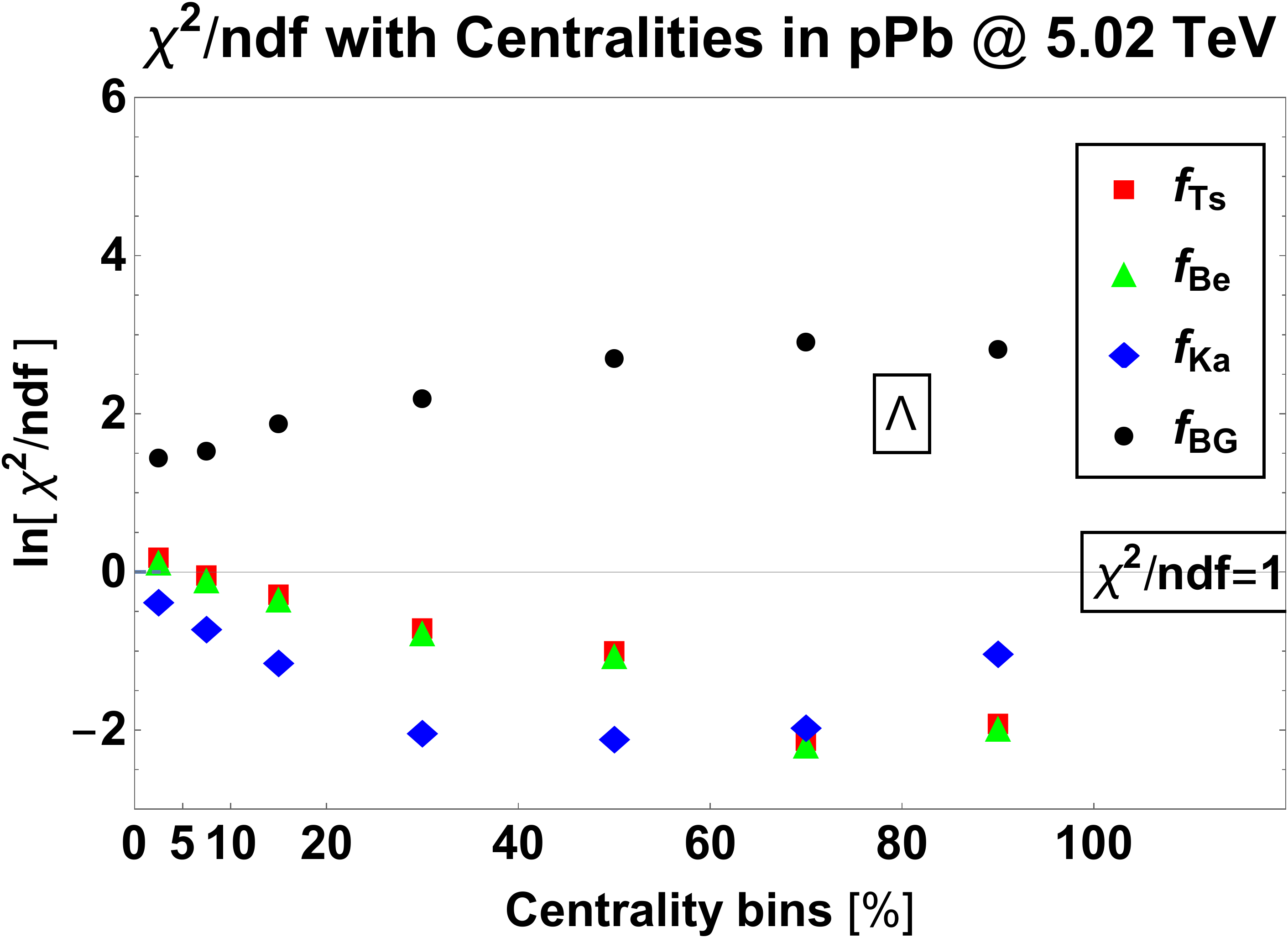}
\includegraphics[width=0.4\linewidth]{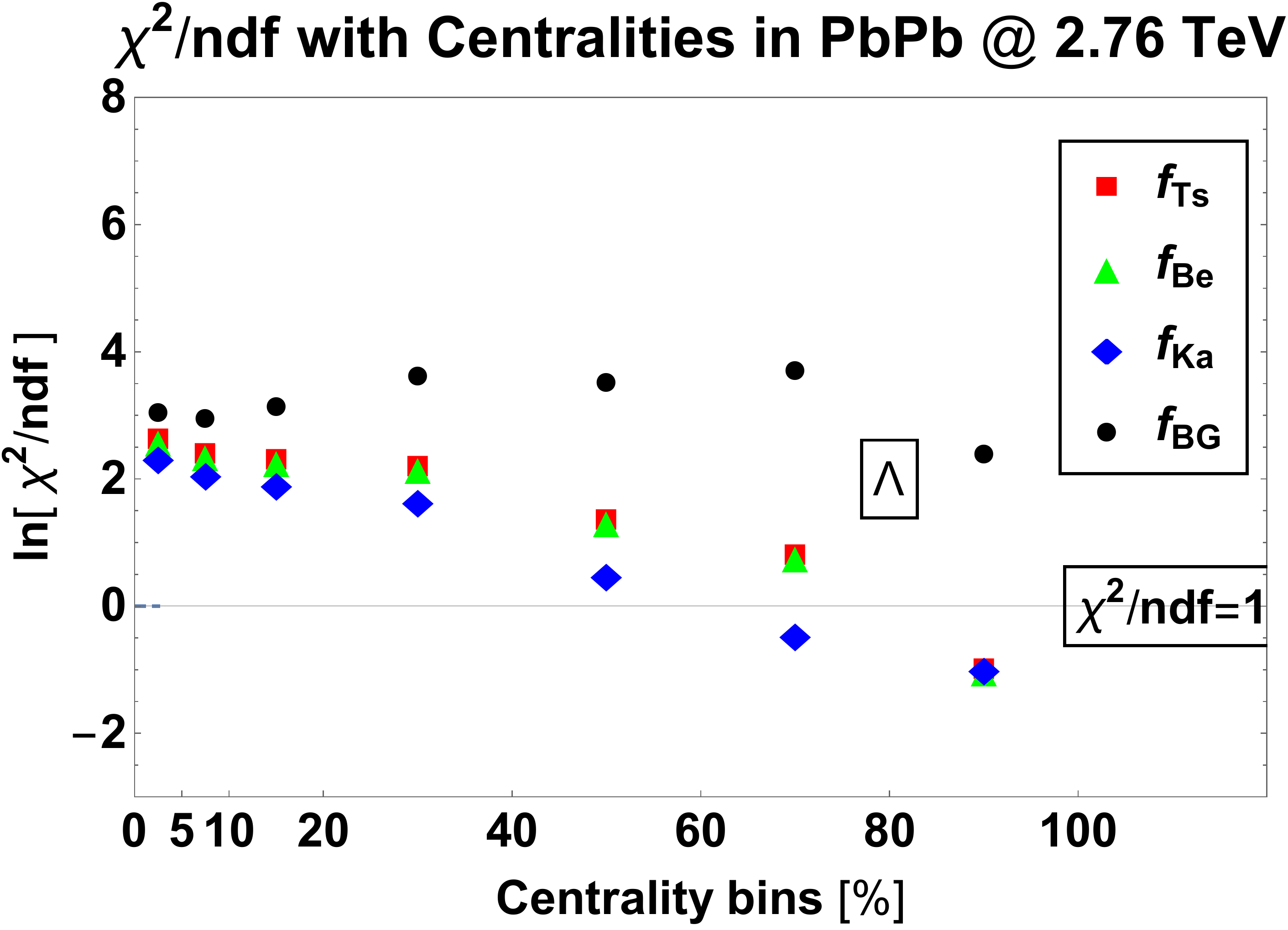}
}
\scalebox{1}[1]{
\includegraphics[width=0.4\linewidth]{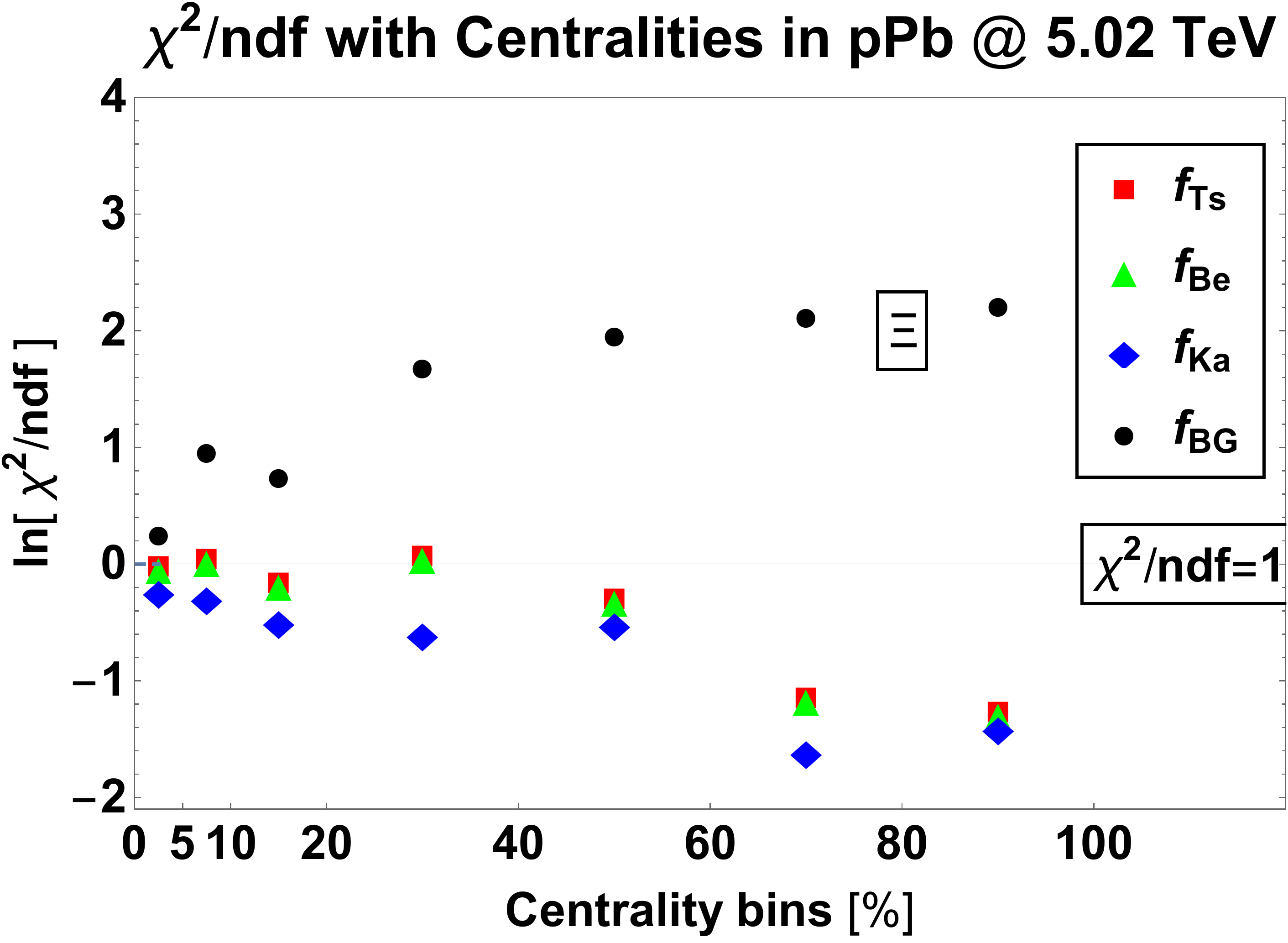}
\includegraphics[width=0.4\linewidth]{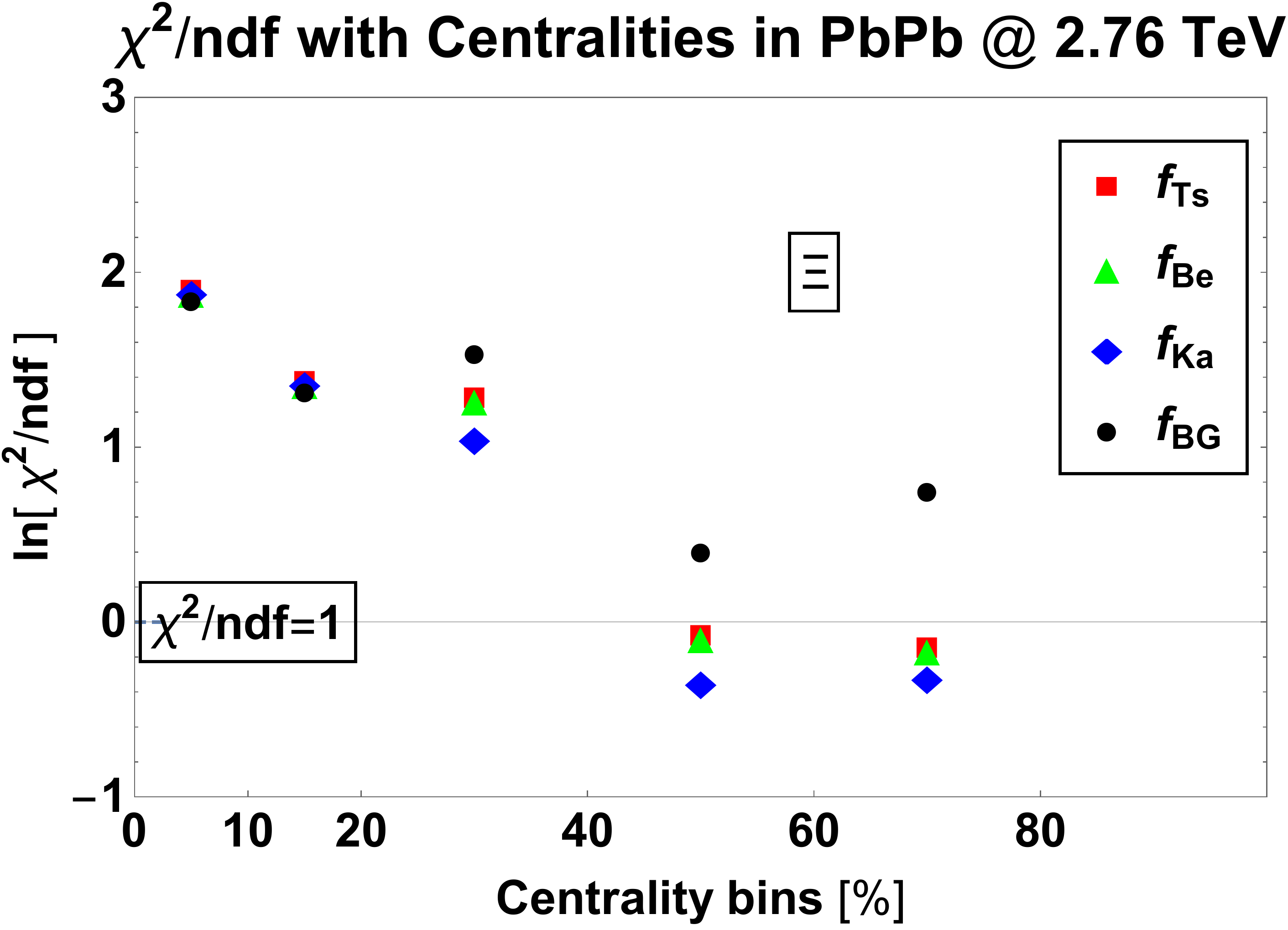}
}
\scalebox{1}[1]{
\includegraphics[width=0.4\linewidth]{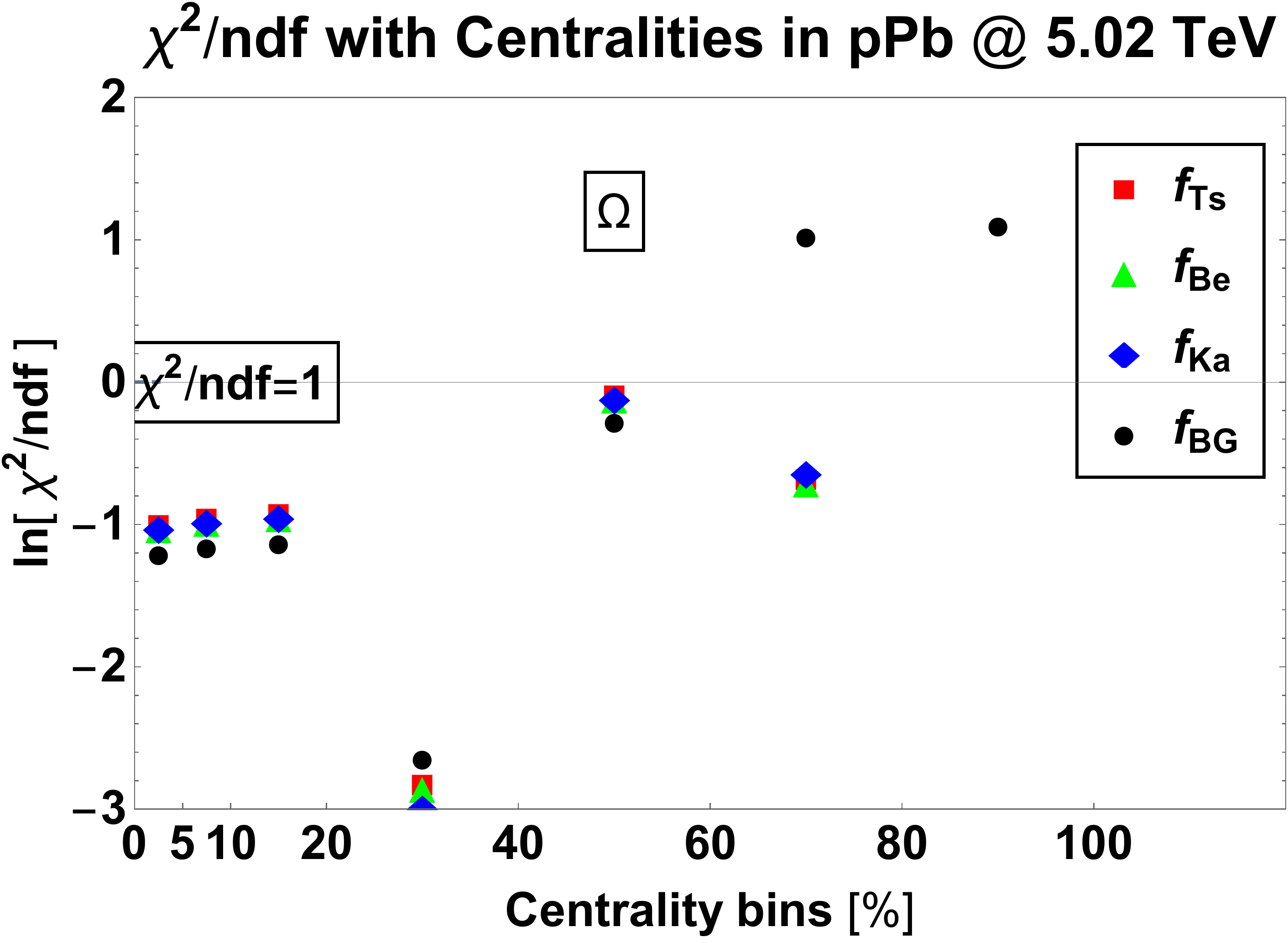}
\includegraphics[width=0.4\linewidth]{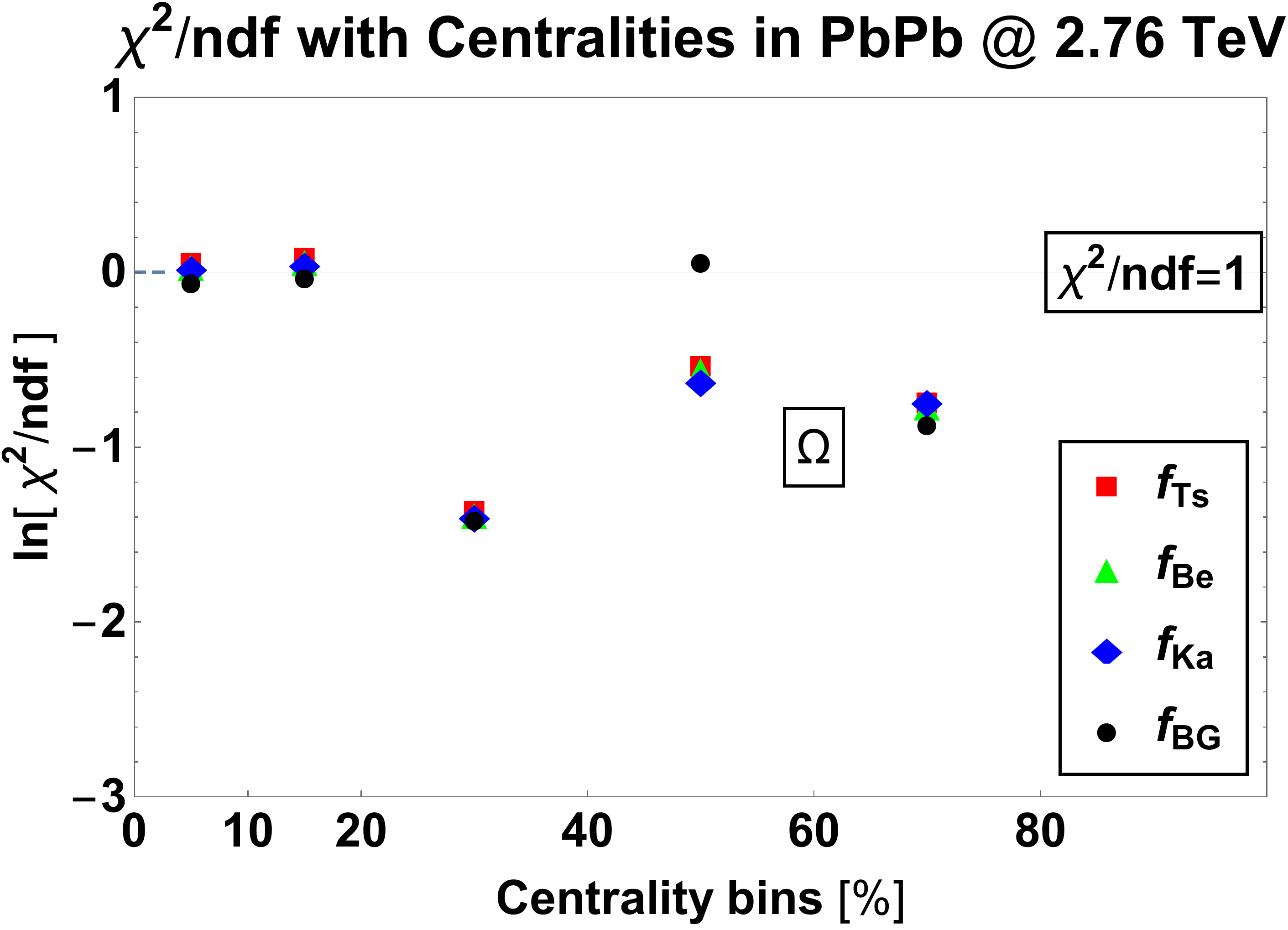}
}
\caption{$\chi^2/ndf$ of all four fittings on $p_T$ spectra of identified hadrons ($\Lambda$, $\Xi$ and $\Omega$) in $pPb$ (left) and $PbPb$ (right) collisions at 5.02 and 2.76 TeV respectively with all centralities. The fitting error $\chi^2/ndf =1$ is also given as a baseline. Note that the $\chi^2/ndf$ are shown as its logarithmic values for comparisons. Exact values are also listed in Tables \ref{tabpAf} and \ref{tabAAf}.}
\label{figxi3}
\end{figure}


\begin{figure}[htb]
\scalebox{1}[1]{
\includegraphics[width=0.45\linewidth]{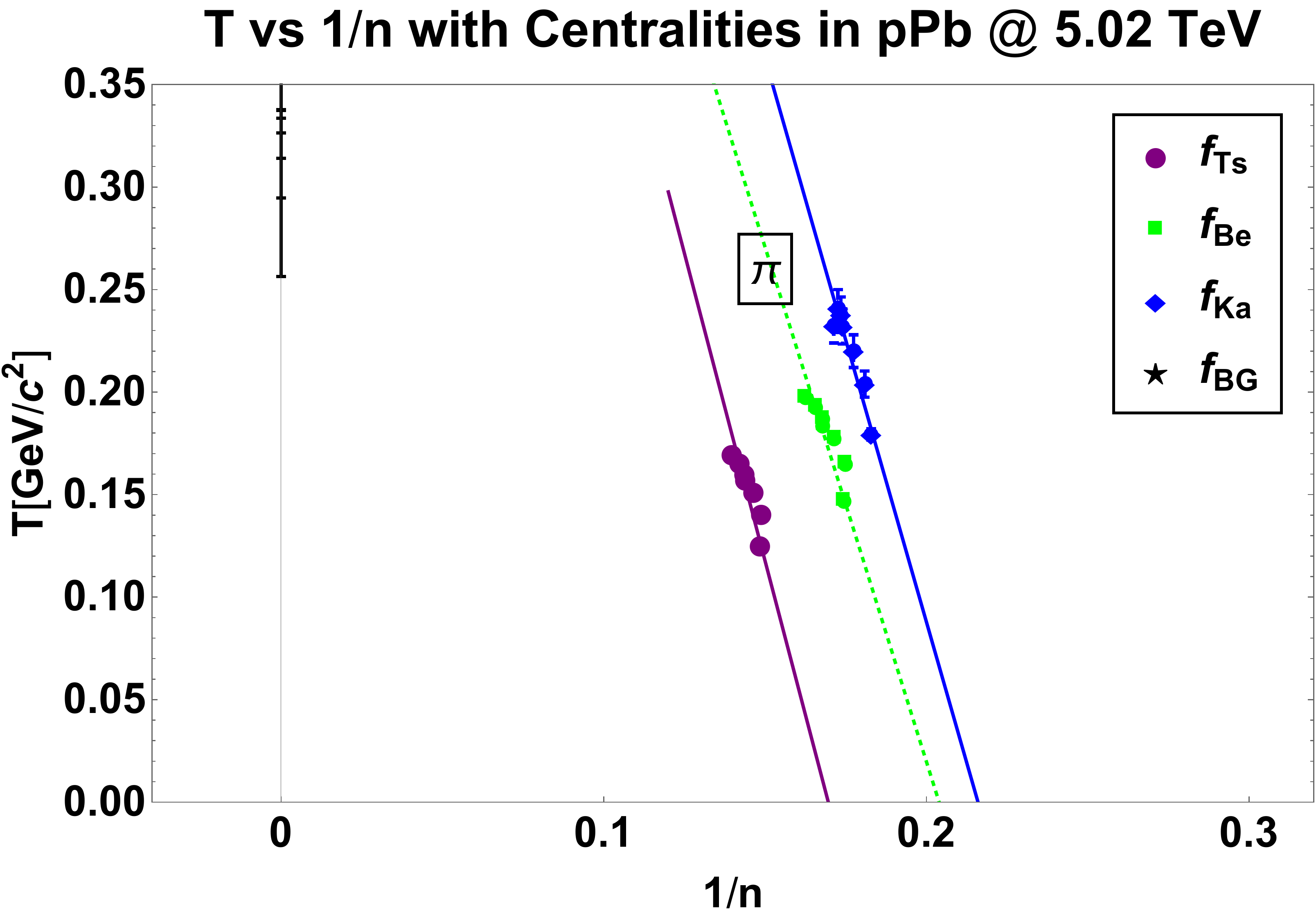}
\includegraphics[width=0.45\linewidth]{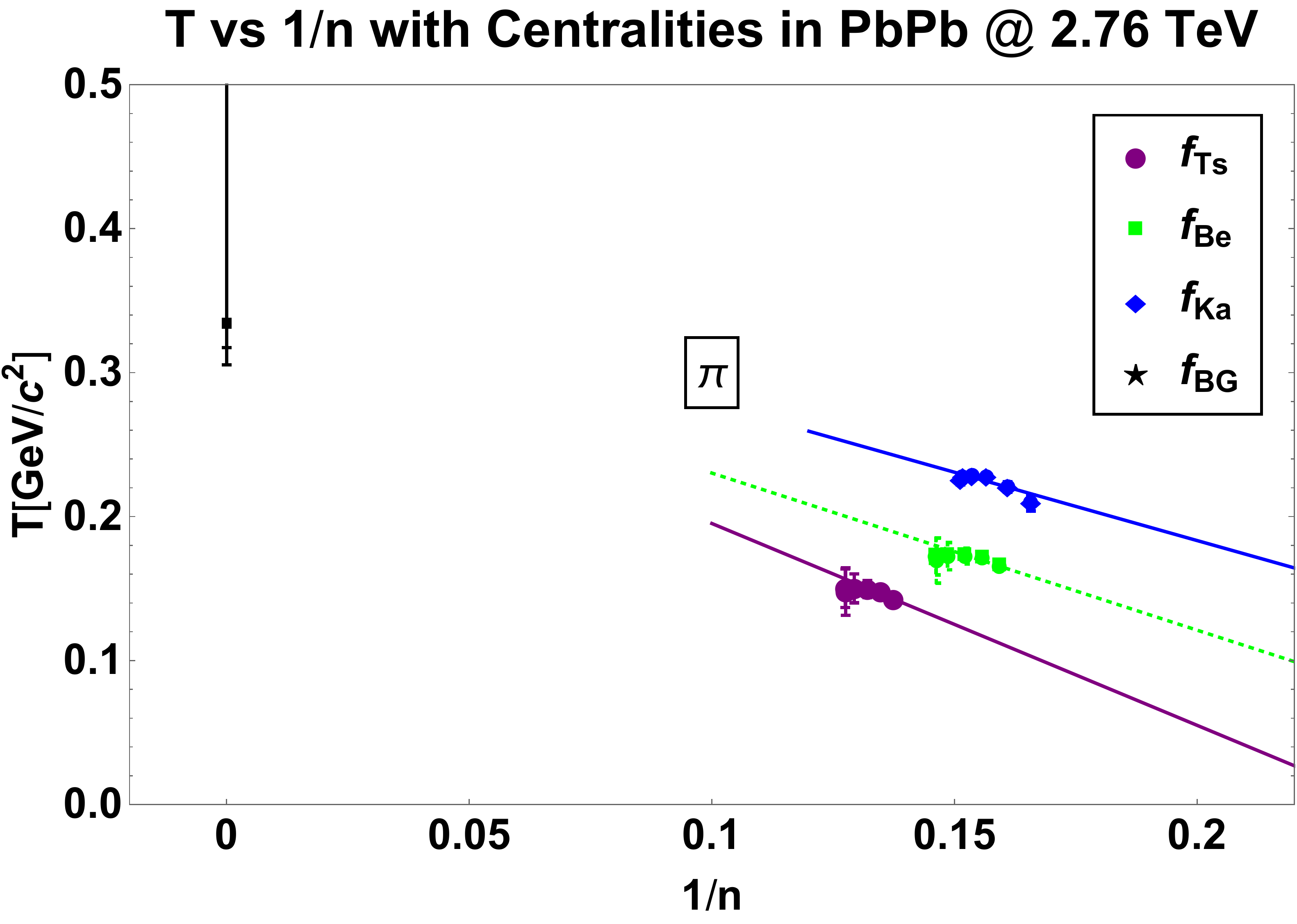}
}
\scalebox{1}[1]{
\includegraphics[width=0.45\linewidth]{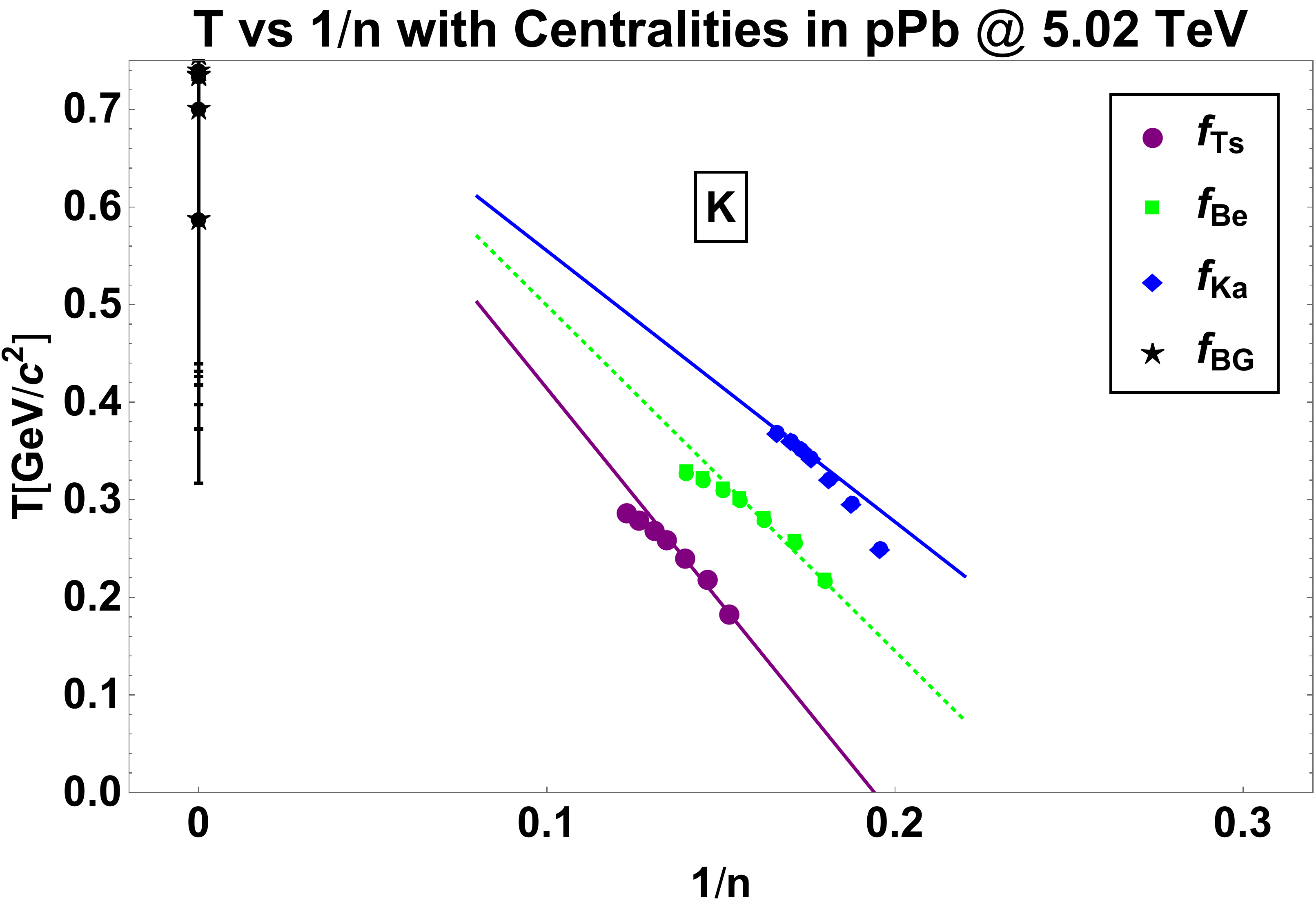}
\includegraphics[width=0.45\linewidth]{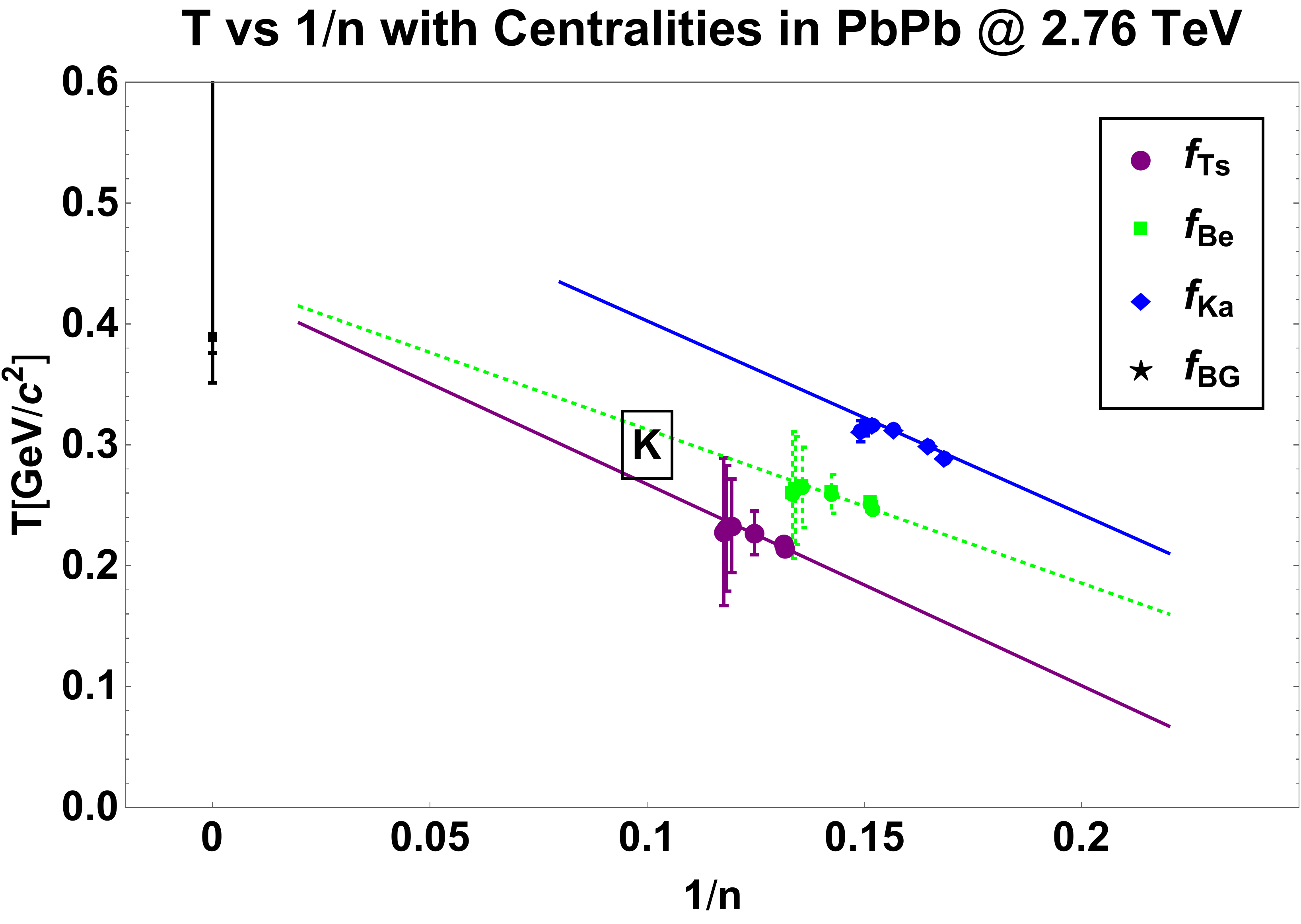}
}
\scalebox{1}[1]{
\includegraphics[width=0.45\linewidth]{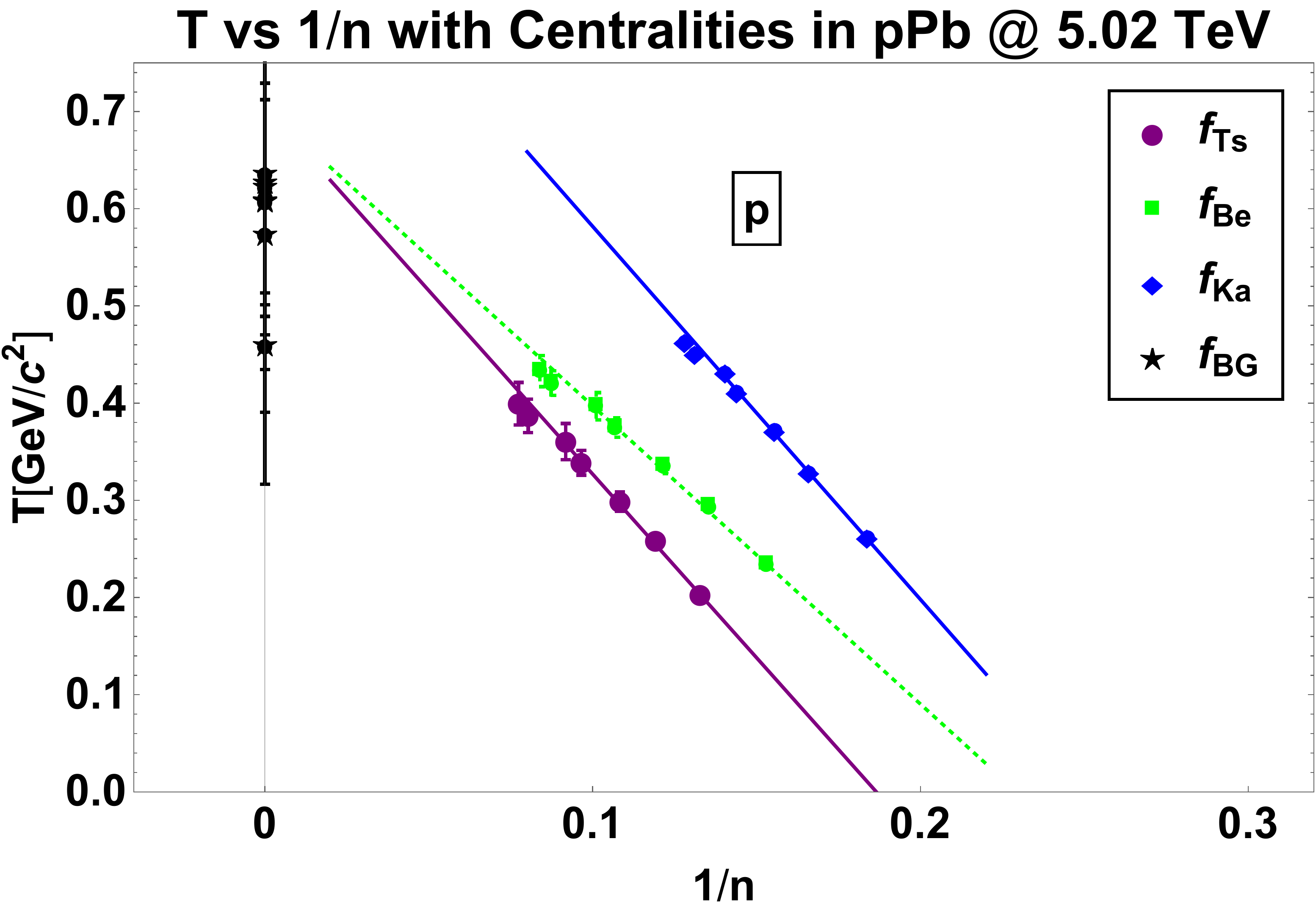}
\includegraphics[width=0.45\linewidth]{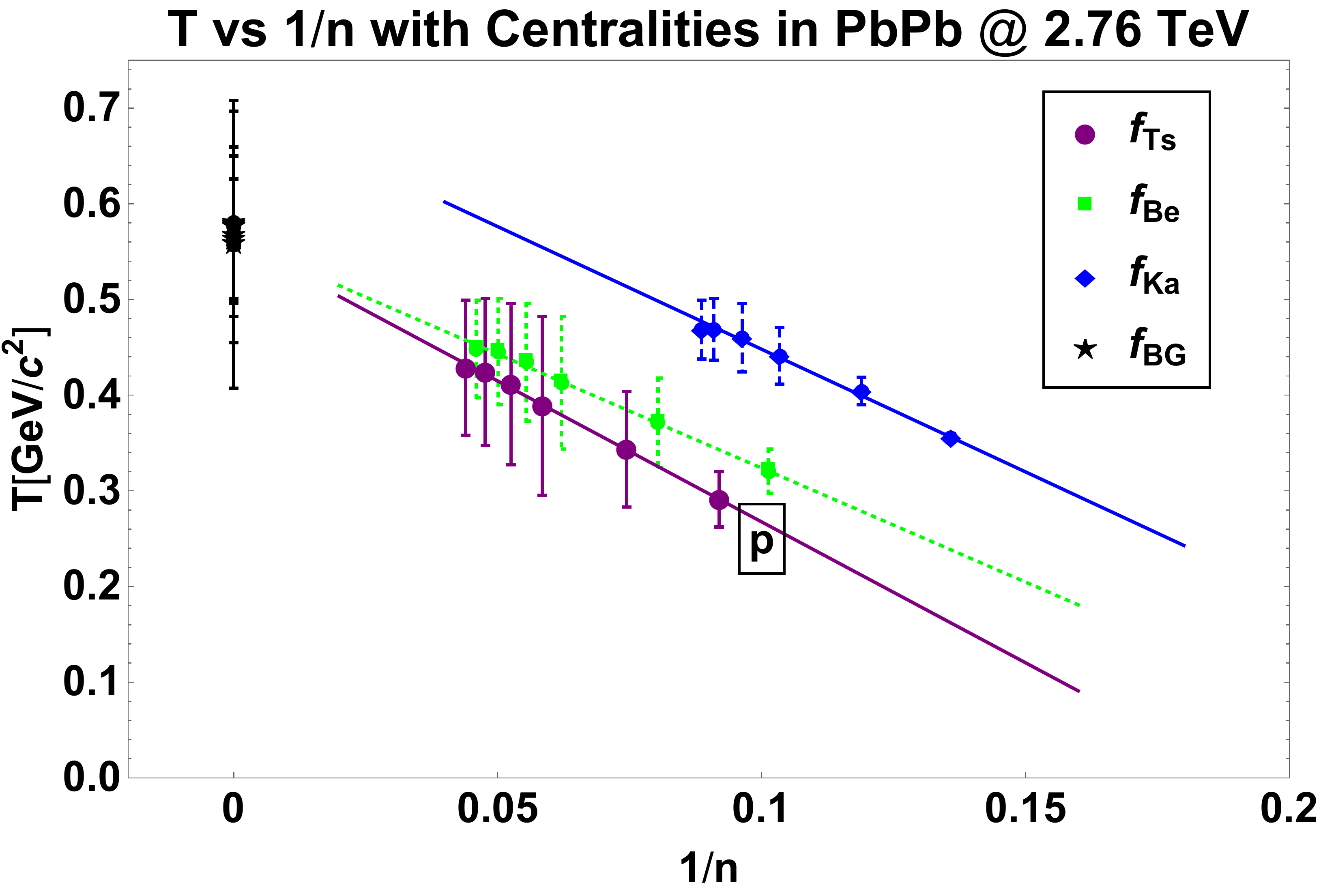}
}
\caption{Dependences of the inverse slope parameter, $T$, on the non-extensive parameter, $1/n$, for kinds of charged particles ($\pi$, $K$ and $p$) in $pPb$ (left) and $PbPb$ (right) collisions at 5.02 and 2.76 TeV respectively with all centralities. The black stars of $1/n=0$ are for the BG results by $f_{BG}$. We could see the obvious linear combination for each fitting formula.}
\label{figTq2}
\end{figure}

\begin{figure}[htb]
\scalebox{1}[1]{
\includegraphics[width=0.45\linewidth]{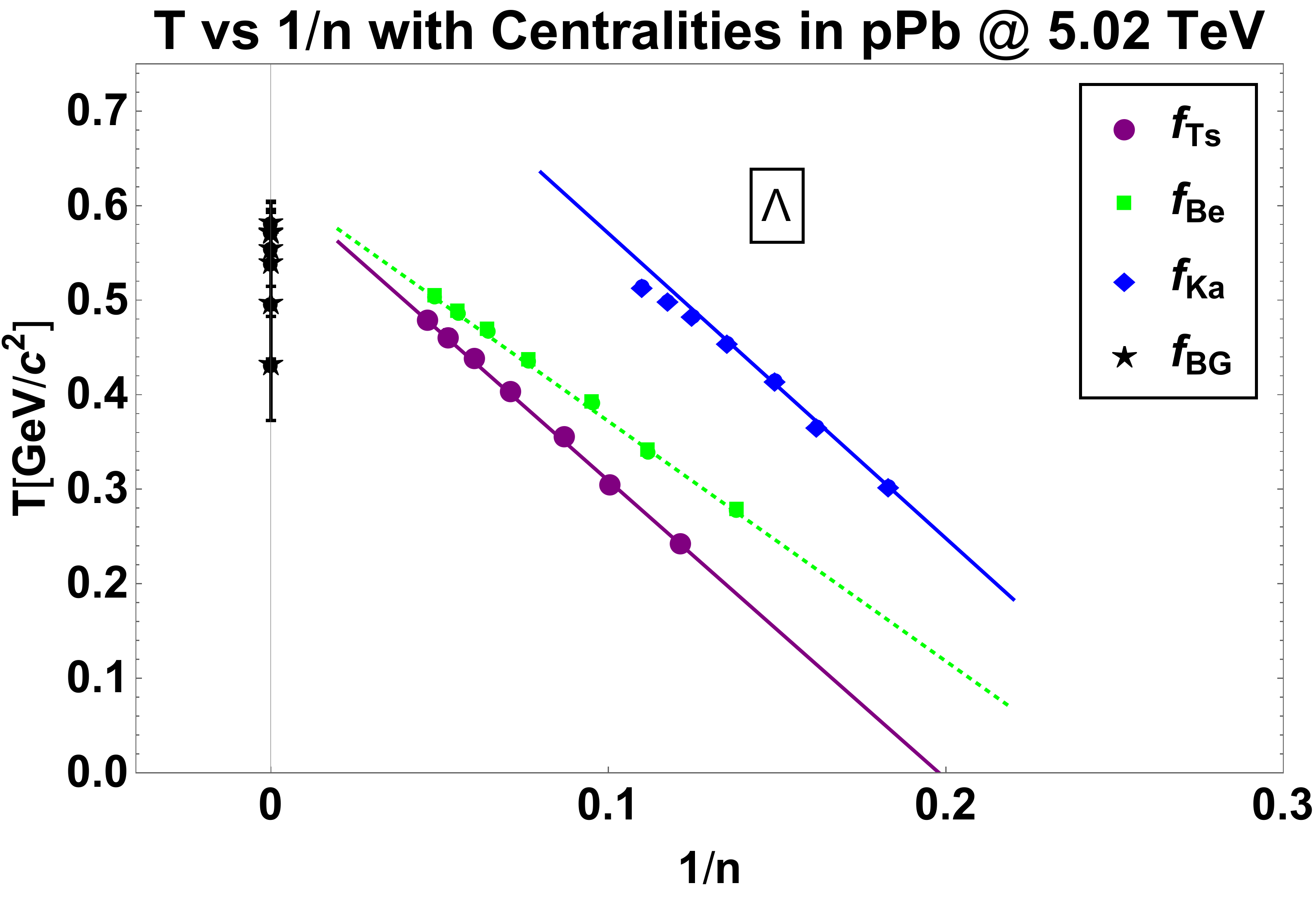}
\includegraphics[width=0.45\linewidth]{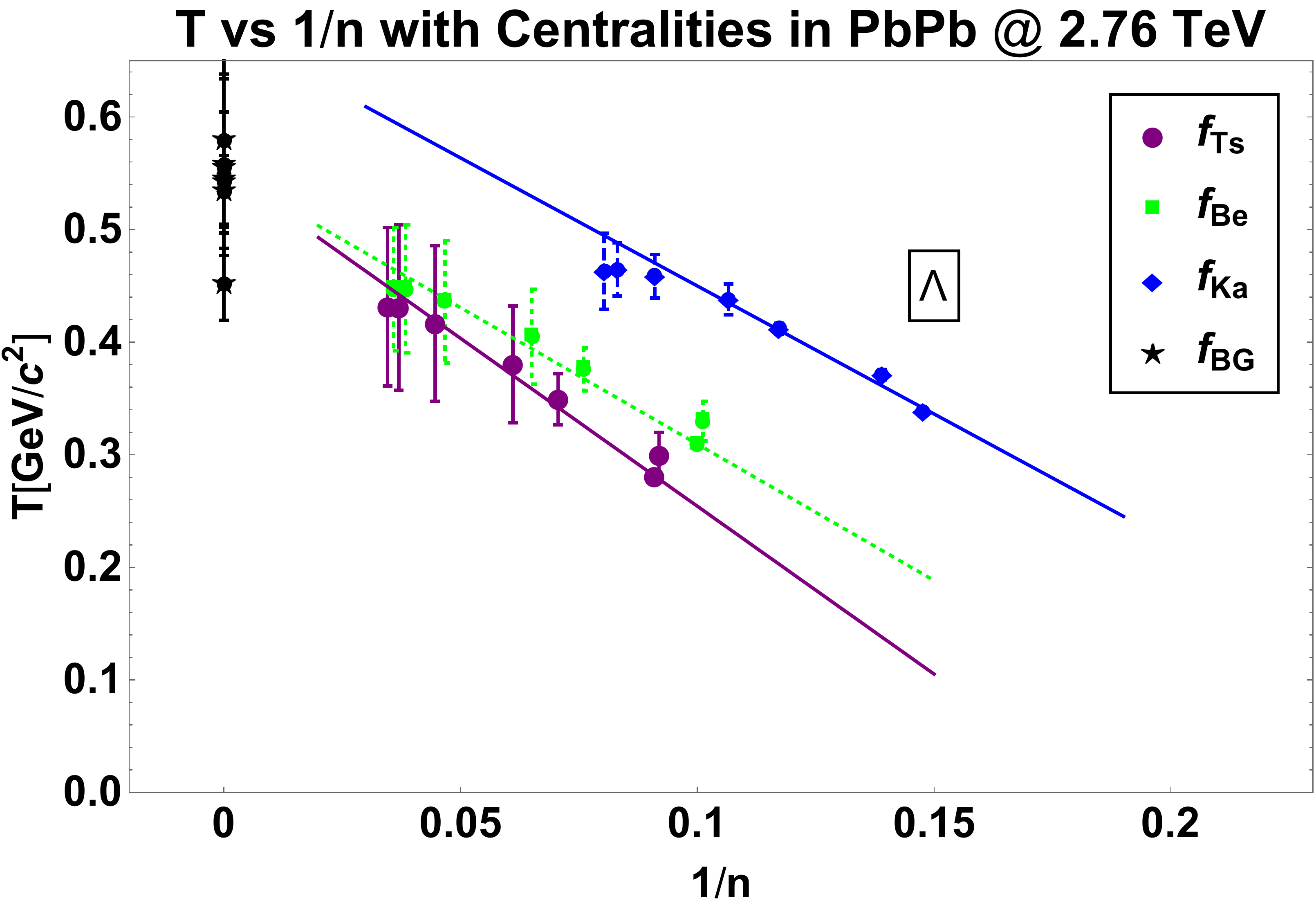}
}
\scalebox{1}[1]{
\includegraphics[width=0.45\linewidth]{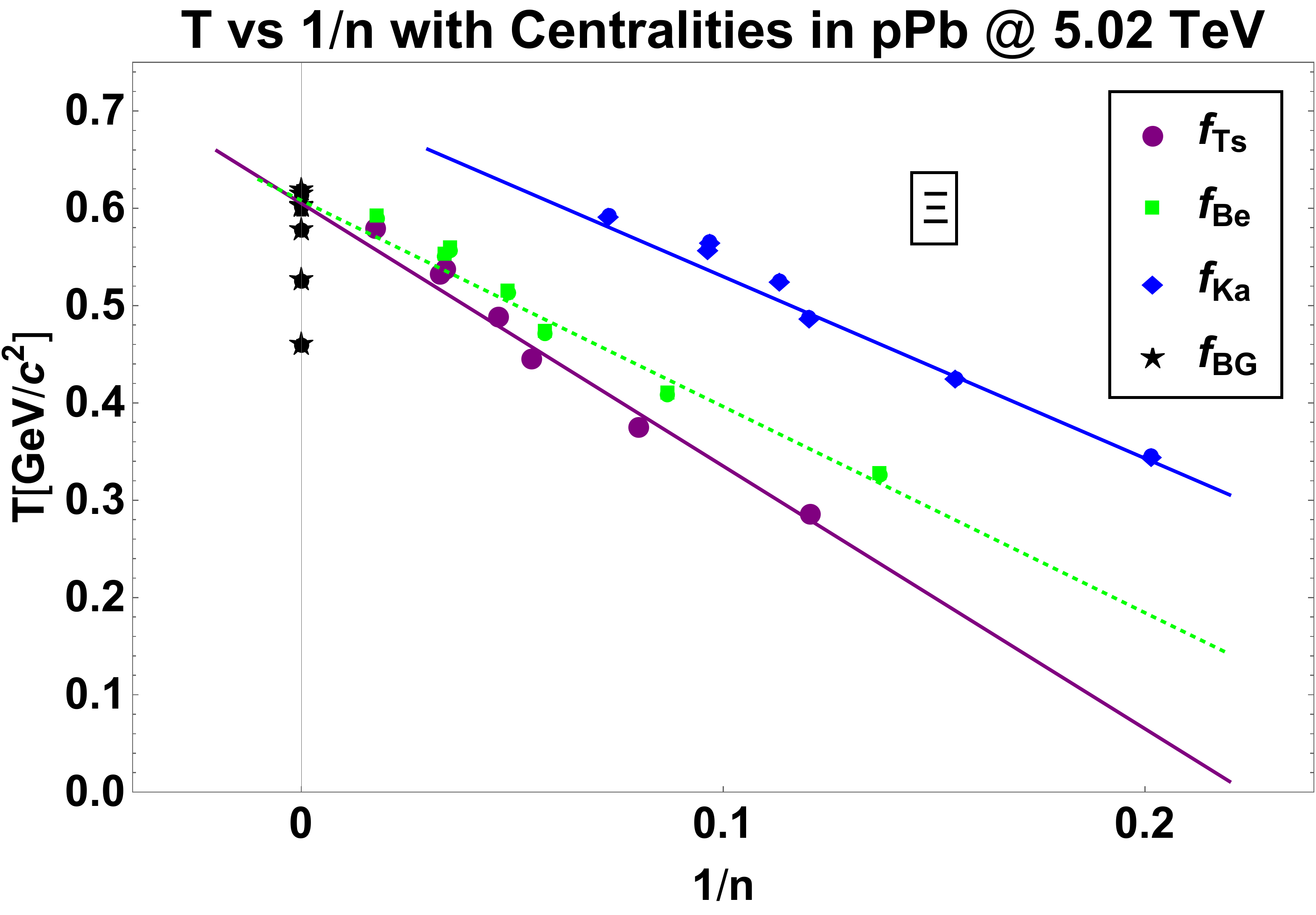}
\includegraphics[width=0.45\linewidth]{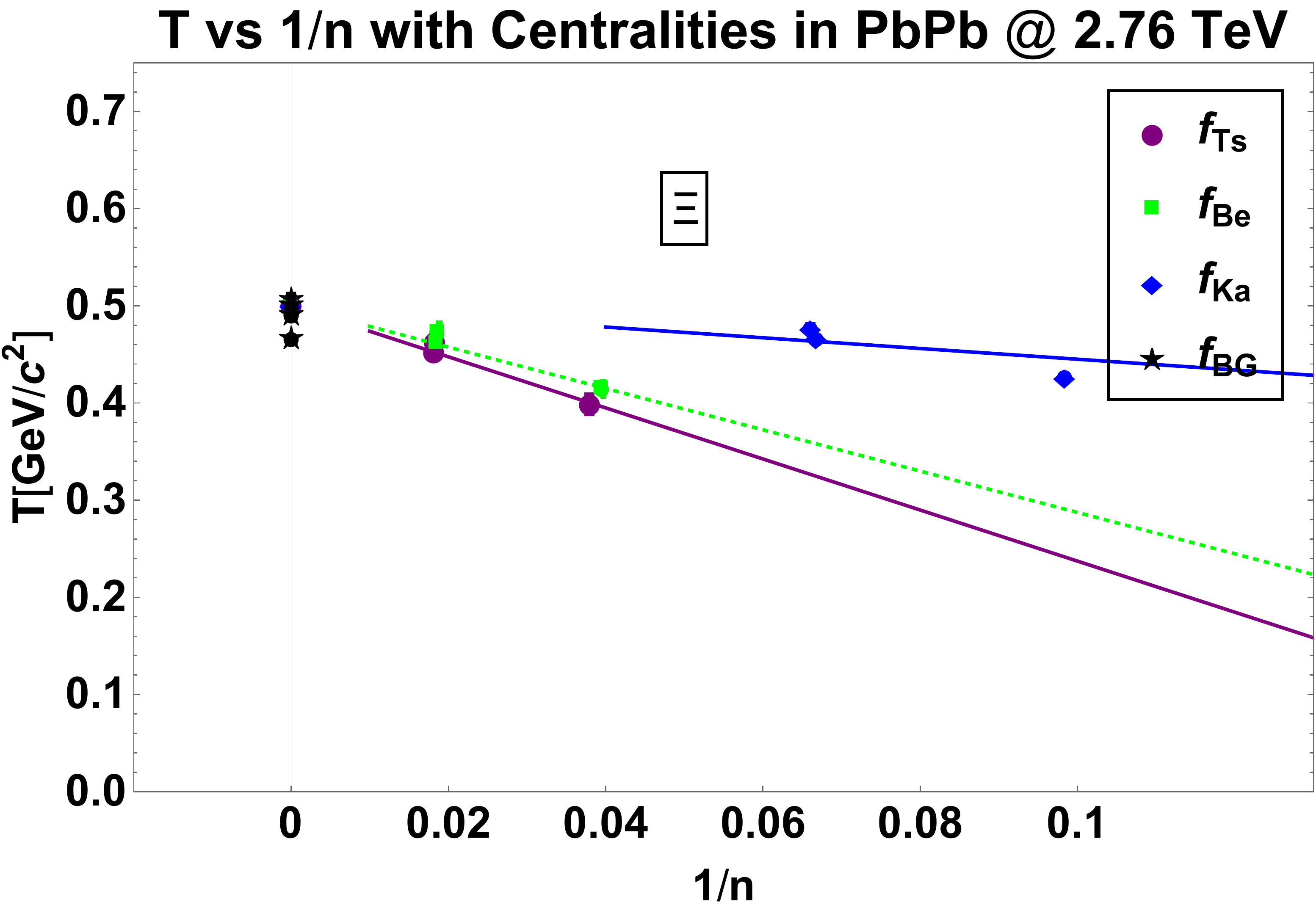}
}
\scalebox{1}[1]{
\includegraphics[width=0.45\linewidth]{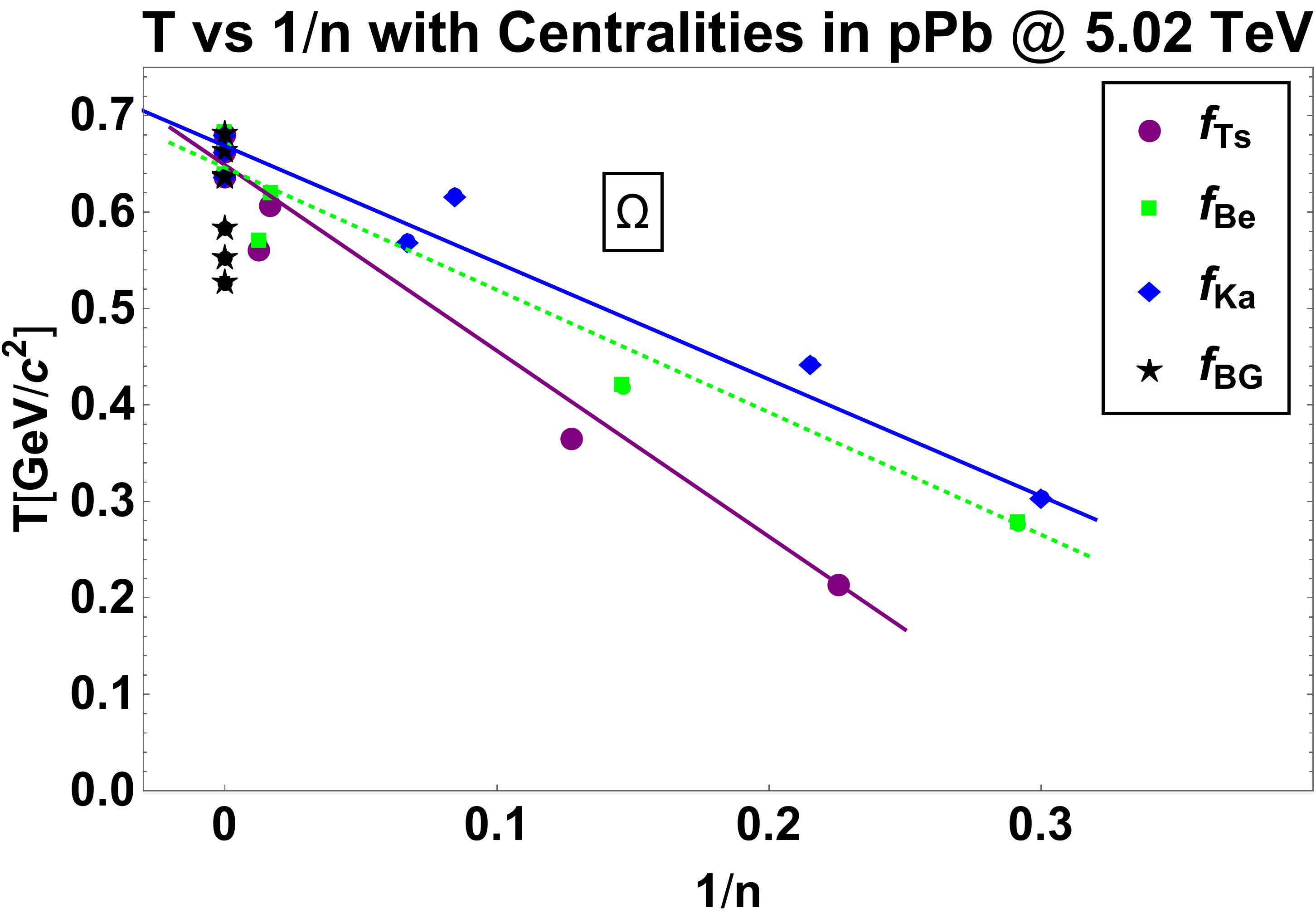}
\includegraphics[width=0.45\linewidth]{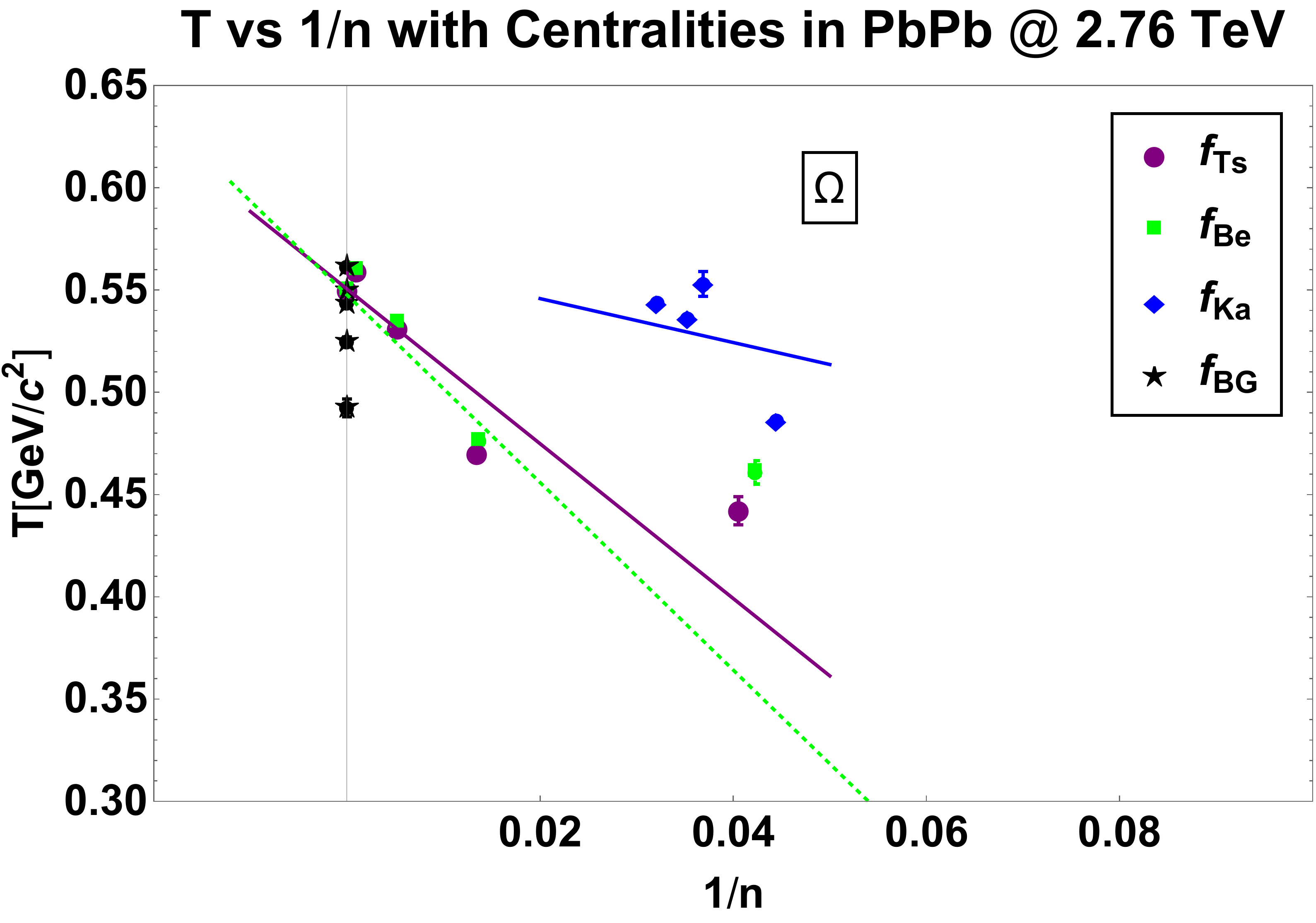}
}
\caption{Dependences of the inverse slope parameter, $T$, on the non-extensive parameter, $1/n$, for kinds of charged particles ($\Lambda$, $\Xi$ and $\Omega$) in $pPb$ (left) and $PbPb$ (right) collisions at 5.02 and 2.76 TeV respectively with all centralities. The black stars of $1/n=0$ are for the BG results by $f_{BG}$. We could see the obvious linear combination for each fitting formula, not for the results of $\Omega$ spectra because of the low number of degree of freedom.}
\label{figTq3}
\end{figure}


In this section, we present the similar researches in heavy-ion collisions. 
All the measured $p_T$ spectra are taken into account of identified charged particles stemming from the $pPb$ collision at 5.02 TeV \cite{pA-1, pA-2, pA-3} and the $PbPb$ collision at 2.76 TeV \cite{AA-1, AA-2, AA-3, AA-4} from the ALICE Collaboration. 
The measured $p_T$ regions for various kinds of hadron spectra are tabulated in Table \ref{tabAApT}.
All centrality bins of data are analyzed for different kinds of charged (anti-)particles: $\pi^\pm$, $K^\pm$, $p(\bar{p})$, $\Lambda(\bar{\Lambda})$, $\Xi^\pm$ and $\Omega^\pm$.

From Fig.\ref{figpT3} to Fig.\ref{figpT8} the transverse momentum spectra of these particles are shown to exhibit a power-law spectral shape at the given $p_T$ range.
Specifically, the usual BG statistics only works well in the small $p_T$ range, $0<p_T<3$ GeV/c.
While all the non-extensive approaches can obtain quite good fitting results over the whole $p_T$ range seen in Table \ref{tabAApT}. 
Without loss of generality, we list the fitting spectra of the most central collisions ($0\sim 5\%$) and most peripheral ones ($60\sim 80\%$) in both $pPb$ and $PbPb$ collisions.
In Tables \ref{tabpApar} and \ref{tabAApar} we present all the fitting parameters, $A_1 \sim A_4$, $T_1\sim T_4$ and $n_1 \sim n_4$, obtained by fitting on the $p_T$ spectra of various hadrons in both $pPb$ and $PbPb$ collisions.
Similar to the case in $pp$ results, we set the Boltzmann\,--\,Gibbs non-extensive parameter $n_4=10^{10}$ since it should be infinity as a matter of fact.
We could see that the first two non-extensive distributions indeed share the same values of the normalization constant, $A_1=A_2$, and the non-extensive parameter, $n_1=n_2+1$, as discussed in $pp$ case.
Note that the results turn to be no big differences for the fittings on data of $\Omega$ in $pPb$ collisions and $\Xi$ and $\Omega$ in $PbPb$ collisions in the centrality bins of $0 \sim 5\%$.
This is probably due to the fact that in this case there are larger multiplicities in the system and it seems to be better described by an exponential function.

The corresponding $\chi^2/ndf$ values are plotted in Fig.\ref{figxi2} and Fig.\ref{figxi3} and listed in Tables \ref{tabpAf} and \ref{tabAAf}.
For all hadron species, we could see that these non-extensive functions, $f_{Ts}$, $f_{Be}$ and $f_{Ka}$, of Eq.(\ref{fit-fun}) do fit the $p_T$ spectra better than the classical BG distribution. 
Lower values of $\chi^2/ndf$ are obtained with the more peripheral collisions and heavier hadrons.
This is probably because of the fact that lower multiplicities and larger masses of particles make it more universal to apply the non-extensive statistics which departs from the classical equilibrium case. 
Note that the usual BG distribution obtained small values of $\chi^2/ndf$ when fitting the spectra of $\Lambda$, $\Xi$ and $\Omega$ in both $pPb$ and $PbPb$ collisions.
It better describes the spectra in $PbPb$ collisions for peripheral collisions than the central ones.
This is somehow acceptable because there exist two different regimes of particle production in heavy-ion collisions:
one is the soft multiparticle production dominant at low transverse momenta, where the spectrum reveals an almost exponential shape close to the BG results;
at high $p_T$, $p_T>3$ GeV/c, on the other hand, they display power-law tails due to the pQCD evolution.
For the $p_T$ spectra of $\Lambda$, $\Xi$ and $\Omega$ in this work, we focus on the small $p_T$ regions because of the limitation of data source.
Even the usual BG statistics, consequently, works well. 
Different from the results in $pp$ collisions, the Kaniadakis $\kappa$-distribution seems to have the minimal values of $\chi^2/ndf$ in heavy-ion collisions typically when the heavier particles are studied such as $\Lambda$, $\Xi$ and $\Omega$.

As shown in Fig.\ref{figTq2} and Fig.\ref{figTq3}, we also analyzed the dependence of the fitting temperature $T$ on the non-extensive parameter $1/n$ for all centralities of each particle in heavy-ion collisions. 
All fitting results obtained by the three non-extensive fitting functions are extended to hold the linear combination of these two parameters with all the centrality bins.
Note that the results obtained by the usual BG distributions are also listed for comparisons but all its non-extensive parameters are vanished, namely $1/n_{BG}=0$. 
This is why all the relations between the temperature $T$ and the non-extensive parameter $1/n$ nearly share the same limiting values for $1/n\to 0$ for the same hadron spectra with different centralities. 
On the other hand, it means that all these three non-extensive approaches indeed share some common properties, which will in turn helps understand them better and deeper.
Please note that a bad linear connection occurs to the case of $\Omega$ in Fig.\ref{figTq3} probably because in these cases the number of degree of freedom (NDF) is quite small, as shown in Tables \ref{tabpAf} and \ref{tabAAf}.

\section{Summary}

In this study we firstly propose to apply the Kaniadakis non-extensive statistics on the transverse momentum spectra of both positive and negative particles in high energy collisions with respect to particle-hole symmetry in quantum statistics.
The $p_T$ spectra of identified charged hadrons in both $pp$ collisions and heavy-ion collisions at various beam energies are studied by this generalized $\kappa$-distribution as well as the previous Tsallis $q$-exponential distribution and its generalized Beck one. 
Compared with the classical Boltzmann\,--\,Gibbs distribution, it is really presented that all of these non-extensive approaches are better applied in the researches on hadron spectra in high energy physics.

In particular, the $\kappa$-exponential function gives out the best $\chi^2/ndf$ for the $p_T$ spectra fittings of heavier particles, $\Lambda$, $\Xi$ and $\Omega$. 
On the other hand, it needs more attention to account for the particle-hole symmetry when studying the systems in heavy-ion collisions.
Since the Tsallis $q$-exponential, cf. Eq.(\ref{q-exp}) fails in this case, the Kaniadakis $\kappa$-exponential statistics also leads to the best fit goodness for all kinds of hadron spectra in heavy-ion collisions where the quantum systems are analyzed within both positive and negative particles. 
Finally we demonstrate the corresponding fitting temperature $T$ performs an almost linear connection with the non-extensive parameter $1/n$ ($1/n=q-1$ for the Tsallis and Beck distributions and $1/n=\kappa$ for the Kaniadakis one).
Note that the linear connections are obtained from different mechanisms: in $pp$ collisions, we analyze the data points for all kinds of hadron species but in the same beam energy; while in $pPb$ and $PbPb$ collisions, parameters are collected from the fitting results of same hadron spectra but different centralities.
To some extent this means these non-extensive approaches have the common properties which will conversely promote the further researches on the non-extensivity of not only the Tsallis statistics but also the Kaniadakis one.

We think that a complete understanding of the physics of non-extensive parameters in these collisions should be solved in general context. 
In this paper we have tried to illuminate it from the dependences of the fitting parameter $T$ on the non-extensive parameter $q$ or $\kappa$. 
These three different non-extensive approaches seem to share the same combinations of the fitting temperature $T$ and the non-extensive parameter $q(\kappa)$, which reminds us to re-think of the physics behind it.
It is then necessary to find out the deeper connections between them and explain what is the physical temperature of the system considered.
Our model, on the other hand, successfully provides another method to investigate on problems met in high energy collisions, even for the case without negative particles considered.
To figure out physical information carried, next we will put emphasis on these fitting parameters. 
More complex systems need to be studied within this $\kappa$-exponential distribution as well as its theoretical researches.

\begin{acknowledgments}

The author hereby shows the best acknowledgments for the Wigner Research Centre for Physics of Hungarian Academy of Sciences, where most of this work was done.
This work has been supported by the Hungarian National Research, Development and Innovation Office (NKFIH) under the contract numbers OTKA K120660 and K123815.
The author also gratefully thanks the funding for the Doctoral Research of ECUT.

\end{acknowledgments}


\end{document}